\documentclass{article}

\usepackage[english]{babel}

\usepackage[a4paper,top=2cm,bottom=2cm,left=3cm,right=3cm,marginparwidth=2cm]{geometry}

\usepackage{setspace}
\usepackage{amsmath}
\usepackage{graphicx}
\usepackage{setspace}
\usepackage{multirow}
\usepackage{longtable}
\usepackage[english]{babel}
\usepackage[utf8]{inputenc}
\usepackage{algorithm}
\usepackage[noend]{algpseudocode}
\usepackage{csquotes}
\usepackage{subcaption}
\usepackage{graphicx}
\usepackage{tikz}
\usepackage{lipsum}
\usepackage{tcolorbox}
\usepackage{tabularx}
\usepackage{booktabs}
\usepackage{listings}
\usepackage{titlesec}
\usepackage{multirow}
\usepackage{amssymb}
\usepackage{xcolor}
\titleformat{\paragraph}
{\normalfont\normalsize\bfseries}{\theparagraph}{1em}{}
\titlespacing*{\paragraph}
{0pt}{3.25ex plus 1ex minus .2ex}{1.5ex plus .2ex}

\newlength{\RoundedBoxWidth}
\newsavebox{\GrayRoundedBox}
   {\setlength{\RoundedBoxWidth}{\dimexpr#1}
    \begin{lrbox}{\GrayRoundedBox}
       \begin{minipage}{\RoundedBoxWidth}}%
   {   \end{minipage}
    \end{lrbox}
    \begin{center}
    \begin{tikzpicture}%
       \draw node[draw=black,fill=black!10,rounded corners,%
             inner sep=2ex,text width=\RoundedBoxWidth]%
             {\usebox{\GrayRoundedBox}};
    \end{tikzpicture}
    \end{center}}

\usepackage{placeins} 
\newcolumntype{Y}{>{\raggedleft\arraybackslash}X}

\usepackage[colorlinks=true, allcolors=blue]{hyperref}

\usepackage{amsfonts}
\usepackage[citestyle=bwl-FU]{biblatex}
\graphicspath{ {./graphs/} }

\newcounter{noteYZctr} 

\newcounter{noteMCctr} 

\newcounter{noteASctr} 

\newcounter{noteSZctr} 

\title{ClusterLOB: Enhancing Trading Strategies by \\Clustering Orders in Limit Order Books}
\author{Yichi Zhang$^{1,2}$, Mihai Cucuringu$^{3,1,2}$, Alexander Y. Shestopaloff$^{4,5}$, Stefan Zohren$^{6}$\\}

\date{
{\fontsize{11.4}{11}\selectfont
\textit{$^{1}$Department of Statistics, University of Oxford\\
        $^{2}$Oxford-Man Institute of Quantitative Finance, University of Oxford\\
        $^{3}$Department of Mathematics, University of California Los Angeles\\
        $^{4}$School of Mathematical Sciences, Queen Mary University of London\\
        $^{5}$Department of Mathematics and Statistics, Memorial University of Newfoundland\\
        $^{6}$Machine Learning Research Group, Department of Engineering, University of Oxford}}
}

\begin{document}

\maketitle
\begin{center}
\textbf{Abstract}
\end{center}

In the rapidly evolving world of financial markets, understanding the dynamics of limit order book (LOB) is crucial for unraveling market microstructure and participant behavior. We introduce \texttt{ClusterLOB}\footnote{The source code for \texttt{ClusterLOB} is available on GitHub: https://github.com/YichiZhang-Oxford/ClusterLOB} as a method to cluster individual market events in a stream of market-by-order (MBO) data into different groups. To do so, each market event is augmented with six time-dependent features. By applying the K-means++ clustering algorithm to the resulting order features, we are then able to assign each new order to one of three distinct clusters, which we identify as directional, opportunistic, and market-making participants, each capturing unique trading behaviors. Our experimental results are performed on one year of MBO data containing small-tick, medium-tick, and large-tick stocks from NASDAQ.
To validate the usefulness of our clustering, we compute order flow imbalances across each cluster within 30-minute buckets during the trading day. We treat each cluster's imbalance as a signal that provides insights into trading strategies and participants' responses to varying market conditions. To assess the effectiveness of these signals, we identify the trading strategy with the highest Sharpe ratio in the training dataset, and demonstrate that its performance in the test dataset is superior to benchmark trading strategies that do not incorporate clustering. We also evaluate trading strategies based on order flow imbalance decompositions across different market event types, including add, cancel, and trade events, to assess their robustness in various market conditions. This work establishes a robust framework for clustering market participant behavior, which helps us to better understand market microstructure, and inform the development of more effective predictive trading signals with practical applications in algorithmic trading and quantitative finance.


\newpage
\tableofcontents
\newpage

\section{Introduction}

Financial markets are complex systems composed of diverse participants, each employing distinct trading strategies that shape market microstructure and liquidity dynamics. Understanding the behavior of these participants is essential for unraveling the underlying mechanisms that drive price formation, order flow dynamics, and market stability. The LOB, a continuously evolving record of all outstanding buy and sell orders, serves as a crucial lens for analyzing these interactions. By examining LOBs, traders, researchers, and regulators can gain insights into market efficiency, trading strategies, and the mechanisms that govern liquidity provision and price discovery.

Traditional approaches to market microstructure analysis primarily focus on aggregated trade data or summary statistics of order flow. However, the emergence of MBO data, which captures every individual market order and its attributes, provides a more granular view of trading behavior. MBO data enables a deeper understanding of how different market participants add, cancel, and trade orders, facilitating the classification of traders based on real-time execution patterns. However, its sheer volume and complexity pose significant challenges for analysis. To address these challenges, researchers have increasingly turned to machine learning techniques. For example, using client order flow data from a large broker, \cite{cont2023analysis} analyzed the properties of variables within this representation. The heterogeneity of the order flow was captured by segmenting clients into distinct clusters, each represented by a prototype. In addition, several studies [\cite{zhang2019deeplob, zhang2021mbo, kolm2023deep, kolm2024improving, briola2024deep}] have employed deep learning methods to extract meaningful patterns and insights from this high-dimensional data.

We introduce \texttt{ClusterLOB} to examine execution behaviour of different market participants using one year of MBO data for a selection of small-tick, medium-tick, and large-tick stocks, where each market order is characterized by six time-dependent features. These features capture variations in execution behavior across different market conditions, enabling a structured analysis of trading dynamics. To classify distinct trading behaviors, we apply the K-means++ clustering algorithm, segmenting daily MBO data into three clusters, which we interpret as directional, opportunistic, and market-making traders, as described in \cite{cartea2025statistical}. Directional traders execute informed trades with significant market impact, opportunistic traders react to short-term price fluctuations, and market-making traders provide liquidity while minimizing inventory risk.

To quantify the impact of these trading behaviors, we compute order flow imbalances across different clusters within 30-minute buckets, generating alpha signals that provide insights into market participants' trading strategies and responses to changing conditions. To evaluate the effectiveness of these signals, we identify the best trading strategy with the highest Sharpe ratio (SR) in the training dataset and compare its performance against benchmark trading strategies that do not incorporate clustering in the test dataset. The results show that the best clustering-driven trading strategy outperforms benchmark trading strategies, leading to improved trading performance and highlighting its predictive power. We also evaluate trading strategies based on order flow imbalance decomposition across different market event types, including add, cancel, and trade events, to assess its robustness in various market conditions.

This study explores both contemporaneous market dynamics, analyzing real-time participant interactions, and forward-looking scenarios, evaluating how cluster-specific behaviors influence future price movements. To the best of our knowledge, this work is the first to establish a robust framework for clustering participant behavior in MBO data, that does not rely on any proprietary data sources (such as exchange or broker's data),  enhancing the understanding of market structure and informing the development of more effective predictive trading strategies with practical applications in algorithmic trading and quantitative finance.

Our main contributions are summarized as follows.

\begin{tcolorbox}
\textbf{Summary of main contributions}.
\begin{enumerate}
    \item 
This study analyzes market participant behavior by applying K-means++ clustering to one year of MBO data for a selection of small-tick, medium-tick, and large-tick stocks. 
    \item 
We classify market participant behaviors into directional, opportunistic, and market-making traders using six time-dependent features, capturing distinct execution patterns in financial markets.

    \item 
By computing order flow imbalances across different clusters within 30-minute buckets, we extract alpha signals that provide actionable insights into trading strategies and
participants' responses.
    \item 
We identify the best trading strategy with the highest SR in the training dataset, and demonstrate that its performance outperforms benchmark trading strategies
that do not incorporate clustering in the test dataset. 
    \item
We evaluate trading strategies based on order flow imbalances decomposition across different market event types, including add, cancel, and trade events, to assess their robustness in various market conditions.

\end{enumerate}

\end{tcolorbox}

\textbf{Paper outline}. This paper is organized as follows. 
Section \ref{sec: Background} provides an overview of the primary concepts, including the LOB, the order flow imbalance, and the order flow imbalance decomposition. In Section \ref{sec: Data_Methodology}, we describe the dataset used in this research, outline the data pre-processing steps, and introduce the independent and dependent variables, performance evaluation metrics, clustering model, and cluster-based trading strategy for the LOB. Section \ref{sec: Result} discusses the empirical results across different event types for small-tick, medium-tick, and large-tick stocks. Finally, we conclude by summarizing our findings and exploring potential directions for future research in Section \ref{sec: Conclusion}.


\section{Background} \label{sec: Background}

In Section \ref{sec: lob}, we provide the mathematical description of the LOB, following the definitions and most of the notation used in \cite{gould2013limit}. We introduce order flow imbalance (OFI) in Section \ref{sec: ofi}, and then decomposed order flow imbalance (Decomposed OFI) in Section \ref{sec: decomposed_ofi}.

\subsection{Limit order books (LOBs)} \label{sec: lob}

First, we introduce the vector \( x = (p_x, q_x, t_x) \) to define a buy and sell order, respectively.

\noindent\textbf{Definition}\quad \textit{An \textnormal{order} \( x = (p_x, q_x, t_x) \), submitted at \textnormal{time} \( t_x \) with \textnormal{price} \( p_x \) and \textnormal{size} \( q_x < 0 \) (\( q_x > 0 \)), is a commitment to \textnormal{\textbf{buy}} \textnormal{(\textbf{sell})} up to \(|q_x|\) units of the traded asset at a price no greater (less) than \( p_x \).}


Second, we introduce the resolution parameters of the LOB: the lot size \( \sigma \) and the tick size \( \pi \).

\noindent\textbf{Definition}\quad \textit{The \textnormal{lot size} \( \sigma \) of the \textnormal{LOB} represents the smallest quantity of the asset that can be traded.
}

All orders must have a \textnormal{size} \( q_x \) such that \( q_x \in \{\pm k\sigma \mid k \in \mathbb{N}^+
\} \), which means all orders must be in whole units of the lot size \( \sigma \), and these units can either be positive or negative, corresponding to selling or buying the asset, respectively.

\noindent\textbf{Definition}\quad \textit{The \textnormal{tick size} \( \pi \) of the \textnormal{LOB} represents the smallest allowable price difference between various orders.
}

All orders must be submitted with prices accurate to some precision \( \pi \). In this paper, we focus on the Nasdaq exchange, where the tick size \( \pi \) is \$0.01.
Next, we define the LOB \( \mathcal{L}(t) \), followed by the buy orders \( \mathcal{B}(t) \) and sell orders \( \mathcal{A}(t) \).

\noindent\textbf{Definition}\quad \textit{An \textnormal{LOB} \( \mathcal{L}(t) \) represents the set of all active orders in a market at \textnormal{time} \( t \), which can be divided into two sets: the set of active \textnormal{buy orders} \( \mathcal{B}(t) \), where \( q_x < 0 \), and the set of active \textnormal{sell orders} \( \mathcal{A}(t) \), where \( q_x > 0 \).
}

Next, we define the best bid price \( b^{1}(t) \) and the best ask price \( a^{1}(t) \), respectively.

\noindent\textbf{Definition}\quad \textit{The \textnormal{best bid price} \( b^{1}(t) \) at \textnormal{time} \( t \), defined as the highest price (level 1) among all active \textnormal{buy orders} \( \mathcal{B}(t) \) at \textnormal{time} \( t \), is given by
}
\begin{equation}
b^{1}(t) := \max\limits_{x\in\mathcal{B}(t)}p_x.
\end{equation}

\noindent\textbf{Definition}\quad \textit{The \textnormal{best ask price} \( a^{1}(t) \) at \textnormal{time} \( t \), defined as the lowest price (level 1) among all active \textnormal{sell orders} \( \mathcal{A}(t) \) at \textnormal{time} \( t \), is given by
}
\begin{equation}
a^{1}(t) := \min\limits_{x\in\mathcal{A}(t)}p_x.
\end{equation}

Subsequently, we establish the \textnormal{bid-side volume} $v^{b}(p,t)$ and the \textnormal{ask-side volume} $v^{a}(p,t)$, which are accessible at \textnormal{price} $p$ and at \textnormal{time} $t$, respectively.

\noindent\textbf{Definition}\quad \textit{The \textnormal{bid-side volume} $v^{b}(p,t)$ available at \textnormal{price} $p$ and at \textnormal{time} $t$, is given by
}
\begin{equation}
v^{b}(p,t) := \sum\limits_{\{x\in\mathcal{B}(t)|p_x = p\}}q_x.
\end{equation}

\noindent\textbf{Definition}\quad \textit{The \textnormal{ask-side volume} $v^{a}(p,t)$ available at \textnormal{price} $p$ and at \textnormal{time} $t$, is given by
}
\begin{equation}
v^{a}(p,t) := \sum\limits_{\{x\in\mathcal{A}(t)|p_x = p\}}q_x.
\end{equation}

\noindent\textbf{Definition}\quad \textit{The \textnormal{state} $\psi(t)$ of the \textnormal{LOB} \( \mathcal{L}(t) \) at \textnormal{time} $t$, defined as the vector containing information about the top 10 (nonempty) levels of ask and bid prices and volumes at \textnormal{time} $t$, is given by}
\begin{equation}
\psi(t) := [a^{1}(t), v^{a}(a^{1}(t),t), b^{1}(t), v^{b}(b^{1}(t),t),\ldots,a^{10}(t), v^{a}(a^{10}(t),t), b^{10}(t), v^{b}(b^{10}(t),t)] \in 	\mathbb{R}^{40}.
\end{equation}
Finally, the bid-ask spread \( s(t) \) and mid-price \( m(t) \) can be calculated from the best ask price \( a^{1}(t) \) and best bid price \( b^{1}(t) \).

\noindent\textbf{Definition}\quad \textit{The \textnormal{bid-ask spread} \( s(t) \) at \textnormal{time} \(t\), is given by
}

\begin{equation}
s(t) := a^{1}(t)-b^{1}(t).
\end{equation}



\noindent\textbf{Definition}\quad \textit{The \textnormal{mid-price} \( m(t) \) at \textnormal{time} \(t\), is given by
}

\begin{equation}
m(t) := \frac{a^{1}(t)+b^{1}(t)}{2}.
\end{equation}

\begin{figure}[!htbp]
\centering
\includegraphics[width=0.64\textwidth,trim=0cm 1.5cm 0cm 1.5cm,clip]{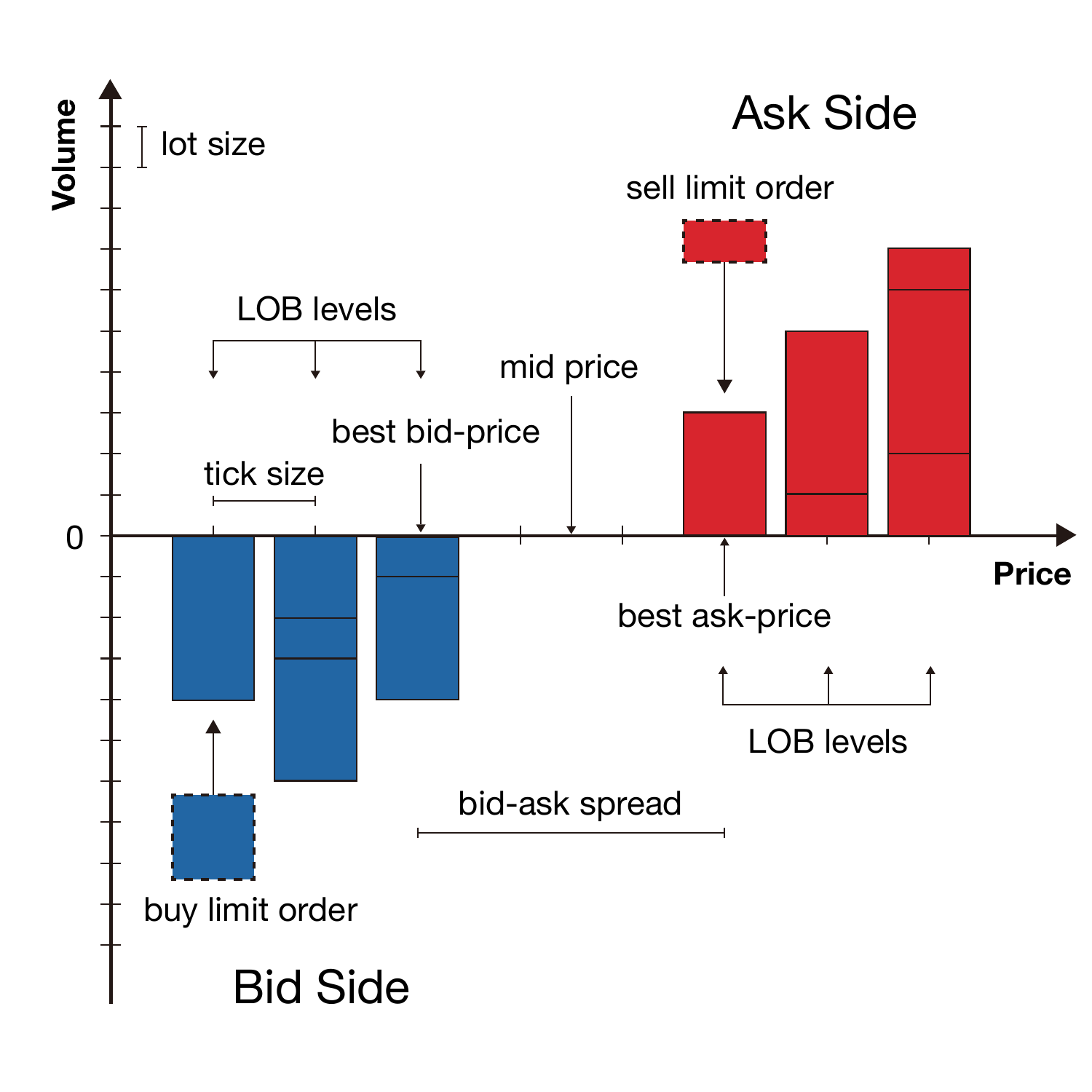}
\caption{Example of a snapshot of the LOB.}
\label{fig: lob}
\end{figure}

Figure \ref{fig: lob} presents a summarized overview of the previously discussed definitions of the LOB, highlighting its structure, key components, and the role of buy and sell limit orders in shaping market dynamics.

\subsection{Order flow imbalance} \label{sec: ofi}

\cite{cont2014price} demonstrated that the OFI, defined as the imbalance between supply and demand at the best bid and ask prices, mainly drives price changes over short time intervals. \cite{xu2018multi} built upon the work of \cite{cont2014price} by conducting a more detailed analysis of the relationship between net order flow at the top $M$ price levels on both sides of the LOB and the contemporaneous changes in the mid-price. Later, \cite{cont2023cross} combined the OFI from the top levels of the limit order book into an integrated OFI single variable that provides a more comprehensive explanation of price impact, in comparison to the best-level OFI. \cite{kolm2023deep} also illustrated that deep learning models trained on OFI exhibit superior performance compared to the majority of models trained directly on order books. Shortly after, \cite{kolm2024improving} predicted future mid-price stock returns across multiple time horizons for individual stocks using deep learning models, with OFI serving as the regression variable. We employ the same definitions of OFI as those in the initial study by \cite{cont2014price}. To further refine our analysis, we introduce two variations of OFI: Size-based OFI and Count-based OFI, defined as follows.




\noindent\textbf{Definition}\quad \textit{Within the \textnormal{time interval} $(t - h, t]$, the best-level \textnormal{Size-based OFI} of \textnormal{LOB} $\mathcal{L}(t)$, denoted by $OFI^{S}(\mathcal{L}(t))$, compares the cumulative OFIs on the order \textnormal{\textbf{size}} of the best ask and bid side over the time interval, is given by
}
%
%
\begin{equation} \label{ofi_s}
OFI^{S}(\mathcal{L}(t)):= L^{b,S}(t) - D^{b,S}(t) + M^{b,S}(t) - L^{a,S}(t) + D^{a,S}(t) - M^{a,S}(t).
\end{equation}

\noindent where $L^{a,S}(t)$ and $L^{b,S}(t)$ represent the total order size of sell and buy orders arriving at the current best ask and bid during the time interval $(t - h, t]$, respectively; $D^{a,S}(t)$ and $D^{b,S}(t)$ denotes the total order size of sell and buy orders cancelled from the current best ask and bid during the same interval, respectively; and $M^{a,S}(t)$ and $M^{b,S}(t)$ indicates the total order size of sell and buy orders traded to the current best ask and bid during the same interval, respectively.


\noindent\textbf{Definition}\quad \textit{Within the \textnormal{time interval} $(t - h, t]$, the best-level \textnormal{Count-based OFI} of \textnormal{LOB} $\mathcal{L}(t)$, denoted by $OFI^{C}(\mathcal{L}(t))$, compares the cumulative OFIs on the order \textnormal{\textbf{count}} of the best ask and bid side over the time interval, is given by
}
\begin{equation} \label{ofi_c}
OFI^{C}(\mathcal{L}(t)):= L^{b,C}(t) - D^{b,C}(t) + M^{b,C}(t) - L^{a,C}(t) + D^{a,C}(t) - M^{a,C}(t).
\end{equation}


\noindent where $L^{a,C}(t)$ and $L^{b,C}(t)$ represent the total order count of sell and buy orders arriving at the current best ask and bid during the time interval $(t - h, t]$, respectively; $D^{a,C}(t)$ and $D^{b,C}(t)$ denotes the total order count of sell and buy orders cancelled from the current best ask and bid during the same interval, respectively; and $M^{a,C}(t)$ and $M^{b,C}(t)$ indicates the total order count of sell and buy orders traded to the current best ask and bid during the same interval, respectively.

\subsection{Order flow imbalance decomposition} \label{sec: decomposed_ofi}

The order flow imbalance decomposition (Decomposed OFI), introduced by \cite{sitaru2023order}, integrates insights into order book event types with order flow information. This work resulted in a notable enhancement in forward-looking predictive scenarios, both statistically and economically. Therefore, we adopt similar ideas as presented in this line of work.

All order book events can be classified into one of the following event types:

\begin{itemize}
    \item \textbf{Submission (add):} Submitting a limit order, which is then placed in the corresponding price level in the LOB.
    \item \textbf{Deletion (cancel):} Deleting (partially or entirely) an existing order, which results in the removal of the order from the LOB.
    \item \textbf{Execution (trade):} Executing a trade, which happens when an ask order matches a bid order (either by a limit or market order), resulting in the removal of both orders from the LOB.
\end{itemize}

Then, we define the Decomposed OFI, which creates a distinct variable for each event type of LOB.

\noindent\textbf{Definition}\quad \textit{Within the \textnormal{time interval} \( (t - h, t] \), the \textnormal{\textbf{add (cancel, trade)}} OFI, denoted by \\$OFI^{S}(e(\mathcal{L}(t))=\textnormal{add})$ \textnormal{(}$OFI^{S}(e(\mathcal{L}(t))=\textnormal{cancel})$, $OFI^{S}(e(\mathcal{L}(t))=\textnormal{trade})$\textnormal{)}, is the sum of order \textnormal{\textbf{size}} of order flow contributions of total corresponding type of events over the time interval, respectively, is given by}
\begin{equation}
\begin{aligned}
&OFI^{S}(e(\mathcal{L}(t))=\textnormal{add}):= L^{b,S}(t) - L^{a,S}(t) , \\
&OFI^{S}(e(\mathcal{L}(t))=\textnormal{cancel}):= -D^{b,S}(t) + D^{a,S}(t) , \\
&OFI^{S}(e(\mathcal{L}(t))=\textnormal{trade}):= M^{b,S}(t) - M^{a,S}(t).
\end{aligned}
\end{equation}

\noindent\textbf{Definition}\quad \textit{Within the \textnormal{time interval} \( (t - h, t] \), the \textnormal{\textbf{add (cancel, trade)}} OFI, denoted by \\$OFI^{C}(e(\mathcal{L}(t))=\textnormal{add})$ \textnormal{(}$OFI^{C}(e(\mathcal{L}(t))=\textnormal{cancel})$, $OFI^{C}(e(\mathcal{L}(t))=\textnormal{trade})$\textnormal{)}, is the sum of order \textnormal{\textbf{count}} of order flow contributions of total corresponding type of events over the time interval, respectively, is given by}

\begin{equation}
\begin{aligned}
&OFI^{C}(e(\mathcal{L}(t))=\textnormal{add}):= L^{b,C}(t) - L^{a,C}(t) , \\
&OFI^{C}(e(\mathcal{L}(t))=\textnormal{cancel}):= -D^{b,C}(t) + D^{a,C}(t) , \\
&OFI^{C}(e(\mathcal{L}(t))=\textnormal{trade}):= M^{b,C}(t) - M^{a,C}(t).
\end{aligned}
\end{equation}

\noindent where $e(\cdot)$ is the indicator function of event type, and \( \text{e}(\mathcal{L}(t)) = \text{add} \) ($\text{e}(\mathcal{L}(t)) = \text{cancel}$, $\text{e}(\mathcal{L}(t)) = \text{trade}$) represents the set of all orders that are add (cancel, trade) events during the time interval \( (t - h, t] \), respectively, and the rest of the variables are defined in the same manner as before.

\section{Data and Methodology} \label{sec: Data_Methodology}

In Section \ref{sec: Data}, we describe the dataset used in this research and outline the data preprocessing steps. Section \ref{sec: Methodology} introduces the independent and dependent variables, the K-means++ clustering algorithm applied in our research, the cluster-based trading strategy, and performance evaluation metrics.

\subsection{Data} \label{sec: Data}
\subsubsection{Data description}

We gather our data through \href{https://lobsterdata.com/}{LOBSTER} by \cite{huang2011lobster}, which uses ITCH data from NASDAQ to reproduce the LOB for any stock on NASDAQ to any specified level. For each active trading day of a selected stock, LOBSTER generates two files: an 'orderbook' file and a 'message' file. The 'orderbook' file records the evolution of the LOB up to the specified number of levels. The 'message' file contains event indicators specifying the type of update that modifies the LOB within the requested price range. Each event is timestamped in seconds after midnight, with a decimal precision ranging from milliseconds to nanoseconds, depending on the selected time period. 

In particular, we analyze 15 stocks spanning six different sectors throughout the entire year of 2021, with the summary information of the stocks provided in Table \ref{tab: data}. The sector classification of stocks is proposed by \href{https://www.nasdaq.com/market-activity/stocks/screener}{NASDAQ}, while the stock's capitalization in 2021 is provided by \url{https://companiesmarketcap.com}. We categorize these stocks into three groups based on their tick size, in line with the work of [\cite{eisler2012price}, \cite{briola2024deep}, \cite{briola2024hlob}]. The \textit{small-tick} group comprises CHTR, GOOG, GS, IBM, MCD, and NVDA, with an average bid-ask spread $\bar{s} \gtrsim 3\pi$. The \textit{medium-tick} group consists of AAPL, ABBV, and PM, with the average bid-ask spread satisfying $1.5\pi \lesssim \bar{s} \lesssim 3\pi$. The \textit{large-tick} group includes CMCSA, CSCO, INTC, MSFT, KO, and VZ, with an average bid-ask spread $\bar{s}$ that meets the condition $\bar{s} \lesssim 1.5\pi$.

\begin{table}[htbp]

\centering
\caption{Summary of 15 stocks categorized by tick size along with their sector and market capitalization.}
\label{tab: data}
\resizebox{\linewidth}{!}{
\begin{tabular}{lllll}
\toprule
\textbf{Tick}&\textbf{Symbol} & \textbf{Name}& \textbf{Sector} & \textbf{Cap (2021)}\\
\midrule
\multirow{6}{*}{\rotatebox[origin=c]{90}{Small}}&
CHTR & Charter Communications, Inc.& Telecommunications & 	\$112.62 B\\
&GOOG & Alphabet Inc. (Google) &Technology &\$1.917 T\\
&GS & Goldman Sachs Group, Inc. &Finance&\$127.61 B \\
&IBM & International Business Machines Corporation &Technology&	\$119.86 B\\
&MCD & McDonald's Corporation &Consumer Discretionary&\$200.31 B\\
&NVDA & NVIDIA Corporation&Technology&\$735.27 B \\
\midrule
\multirow{3}{*}{\rotatebox[origin=c]{90}{Medium}}&
AAPL & Apple Inc. &Technology&\$2.901 T\\
&ABBV & AbbVie Inc. &Health Care&\$239.37 B\\
&PM & Philip Morris International Inc. &Health Care&\$147.89 B\\
\midrule
\multirow{6}{*}{\rotatebox[origin=c]{90}{Large}}&
CMCSA & Comcast Corporation & Telecommunications & \$229.95 B\\
&CSCO & Cisco Systems, Inc. & Telecommunications&\$267.26 B \\
&INTC & Intel Corporation &Technology&\$209.60 B\\
&MSFT & Microsoft Corporation &Technology &\$2.522 T\\
&KO & Coca-Cola Company&Consumer Staples&\$255.75 B \\
&VZ & Verizon Communications Inc.&Telecommunications&\$218.11 B \\
\bottomrule
\end{tabular}}
\end{table}

\subsubsection{Data pre-processing} \label{sec: pre-processing}
In our analysis, we exclusively consider complete trading days in 2021, excluding weekends, public holidays, and any days with missing data in LOBSTER. For each trading day, we only analyze data from 09:30 am to 16:00 pm, aligning with NASDAQ's opening hours. We also filter out any orders related to auction trades and trading halts, corresponding to event types 6 and 7 in LOBSTER, respectively. To avoid memory leaks, for each stock, we parallelize the processing of data day by day using the Python library \texttt{\href{https://joblib.readthedocs.io/en/stable/}{joblib}}. This is executed on an Intel(R) Xeon(R) Gold 6140 CPU @ 2.30GHz with 72 cores and a total memory of 502.51 GB.

\subsection{Methodology} \label{sec: Methodology}

\subsubsection{Independent variables}

In Section \ref{sec: feature_engineering}, we outline the definition for creating six time-dependent features for new orders on each trading day, and then discuss the forward rolling normalization of these features.

\paragraph{Feature engineering} \label{sec: feature_engineering}

We distinguish the new order by creating six time-dependent features on each trading day, which captures key aspects of order dynamics, such as arrival patterns, execution timing, and price impact. By leveraging these engineered features, we apply K-means++ clustering (Algorithm \ref{K-means++}) to classify orders and systematically identify distinct market participant trading behaviors. Below are the definitions for the displayed features.

\noindent\textbf{Definition}\quad \textit{The available \textnormal{bid-side (ask-side) volume} ${V}(p_x,t_x)$ of the new \textnormal{order} $x$ with \textnormal{price} $p_x$ and at \textnormal{time} $t_x$ is given by
}
\begin{equation}
    V(p_x,t_x):= \left\{
                     \begin{array}{@{}l@{\thinspace}l}
                       v^{b}(p_x,t_x)  & \quad \textnormal{bid-side},\\
                       v^{a}(p_x,t_x)  & \quad \textnormal{ask-side}.\\
                     \end{array}
                   \right.
\end{equation}
\noindent\textbf{Definition}\quad \textit{The \textnormal{time} elapsed between the \textnormal{time} \( t_x \) of the new \textnormal{order} \( x \) and the \textnormal{time} \( t_m \) of the \textnormal{mid-price} \( m(t_m) \) last changing, denoted by \( T^m(t_x) \), is given by}
\begin{equation}
T^m(t_x) := t_x-t_m.
\end{equation}
\noindent\textbf{Definition}\quad \textit{The \textnormal{time} elapsed between the \textnormal{time} \( t_x \) of the new \textnormal{order} \( x \) with \textnormal{price} \( p_x \) and the \textnormal{time} \( t_{x_1} \) of the first \textnormal{order} \( x_1 \), which \textnormal{\textbf{firstly}} arrived at the same \textnormal{price} \( p_x \), denoted by \( T^{1}(t_x) \), is given by}
\begin{equation}
T^{1}(t_x) := t_x-t_{x_1}.
\end{equation}
\noindent\textbf{Definition}\quad \textit{The \textnormal{time} elapsed between the \textnormal{time} \( t_x \) of the new \textnormal{order} \( x \) with \textnormal{price} \( p_x \) and the \textnormal{time} \( t_{x^{'}} \) of the previous \textnormal{order} \( x^{'} \), which \textnormal{\textbf{previously}} arrived at the same \textnormal{price} \( p_x \), denoted by \( T^{'}(t_x) \), is given by}
\begin{equation}
T^{'}(t_x) := t_x-t_{x^{'}}.
\end{equation}
\noindent\textbf{Definition}\quad \textit{The sum of the available \textnormal{bid-side (ask-side) volume} $v(p_x,t_x)$ of the new \textnormal{\textbf{buy} (\textbf{sell}) order} \( x \) with \textnormal{price} \( p_x \) and at \textnormal{time} \( t_x \) until the \textnormal{volume of the best bid price} $v(b^{1}(t_x),t_x)$ \textnormal{(volume of the best ask price} $v(a^{1}(t_x),t_x)$\textnormal{)}, denoted by \textnormal{same book side} \(\textnormal{SBS}(p_x,t_x) \), is given by
}
\begin{equation}
SBS(p_x,t_x) := \left\{
 \begin{array}{@{}l@{\thinspace}l}
   \sum\limits_{p=p_x}^{b^{1}(t_x)}\hspace{3.3mm}v(p,t_x)  & \quad \textnormal{bid-side},\\
   \sum\limits_{p=a^{1}(t_x)}^{p_x}v(p,t_x)  & \quad \textnormal{ask-side}.\\
 \end{array}
\right.
\end{equation}
\noindent\textbf{Definition}\quad \textit{The sum of the available \textnormal{bid-side (ask-side) volume} $v(p_{x}',t_x)$ of the new \textnormal{\textbf{sell} (\textbf{buy}) order} \( x \) with \textnormal{price} \( p_x \) and at \textnormal{time} \( t_x \) until the \textnormal{volume of the best bid price} $v(b^{1}(t_x),t_x)$ \textnormal{(volume of the best ask price} $v(a^{1}(t_x),t_x)$\textnormal{)}, denoted by \textnormal{opposite book side} \(OBS(p_x,t_x) \), is given by
}

\begin{equation}
OBS(p_x,t_x) := \left\{
 \begin{array}{@{}l@{\thinspace}l}
   \sum\limits_{p=a^{1}(t_x)}^{p_{x}'}v(p,t_x)   & \quad \textnormal{bid-side},\\
   \sum\limits_{p=p_{x}'}^{b^{1}(t_x)}\hspace{3.3mm}v(p,t_x)  & \quad \textnormal{ask-side}.\\
 \end{array}
\right.
\end{equation} 

\noindent where \( p_x \) and \( p_{x}' \) are symmetric with respect to the mid-price $m(t_x)$.

\noindent\textbf{Definition}\quad \textit{The six time-dependent \textnormal{features} $F(p_x,t_x)$ of the new \textnormal{\textbf{buy} (\textbf{sell}) order} \( x \) with \textnormal{price} \( p_x \) and at \textnormal{time} \( t_x \), are encapsulated in the six-dimensional feature vector, is given by}
\begin{equation}
F(p_x,t_x) := [V(p_x,t_x), T^m(t_x), T^{1}(t_x), T^{'}(t_x), SBS(p_x,t_x), OBS(p_x,t_x)]  \in \mathbb{R}^{6}.
\end{equation}

\paragraph{Normalization}  \label{sec: normalization}
To ensure proper clustering across different stocks, we apply a forward-rolling normalization technique to the six time-dependent features. This approach standardizes each feature dynamically within a forward moving window, capturing local variations in trading behavior rather than relying on static global normalization. To avoid look-ahead bias, we calculate forward-rolling normalization for a new order using a sliding window of the most recent 100 orders to eliminate the influence of future data on historical calculations. Also, this method helps mitigate differences in scale across stocks and ensures that clustering is driven by relative feature variations rather than absolute magnitudes. By applying forward-rolling normalization, we enhance the robustness of the clustering process, allowing for a more accurate classification of trading behaviors across diverse market conditions.

\noindent\textbf{Definition}\quad \textit{The forward-rolling normalization of each \textnormal{feature} $f(p,t)_{i} \in [V(p,t), T^m(t), T^{1}(t), T^{'}(t), $\\$SBS(p,t), OBS(p,t)]$ $(i = 1,\ldots, 6)$, denoted by $\mathcal{N}(f(p,j)_{i})$, dynamically within a moving window of \textnormal{size} \( w \), is given by
}

\begin{equation}
\begin{aligned}
&MA(f(p,t)_{i}) = \frac{1}{w} \sum_{j=t-w+1}^{t} f(p,j)_{i} , \\
&STD(f(p,t)_{i}) = \sqrt{\frac{1}{w} \sum_{j=t-w+1}^{t} (f(p,j)_{i} - MA(f(p,t)_{i}))^2}, \\
&\mathcal{N}(f(p,j)_{i}) = \frac{f(p,j)_{i} - MA(f(p,t)_{i})}{STD(f(p,t)_{i})}.
\end{aligned}
\end{equation}

\noindent where \( MA(f(p,t)_{i}) \) and \( STD(f(p,t)_{i}) \) are the forward-rolling mean and the forward-rolling standard deviation of $f(p,t)_{i}$ at time \( t \), respectively.

\noindent\textbf{Definition}\quad \textit{Within the \textnormal{time interval} $(t - h, t]$, during which the \textnormal{LOB} $\mathcal{L}(t)$ contains at least $w$ orders, the forward rolling normalization of the six time-dependent features of the \textnormal{LOB} $\mathcal{L}(t)$, denoted by $\mathcal{N}(F(\mathcal{L}(t))) \in \mathbb{R}^{(|\mathcal{L}(t)|-w+1)\times6}$, represents all order features dynamically normalized within a moving window of size $w$ over the time interval.
}

\noindent\textbf{Definition}\quad \textit{Within the \textnormal{time interval} $(t - h, t]$, during which the \textnormal{LOB} $\mathcal{L}(t)$ for \textnormal{\textbf{add (cancel, trade)}} events contains at least $w$ orders, the forward rolling normalization of the six time-dependent features of the \textnormal{LOB} $\mathcal{L}(t)$ for \textnormal{\textbf{add (cancel, trade)}} events, denoted by $\mathcal{N}(F(e(\mathcal{L}(t))=\textnormal{add})) \in \mathbb{R}^{(|e(\mathcal{L}(t))=\textnormal{add}|-w+1)\times6}$ \textnormal{(}$\mathcal{N}(F(e(\mathcal{L}(t))=\textnormal{cancel})) \in \mathbb{R}^{(|e(\mathcal{L}(t))=\textnormal{cancel}|-w+1)\times6}$, $\mathcal{N}(F(e(\mathcal{L}(t))=\textnormal{trade})) \in \mathbb{R}^{(|e(\mathcal{L}(t))=\textnormal{trade}|-w+1)\times6}$\textnormal{)}, represents all order features dynamically normalized within a moving window of size $w$ corresponding type of events over the time interval, respectively.
}

\subsubsection{Dependent variable} \label{sec: dependent}

We analyze three categories of returns as our dependent variable. Each trading day $T$, the trading hours from 9:30 am to 16:30 pm are segmented into thirteen 30-minute buckets $\Delta_i$ ($i = 1, \dots, 13$). Then, we define the contemporaneous return, denoted by $CONR(\Delta_i, T)$, the future return in the next bucket, denoted by $FRNB(\Delta_i, T)$, and the future return in the end bucket, denoted by $FREB(\Delta_i, T)$, for each bucket $\Delta_i$ as follows.

\noindent\textbf{Definition}\quad \textit{The $CONR(\Delta_i, T)$ for each \textnormal{bucket} \( \Delta_i \) on \textnormal{day} $T$, is given by
}

\begin{equation}
CONR(\Delta_i, T) := log \left(\frac{m(t_{i})}{m(t_{i-1})}\right) \quad i = 1, \dots, 13.
\end{equation} 

\noindent where $m(t_{i})$ and $m(t_{i-1})$ are the open and close mid-prices for the current bucket $\Delta_i$ on day $T$.

\noindent\textbf{Definition}\quad \textit{The $FRNB(\Delta_i, T)$ for each \textnormal{bucket} \( \Delta_i \) on \textnormal{day} $T$, is given by
}

\begin{equation}
    FRNB(\Delta_i, T):= log\left(\frac{m(t_{i+1})}{m(t_{i})}\right)   \quad i = 1, \dots, 12.
\end{equation}

\noindent where $m(t_{i})$ and $m(t_{i+1})$ are the close mid-prices for the current bucket $\Delta_i$ and the next bucket $\Delta_{i+1}$ on day $T$, respectively.

\noindent\textbf{Definition}\quad \textit{The $FREB(\Delta_i, T)$ for each \textnormal{bucket} \( \Delta_i \) on \textnormal{day} $T$, is given by
}

\begin{equation}
    FREB(\Delta_i, T):= log\left(\frac{m(t_{13})}{m(t_{i})}\right)  \quad i = 1, \dots, 12.
\end{equation}

\noindent where $m(t_{i})$ and $m(t_{13})$ are the close mid-prices for the current bucket $\Delta_i$ and the end bucket $\Delta_{13}$ on day $T$, respectively.

The following schematic diagrams illustrate the structures and relationships of $CONR(\Delta_i, T)$, $FRNB(\Delta_i, T)$, and $FREB(\Delta_i, T)$.

\begin{figure}[!htbp]
\centering
\includegraphics[width=0.9\textwidth,trim=0cm 0cm 0cm 0cm,clip]{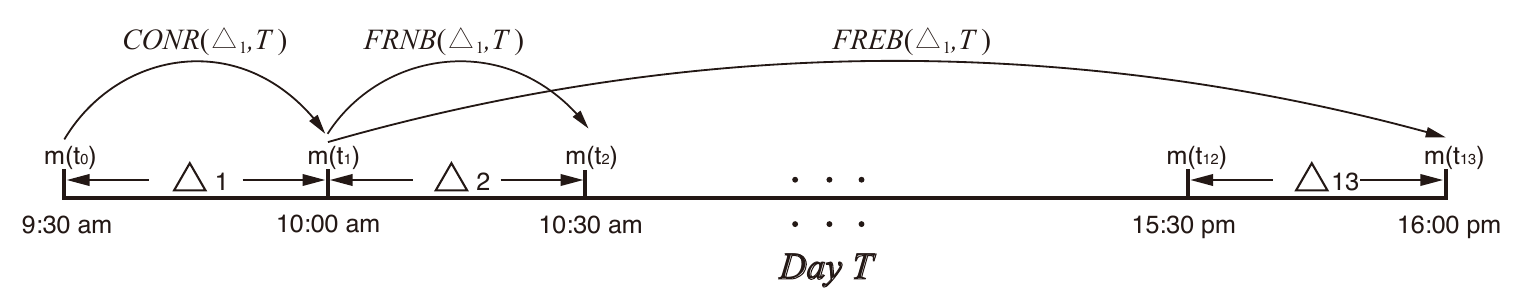}
\caption{An illustration of the return calculations for $CONR(\Delta_1, T)$, $FRNB(\Delta_1, T)$, and $FREB(\Delta_1, T)$ on a single trading day $T$. Here, we use the first bucket $\Delta_1$ as an example to demonstrate the computation process.}
\label{fig: returns}
\end{figure}


\newpage
\subsubsection{Clustering model} \label{sec: clustering_model}

This section outlines the K-means++ clustering algorithm presented by \cite{arthur2006k}, which is employed in our research. Clustering is a fundamental unsupervised learning technique used to group similar observations based on their underlying structure. The K-means++ clustering algorithm is an enhancement of the standard K-means clustering algorithm, designed to improve the initialization process of cluster centroids. Unlike traditional K-means clustering algorithm, which initializes centroids randomly, K-means++ clustering algorithm strategically selects initial centroids in a manner that reduces the likelihood of poor local minima, thereby improving convergence speed and cluster quality. This version of the algorithm also provides an expected O($\log k$)-approximation theoretical guarantee relative to the optimal k-means cost. For completeness, we provide a summary of the K-means++ clustering in Algorithm \ref{K-means++} below.

\begin{algorithm}[!htbp] 
\caption{\textbf{\small : K-means++ clustering}}
\label{K-means++}
\hspace*{\algorithmicindent} \textbf{Input:} Input matrix $U$, the total number of clusters $K$. \\
\hspace*{\algorithmicindent} \textbf{Output:} Clusters $\phi_1, \phi_2, ..., \phi_K$. 
\begin{algorithmic}[1]
\State Randomly select an initial center $\phi_1$ from $U$.
\State \textbf{Repeat} for $i \in \{1,2, \ldots, K-1, K$\}. 
Select the next center $\phi_i=x \in U$ with the probability
$$
P(x)=\frac{Z(x)^2}{\sum\limits_{x^{\prime} \in U} Z\left(x^{\prime}\right)^2},
$$
where $x^{\prime}$ is the closest center that has already been chosen and $Z\left(x^{\prime}\right)$ is the distance to that center.
\State Continue with the standard K-means algorithm.
\end{algorithmic}
\end{algorithm}

\subsubsection{ClusterLOB} \label{sec: ClusterLOB}

In this section, we introduce \texttt{ClusterLOB}, a methodology designed to enhance trading performance by leveraging clustering techniques. Specifically, we classify market participants into three distinct clusters representing directional, opportunistic, and market-making traders, which allows for a more nuanced approach to market microstructure modelling and signal generation.

To assess the effectiveness of \texttt{ClusterLOB}, we conduct experiments across small-tick, medium-tick, and large-tick stocks separately, as shown in Table \ref{tab: data}. The rationale behind this classification is that tick size significantly influences market microstructure properties, which in turn influences liquidity provision and order placement strategies. By analyzing different tick size groups, we evaluate whether \texttt{ClusterLOB} consistently improves performance across various market conditions.

We preprocess the data following the procedure outlined in Section \ref{sec: pre-processing}. The dataset comprises a full year of MBO data in 2021. To ensure robust model training and evaluation, we divide the dataset into a training dataset, which includes trading day data from January 1, 2021, to June 30, 2021, and a test dataset, which includes trading day data from July 1, 2021, to December 31, 2021.

In the training dataset for each tick size groups, we create six time-dependent features for each stock, denoted as $V$, $T^m$, $T^{1}$, $T^{'}$, $SBS$, and $OBS$, as described in Section \ref{sec: feature_engineering}, to capture essential order flow characteristics. To ensure feature consistency across different stocks, we apply forward-rolling normalization, as outlined in Section \ref{sec: normalization}. The normalized six-month feature dataset is then used as input for the K-means++ clustering Algorithm \ref{K-means++}. We set $K=3$ as the number of clusters, which we retrospectively interpret as three trader types: directional, opportunistic, and market-making. One critical challenge in applying clustering to multiple stocks is ensuring that cluster labels remain consistent across different stocks. A naive application of K-means++ may lead to inconsistent cluster assignments, where labels do not align across different stocks. To address this issue, we employ a base initialization algorithm \ref{initialization}, as demonstrated below. By using this base initialization scheme, we maintain a consistent mapping of cluster labels across stocks, ensuring that clusters are interpretable and comparable.

\begin{algorithm}[!htbp] 
\caption{\textbf{\small : Base initialization}}
\label{initialization}
\begin{algorithmic}[1]
\State Select a randomly chosen stock from the training dataset as the reference stock.
\State Apply K-means++ clustering to this reference stock's normalized six-month feature dataset and store the resulting cluster centroids. 
\State For all remaining stocks, initialize the K-means++ clustering using the centroids obtained from Step 2 rather than using random initialization.
\end{algorithmic}
\end{algorithm}

After clustering the training dataset, each order is assigned a corresponding cluster label, allowing us to analyze order flow patterns within each cluster. For each stock in the dataset, we calculate two OFIs (Size-based OFI and Count-based OFI) at the cluster level, which are formally introduced in Section \ref{sec: ofi}, aggregating them within 30-minute buckets throughout the trading day for the entire training dataset. This ensures that two OFIs are measured over short-term trading windows, allowing for an assessment of how different types of market participants interact with price movements over time. These calculations are performed consistently across all stocks in the training dataset to maintain comparability.

To systematically classify trading behaviors, for each stock, we assess the correlation between two OFIs and three categories of returns (CONR, FRNB, and FREB) for the entire training dataset, as defined in Section \ref{sec: dependent}. The correlation structure provides insights into how each cluster interacts with price changes and helps us define the type of traders. The clusters are defined based on the following correlation patterns.

\begin{itemize}
    \item Cluster $\phi_1$: The cluster that exhibits the highest correlation between two OFIs and CONR is identified as the directional trading class. This classification is based on the premise that traders in this cluster execute aggressive orders that significantly impact price movements within the same bucket.
    \item Cluster $\phi_2$: The cluster that has the highest correlation between  the two OFIs and FREB is classified as the opportunistic trading class. Traders in this cluster strategically time their order executions based on short-term predictive signals, seeking to capitalize on market inefficiencies.
    \item Cluster $\phi_3$: The remaining cluster, whose two OFIs do not exhibit strong correlations with CONR or FREB, is categorized as the market-making trading class. These traders primarily provide liquidity, maintaining bid-ask spreads while minimizing exposure to directional market movements.
    \item Cluster $\phi_*$: A benchmark scenario in which no clustering is applied.
\end{itemize}

Although we maintain a consistent clustering methodology across different stocks, the classification of traders may vary slightly between stocks due to differences in correlation structures. This variability may arise because the relationship between two OFIs and price movements is not uniform across all securities. To ensure a more robust and stable classification, we aggregate the trading class labels across stocks using mode estimation. This approach determines the most frequently assigned trading class for each cluster across all stocks. To further enforce consistency and facilitate meaningful comparisons across different tick sizes of stocks, we systematically adjust cluster labels as needed. Without loss of generality, we reassign cluster labels until they satisfy the following standardized classification: Cluster $\phi_1$ as the directional trading class, Cluster $\phi_2$ as the opportunistic trading class, and Cluster $\phi_3$ as the market-making trading class.

For each stock, after identifying the traders in the training dataset, we compute profit and loss (PnL) based on Size-based OFI $OFI^{S}(\phi_i, \Delta_j, T)$ across different clusters $\phi_i$ within \textnormal{bucket} \( \Delta_j \) for $FRNB(\Delta_j, T)$ on day $T$ as follows.

\noindent\textbf{Definition}\quad \textit{The PnL based on \textnormal{Size-based OFI} $OFI^{S}(\phi_i, \Delta_j, T)$ across different \textnormal{clusters} $\phi_i$ within \textnormal{bucket} \( \Delta_j \) for $FRNB(\Delta_j, T)$ on \textnormal{day} $T$, is given by
}

\vspace{-3mm}
\begin{equation}
    {pnl}^{S}_{FRNB}(\phi_i, T):= \sum_{j=1}^{12} sign(OFI^{S}(\phi_i, \Delta_j, T))\cdot FRNB(\Delta_j, T)  \quad i = 1, \dots, 3.
\end{equation}

\noindent where $sign(\cdot)$ is the sign function, $OFI^{S}(\phi_i, \Delta_j, T)$ represents Size-based OFI across different clusters $\phi_i$ within \textnormal{bucket} \( \Delta_j \) on \textnormal{day} $T$, respectively.

Next, we compute the equal-weighted PnL by aggregating the PnL of all stocks on each trading day in the training dataset as follows.

\noindent\textbf{Definition}\quad \textit{The \textnormal{equal-weighted PnL}, denoted by ${PNL}^{S}_{FRNB}(\phi_i)$, by averaging the ${pnl}^{S}_{FRNB}(\phi_i, T)$ of all stocks on each trading day, is given by
}

\vspace{-4mm}
\begin{equation}
    {PNL}^{S}_{FRNB}(\phi_i):= [\mathbb{E}[{pnl}^{S}_{FRNB}(\phi_i, 1)], \dots, \mathbb{E}[{pnl}^{S}_{FRNB}(\phi_i, 126)]] \in \mathbb{R}^{126} \quad i = 1, \dots, 3.
\end{equation}

Then, we rescale the equal-weighted PnL by their volatility to target equal risk assignment, and set our annualized volatility target $\sigma_{\text{tgt}}$ to be 0.15.

\noindent\textbf{Definition}\quad \textit{The \textnormal{rescaled equal-weighted PnL}, denoted by ${\widehat{PNL}}^{S}_{FRNB}(\phi_i)$, with the annualized \textnormal{volatility target} $\sigma_{tgt}$ set to 0.15, is given by
}

\begin{equation}
    {\widehat{PNL}}^{S}_{FRNB}(\phi_i):= \frac{\sigma_{tgt}}{STD({PNL}^{S}_{FRNB}(\phi_i))\cdot\sqrt{252}}\cdot{PNL}^{S}_{FRNB}(\phi_i) \quad i = 1, \dots, 3.
\end{equation}

To determine the optimal strategy among the three clusters, we evaluate the annualized SR, which measures the risk-adjusted return of each strategy.

\noindent\textbf{Definition}\quad \textit{The annualized \textnormal{SR}, denoted by $SR({\widehat{PNL}}^{S}_{FRNB}(\phi_i))$, measures the risk-adjusted return of a portfolio, is given by}

\begin{equation}
SR({\widehat{PNL}}^{S}_{FRNB}(\phi_i)) = \frac{\mathbb{E}[{\widehat{PNL}}^{S}_{FRNB}(\phi_i)]}{STD({\widehat{PNL}}^{S}_{FRNB}(\phi_i))}\cdot \sqrt{252} \quad i = 1, \dots, 3.
\end{equation}

By a similar definition, we compute the annualized SR, denoted by $SR({\widehat{PNL}}^{C}_{FRNB}(\phi_i))$, based on Count-based OFI \( OFI^{C}(\phi_i, \Delta_j, T) \) across different clusters \( \phi_i \) within \textnormal{bucket} \( \Delta_j \) for \( FRNB(\Delta_j, T) \). Then, we select the highest SR, \( \max\{SR({\widehat{PNL}}^{S}_{FRNB}(\phi_i)), SR({\widehat{PNL}}^{C}_{FRNB}(\phi_i))\}\) ($i = 1, \dots, 3$), which is identified as the best trading strategy in the training dataset.

After selecting the best trading strategy, we follow the same process to create six time-dependent features and apply forward rolling normalization on each trading day. We then use the K-means++ clustering model trained on the training dataset to predict the cluster assignments of orders in the test dataset. After this step, we compute the two OFIs only for the orders belonging to the same cluster as the best trading strategy. Finally, we conduct a comparative evaluation against a benchmark trading strategy that does not incorporate clustering and thus relies on the entire order flow. The benchmark strategy follows a standard approach where trading signals are generated based on Size-based OFI $OFI^{S}(\phi_*, \Delta_j, T)$ and Count-based OFI $OFI^{C}(\phi_*, \Delta_j, T)$ within \textnormal{bucket} \( \Delta_j \) for $FRNB(\Delta_j, T)$ directly, \textnormal{without} using different clusters $\phi_i$ on each trading day $T$ in the test dataset. 

By a similar process as above, we also run the clustering experiment for FREB. First, we select the highest SR, \( \max\{SR({\widehat{PNL}}^{S}_{FREB}(\phi_i)), SR({\widehat{PNL}}^{C}_{FREB}(\phi_i))\}\) ($i = 1, \dots, 3$), from strategies based on two OFIs, as the best trading strategy in the training dataset, and then evaluate its performance by comparing it with a benchmark trading strategy that does not incorporate clustering in the test dataset. We summarize the \texttt{ClusterLOB} Algorithm \ref{ClusterLOB} as follows, detailing the key steps involved in its implementation and evaluation. 

\begin{algorithm}[!htbp] 
\caption{\textbf{\small : \texttt{ClusterLOB}}}
\label{ClusterLOB}
\textbf{Train:}
\begin{algorithmic}[1]
\State For each stock in the same tick size group:
\begin{itemize}
    \item Create six time-dependent features and apply forward rolling normalization on each trading day.
    \item Apply K-means++ clustering with \( K = 3 \) based on six time-dependent features following Algorithm \ref{initialization}, and assign each order a cluster label.
    \item Compute two OFIs (Size-based OFI and Count-based OFI) across different clusters within 30-minute buckets, and calculate correlations between two OFIs and three returns (CONR, FRNB, and FREB) for the entire training dataset.
    \item Define clusters: 
        \begin{itemize}
            \item The highest correlation with CONR as the directional trading class.
            \item The highest correlation with FREB as the opportunistic trading class.
            \item The remaining as the market-making trading class.
        \end{itemize}
\end{itemize}    
\State Apply mode estimation across stocks to ensure stable trading class assignments, and reassign labels for consistency.
\State Select the highest SR from the strategies based on two OFIs as the best trading strategy for the FRNB and FREB.
\end{algorithmic}
\textbf{Test:}
\begin{algorithmic}[1]
\State For each stock in the same tick size group:
\begin{itemize}
    \item Create six time-dependent features and apply forward rolling normalization on each trading day.
    \item Apply K-means++ clustering model with \( K = 3 \) trained on the training dataset to predict the cluster assignments of orders based on six time-dependent features.
    \item Compute two OFIs only for the orders belonging to the same cluster as the best trading strategy. 
\end{itemize} 
\State Compare the best trading strategy with a benchmark trading strategy
that does not incorporate clustering for the FRNB and FREB.
\end{algorithmic}
\end{algorithm}

\begin{figure}[!htbp]
\centering
\includegraphics[width=0.83\textwidth,trim=0cm 0cm 0cm 0cm,clip]{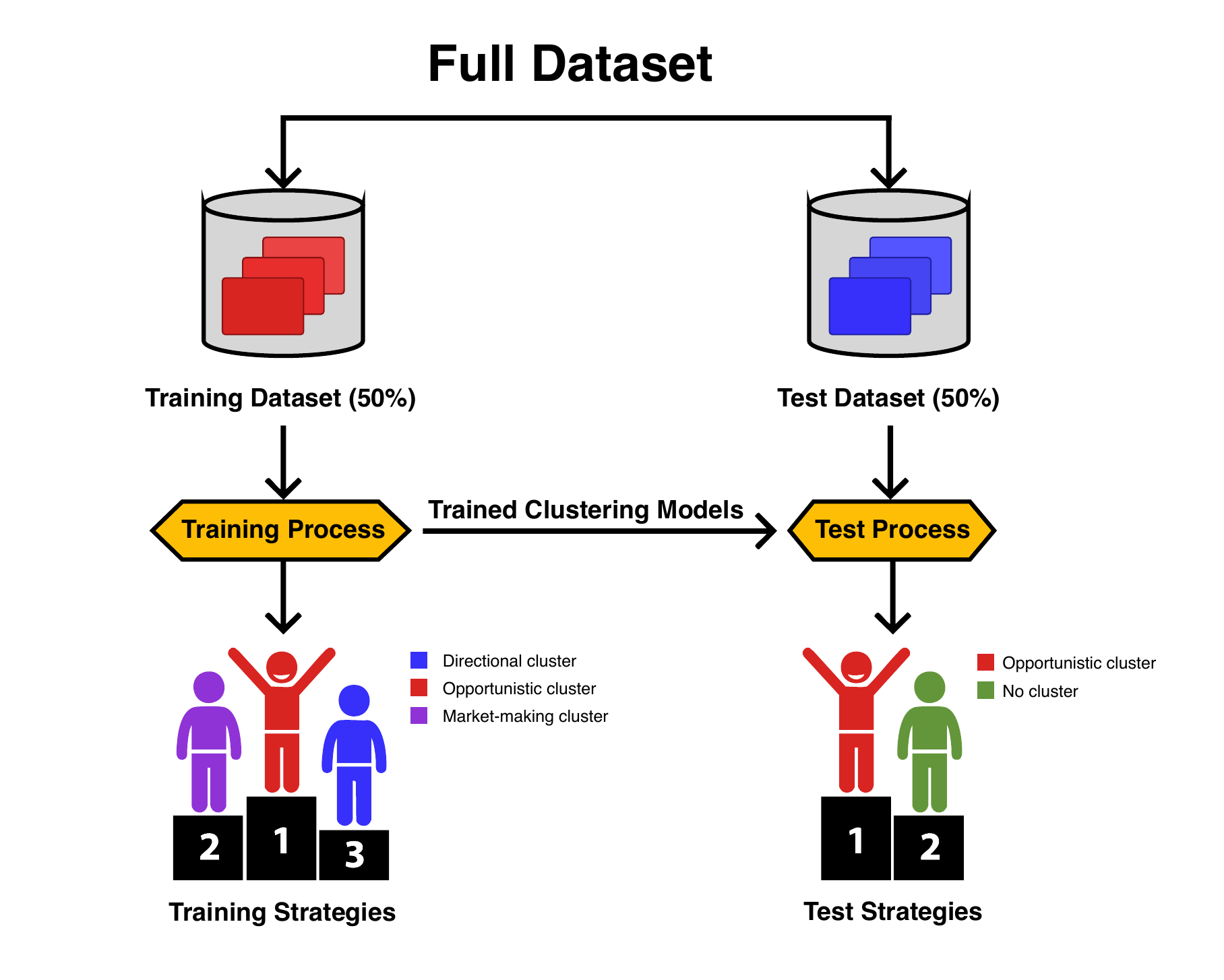} 
\caption{Schematic diagram of the \texttt{ClusterLOB}: We identify the best trading strategy with the highest SR in the training dataset, and demonstrate that its performance outperforms benchmark trading strategies
that do not incorporate clustering in the test dataset. }
\label{fig: ClusterLOB}
\end{figure}

We also evaluate trading strategies based on the decomposed OFI, defined in Section \ref{sec: decomposed_ofi}, across different market event types, including add, cancel, and trade events, to assess its robustness in various market conditions. This analysis helps determine how the \texttt{ClusterLOB} adapts to different order flow dynamics and execution patterns. In addition, we provide a schematic diagram of the \texttt{ClusterLOB} in Figure \ref{fig: ClusterLOB}, which visually illustrate its overall structure and the interactions between its main components.

\newpage
\subsubsection{Performance evaluation}

To evaluate the \texttt{ClusterLOB}, we compute a set of annualized performance metrics, ensuring consistency with the evaluation framework adopted in recent literature [\cite{lim2019enhancing, wood2022slow, tan2023spatio, poh2022transfer, wood2021trading, wood2021slow, liu2023multi, liu2023deep}]. These metrics provide a comprehensive view of the model's effectiveness across profitability, risk management, and trading performance.

\begin{itemize}
  \item \textbf{Profitability}: annualized expected excess return (E[Returns]), hit rate.
  \item \textbf{Risk}: volatility, downside deviation, maximum drawdown.
  \item \textbf{Performance}: SR, Sortino ratio, Calmar ratio, average profit / average loss, PnL per trade (PPT).
\end{itemize}


\section{Empirical Results} \label{sec: Result}

In this section, we first illustrate the empirical results using Comcast Corporation (CMCSA) as an example. We then discuss the cumulative PnL results for FRNB and FREB across all events, specifically focusing on small-tick stocks. Additionally, we provide a comparative analysis of the results from FRNB and FREB across all event types, segmented by small-tick, medium-tick, and large-tick stocks. Finally, detailed results for the add, cancel, and trade event strategies are presented in the Appendix \ref{appendix}.

Figure \ref{fig: CMCSA_corr} presents the correlation analysis of CMCSA between two OFIs and CONR, FRNB, and FREB across the training and test datasets. 

{\textbullet} Cluster $\phi_1$ is strongly associated with directional traders. This is evidenced by the substantial positive correlation between the two OFIs and the CONR, indicating that activity within this cluster exerts meaningful market impact. The correlation with forward-looking returns FRNB and FREB, however, is comparatively weaker, suggesting that the price impact induced by these traders is largely absorbed within the same bucket, often followed by partial price reversion. This aligns with typical directional strategies that aim to capitalize on short-term price inefficiencies but are less predictive of longer-horizon returns. 

{\textbullet}  Cluster $\phi_2$, in contrast, captures the behavior of opportunistic traders. Two OFIs associated with this cluster display relatively weaker contemporaneous correlations but exhibit a higher degree of alignment with future returns, particularly FREB. This suggests that traders in this group are able to anticipate short-term price movements before they materialize, reflecting a high degree of intraday alpha generation. Their activity is more predictive than reactive, and thus more valuable from a forecasting perspective. 

{\textbullet} Cluster $\phi_3$ reflects market-making behavior. Here, the two OFIs show minimal correlation with all three return measures across both datasets. This indicates that these traders operate in a manner that minimizes market impact, providing liquidity without inducing or reacting to substantial price movement. 

Importantly, the correlation structures observed in the test set closely mirror those in the training set. This consistency suggests that the latent structure identified by \texttt{ClusterLOB} is stable and generalizable across different time periods. In particular, the roles of the three clusters—directional, opportunistic, and market-making—remain distinguishable out-of-sample, supporting the robustness of the model in uncovering economically interpretable trading behaviors.

\begin{table}[h!] 
  \centering
    \centering
    \begin{tabular}{p{0.1cm}p{6.8cm}p{6.8cm}}
        \toprule
&\multicolumn{1}{c}{\textbf{Training}} &  \multicolumn{1}{c}{\textbf{Test}} \\
        \midrule
    \vspace{-5cm}
    \multirow{1}{*}{\rotatebox[origin=c]{90}{\textbf{Correlation}}}
    &
      \includegraphics[width=\linewidth,trim=0cm 0cm 0cm 0cm,clip]{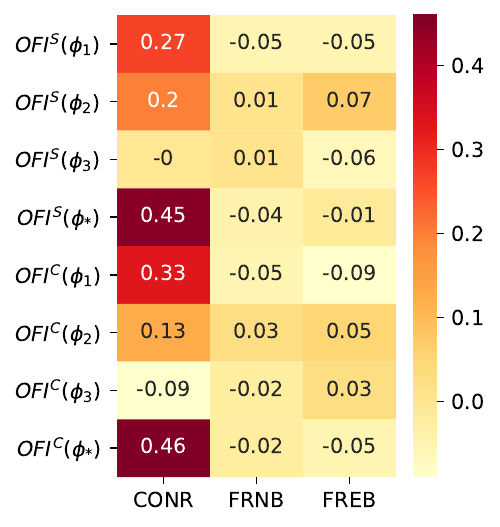}
    &
          \includegraphics[width=\linewidth,trim=0cm 0cm 0cm 0cm,clip]{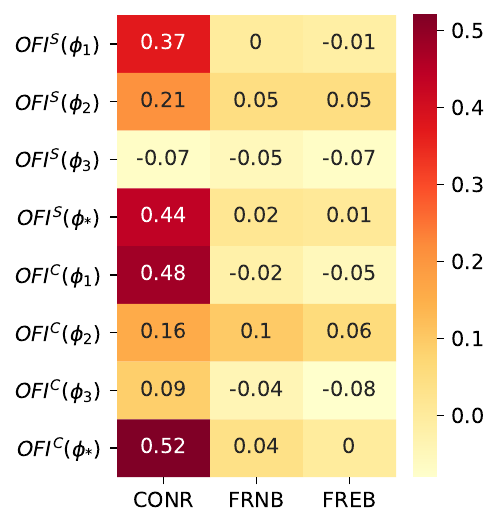}
          \\
          \bottomrule
        \end{tabular}
\captionof{figure}{CMCSA: correlation analysis between two OFIs and CONR, FRNB, and FREB  across training and test datasets.}
\label{fig: CMCSA_corr}
\end{table}


In Figure~\ref{fig: CMCSA_mean_median}, we present heatmaps illustrating the mean and median values of six time-dependent features across the identified clusters in both the training and test datasets. These features include $V$, $T^m$, $T^{1}$, $T^{'}$, $SBS$, and $OBS$, as shown in Section \ref{sec: feature_engineering}. The heatmaps offer a comprehensive view of the feature distributional characteristics within each cluster, enabling an interpretation of the trading behaviors encoded by the \texttt{ClusterLOB}.
First, we observe strong alignment between the mean and median patterns within each dataset, indicating that the feature distributions are relatively symmetric and free from extreme outliers. This consistency across statistical moments reinforces the stability and reliability of the cluster assignments. For instance, cluster $\phi_3$ consistently exhibits the highest values for $SBS$, $OBS$, $T^1$, and $T^{'}$, in both mean and median estimates, across training and test sets. Cluster $\phi_2$, by contrast, demonstrates the highest value of $T^m$, and Cluster $\phi_1$ maintains relatively moderate values across most features in the training and test datasets. Importantly, the feature-level patterns are highly preserved when transitioning from the training to the test set, which provides further empirical support for the robustness and generalization ability of \texttt{ClusterLOB}. The model is not only able to identify economically meaningful trading behaviors in-sample but also replicates these behavioral patterns reliably in out-of-sample data. This cross-period consistency in both mean and median statistics underscores the interpretability and stability of the learned clusters.

\begin{table}[h!] 
  \centering
    \centering
    \begin{tabular}{p{0.1cm}p{6.8cm}p{6.8cm}}
        \toprule
&\multicolumn{1}{c}{\textbf{Training}} &  \multicolumn{1}{c}{\textbf{Test}} \\
        \midrule
    \vspace{-3cm}
    \multirow{1}{*}{\rotatebox[origin=c]{90}{\textbf{Mean}}}
    &
      \includegraphics[width=\linewidth,trim=0cm 0cm 0cm 0cm,clip]{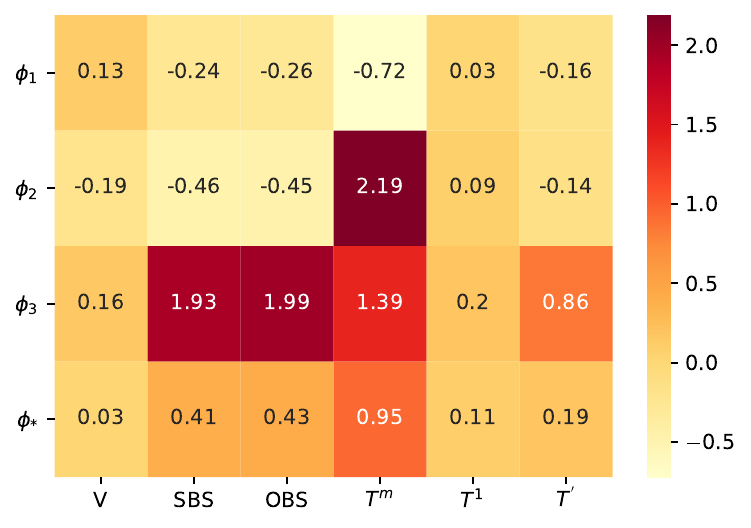}
    &
          \includegraphics[width=\linewidth,trim=0cm 0cm 0cm 0cm,clip]{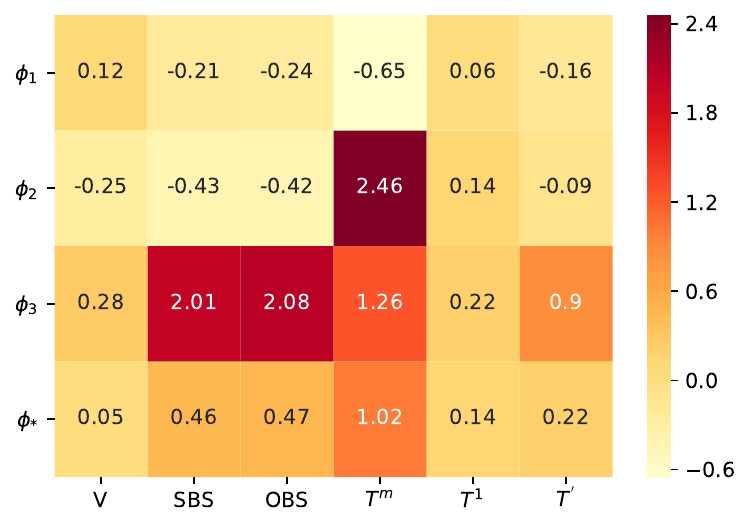}
          \\
            \midrule
    \vspace{-3cm}
    \multirow{1}{*}{\rotatebox[origin=c]{90}{\textbf{Median}}}
    &
      \includegraphics[width=\linewidth,trim=0cm 0cm 0cm 0cm,clip]{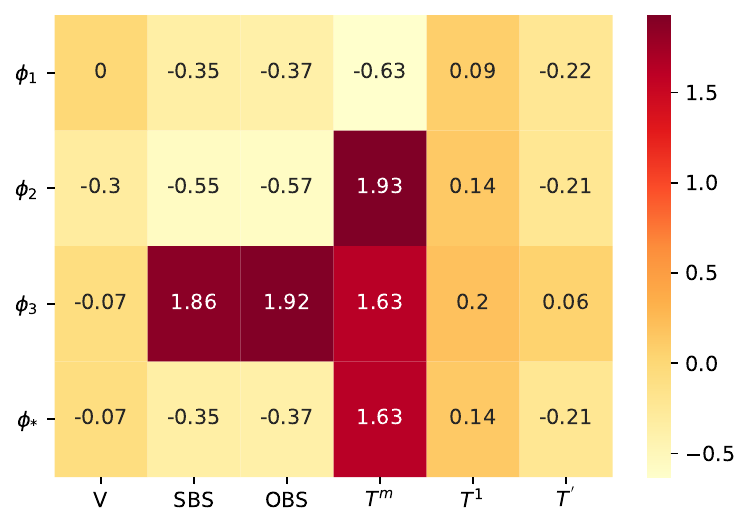}
    &
          \includegraphics[width=\linewidth,trim=0cm 0cm 0cm 0cm,clip]{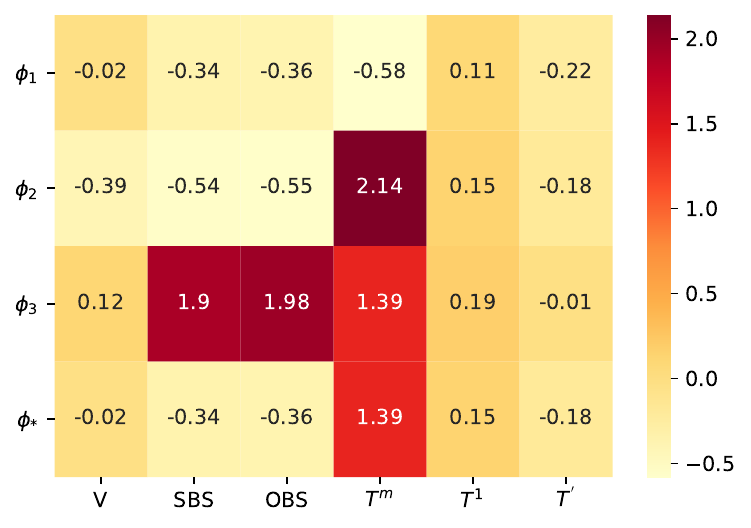}
          \\
          \bottomrule
        \end{tabular}
\captionof{figure}{CMCSA: heatmap of mean and median values of six time-dependent features across different clusters in training and test datasets.}
\label{fig: CMCSA_mean_median}
\end{table}

After confirming that the \texttt{ClusterLOB} produces consistent clustering behavior across both the training and test datasets, we now turn our attention to evaluating its out-of-sample performance. In particular, the opportunistic cluster achieves the highest SR in the training dataset with respect to both FRNB and FREB, motivating an analysis of its performance across all events for small-tick stocks in the test dataset.

Figure~\ref{fig: small_tick_top_FRNB_FREB} presents the cumulative PnL curves for FRNB and FREB in the test dataset, each rescaled to a target volatility of 15\%. The top panel shows the performance of OFI signals in forecasting FRNB. Here, we observe that the $OFI^S$ signal derived from the opportunistic cluster $\phi_2$ clearly outperforms the benchmark models that use unclustered OFIs. The cumulative return steadily increases over the second half of 2021, achieving a SR of 1.34 and a PPT of 0.66. In contrast, the baseline model without clustering $OFI^S(\phi_*)$ yields a lower Sharpe ratio of 0.6 and a PPT of 0.3, while $OFI^C(\phi_*)$  results in negative performance, with a Sharpe ratio of -0.78 and PPT of -0.39. These results provide strong evidence that behaviorally informed clustering enhances the predictive power of OFIs for short-horizon returns.

The bottom panel of Figure~\ref{fig: small_tick_top_FRNB_FREB} displays the performance on FREB, which captures longer-horizon return movements. Once again, the opportunistic cluster $\phi_2$ achieves superior performance relative to the no-cluster benchmarks, albeit with smaller returns. Specifically, the opportunistic cluster $\phi_2$ records a modest SR of 0.23 and a PPT of 0.11, while both unclustered benchmarks remain sharply negative, with SRs below $-2.35$ and large cumulative drawdowns. The fact that the opportunistic cluster $\phi_2$ retains some predictive power even at longer horizons is particularly notable, given that OFIs are typically most effective for short-term forecasting. Together, these findings demonstrate that \texttt{ClusterLOB} not only provides a meaningful decomposition of trader behavior, but also translates into substantial 
economic gains in predictive trading strategies.

\begin{table}[h!]
\centering

  \centering
    \begin{tabular}{p{0.1cm}p{13.8cm}}
        \toprule
&\multicolumn{1}{c}{\textbf{All events}}  \\
        \midrule
    \vspace{-4cm}
    \multirow{1}{*}{\rotatebox[origin=c]{90}{\textbf{FRNB}}}
    &

      \includegraphics[width=\linewidth,trim=0cm 0cm 0cm 0cm,clip]{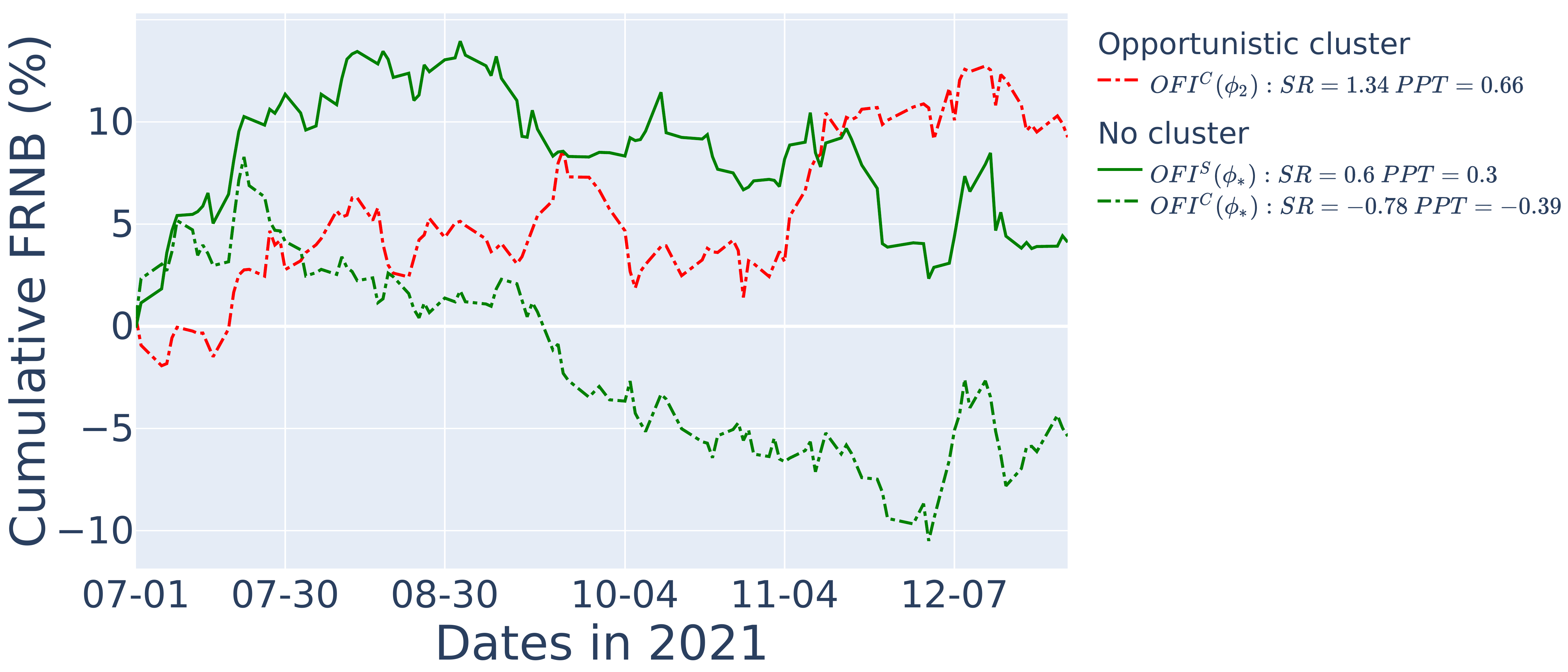}\\
       \midrule
    \vspace{-4cm}
    \multirow{1}{*}{\rotatebox[origin=c]{90}{\textbf{FREB}}}
    &
      
          \includegraphics[width=\linewidth,trim=0cm 0cm 0cm 0cm,clip]{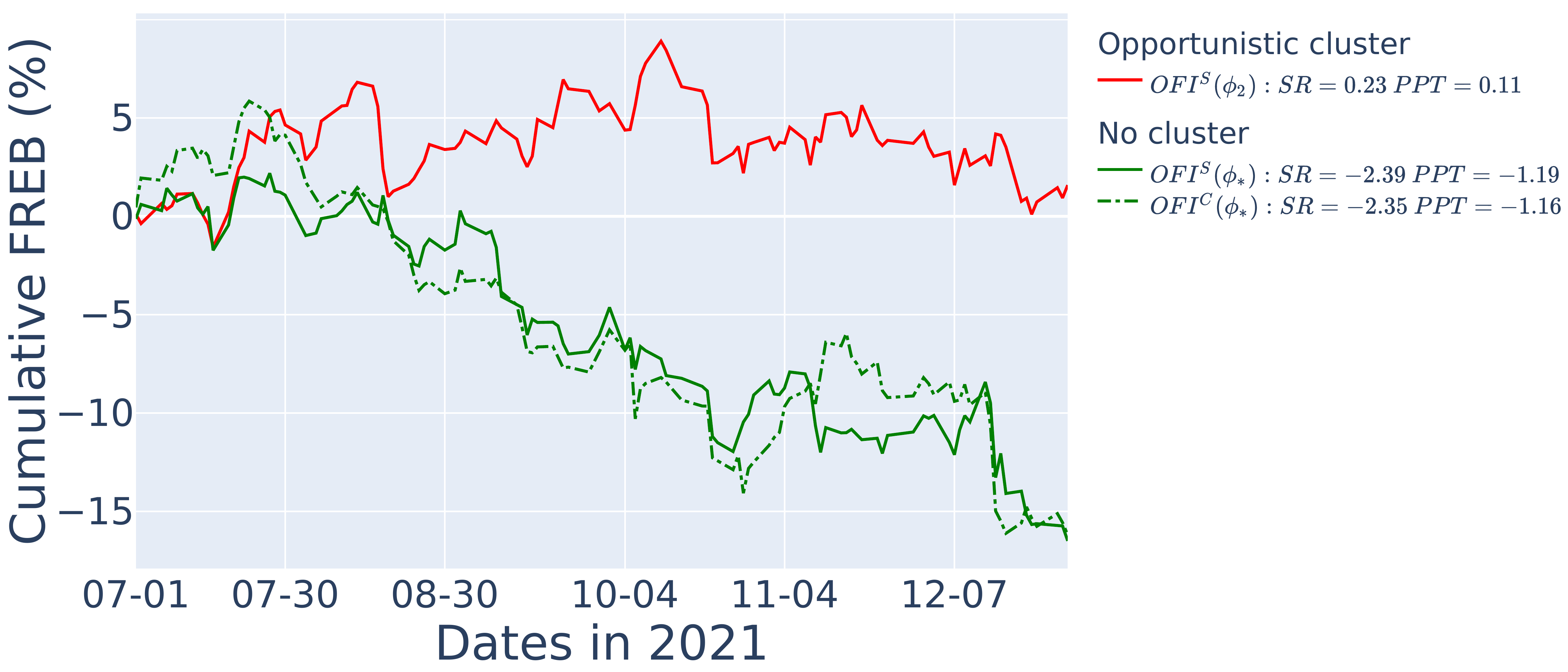}
          \\
\bottomrule
    \end{tabular}
\captionof{figure}{Cumulative PnL of \textbf{FRNB} and \textbf{FREB} across \textbf{all events} for small-tick stocks in the test dataset, rescaled to a target volatility of 15\%.}
\label{fig: small_tick_top_FRNB_FREB}
\end{table}

Table~\ref{tab: FRNB_all} reports the performance metrics for FRNB across all events, separated by small-tick, medium-tick, and large-tick stocks in the test dataset. For each tick-size group, we compare OFIs derived from \texttt{ClusterLOB} against benchmark models without clustering. The highest SR within each group is highlighted.

In the small-tick group, the opportunistic cluster $\phi_2$ achieves the best overall performance, with the highest SR of 1.34, a Sortino ratio of 2.226, and a Calmar ratio of 2.831. It also has the highest expected return 0.201 and a strong hit rate 0.569. In contrast, the benchmark models without clustering yield significantly lower or even negative Sharpe ratios, such as -0.779 for $OFI^C(\phi_*)$. These results provide strong evidence that behaviorally informed clustering significantly improves returns for small-tick stocks. For medium-tick stocks, the directional cluster $\phi_1$ delivers the best performance, achieving a SR of 1.248, slightly below that of the opportunistic cluster $\phi_2$ in the small-tick group. This cluster also yields high values for other performance metrics, such as a Sortino ratio of 2.125 and a PPT of 0.619. This suggests that directional strategies tend to be more effective for assets with slightly wider spreads and lower quote update frequencies, where market impact may persist longer. In the large-tick group, the opportunistic cluster $\phi_2$ again yields the highest SR 1.548, outperforming both no-cluster benchmarks. Although the expected return 0.232 and PPT 0.768 are comparable to those in smaller tick group, the improvement in risk-adjusted metrics is especially noteworthy. The large-tick structure may offer more persistent price inefficiencies that opportunistic traders can exploit, while market-making strategies become less effective due to larger tick sizes and reduced quote granularity.

\begin{table}[h!]
\caption{Performance metric results for \textbf{FRNB} across \textbf{all events} for small-tick, medium-tick, and large-tick stocks in the test dataset. The highest SR within each tick-size group is highlighted.}
\label{tab: FRNB_all}
\resizebox{\linewidth}{!}{
\begin{tabular}{l|lll|lll|lll}

\toprule
\multicolumn{1}{l}{\textbf{All events}}  &\multicolumn{3}{c}{\textbf{Small-tick stocks}}  &  \multicolumn{3}{c}{\textbf{Medium-tick stocks}} &  \multicolumn{3}{c}{\textbf{Large-tick stocks}}  \\
\cmidrule(lr){2-4} \cmidrule(lr){5-7}\cmidrule(lr){8-10}
\multicolumn{1}{l}{\textbf{FRNB}}  &\multicolumn{1}{c}{\textbf{Opportunistic}}  &  \multicolumn{2}{c}{\textbf{No cluster}} &  \multicolumn{1}{c}{\textbf{Directional}}  &  \multicolumn{2}{c}{\textbf{No cluster}} &\multicolumn{1}{c}{\textbf{Opportunistic}}  &  \multicolumn{2}{c}{\textbf{No cluster}} \\
\cmidrule(lr){2-2} \cmidrule(lr){3-4}\cmidrule(lr){5-5}\cmidrule(lr){6-7}\cmidrule(lr){8-8}\cmidrule(lr){9-10}

   & $OFI^{C}(\phi_{2})$ & $OFI^{S}(\phi_{*})$ & $OFI^{C}(\phi_{*})$ & $OFI^{S}(\phi_{1})$ & $OFI^{S}(\phi_{*})$ & $OFI^{C}(\phi_{*})$ & $OFI^{C}(\phi_{2})$ & $OFI^{S}(\phi_{*})$ & $OFI^{C}(\phi_{*})$ \\
\midrule
                        
E{[}Return{]}           & 0.201  & 0.089  & -0.117 & 0.187  & -0.07  & -0.089 & 0.232 & -0.002 & 0.141  \\
Volatility              & 0.15   & 0.15   & 0.15   & 0.15   & 0.15   & 0.15   & 0.15  & 0.15   & 0.15   \\
Downside deviation      & 0.09   & 0.116  & 0.078  & 0.088  & 0.078  & 0.093  & 0.099 & 0.088  & 0.082  \\
Maximum drawdown        & -0.071 & -0.112 & -0.173 & -0.108 & -0.129 & -0.161 & -0.06 & -0.132 & -0.057 \\
Sortino ratio           & 2.226  & 0.773  & -1.505 & 2.125  & -0.897 & -0.964 & 2.344 & -0.021 & 1.734  \\
Calmar ratio            & 2.831  & 0.798  & -0.675 & 1.733  & -0.543 & -0.555 & 3.869 & -0.014 & 2.482  \\
Hit rate                & 0.569  & 0.534  & 0.414  & 0.517  & 0.466  & 0.457  & 0.552 & 0.491  & 0.509  \\
Avg. profit / avg. loss & 0.94   & 0.963  & 1.256  & 1.159  & 1.059  & 1.075  & 1.053 & 1.033  & 1.128  \\
PnL per trade           & 0.665  & 0.296  & -0.386 & 0.619  & -0.232 & -0.296 & 0.768 & -0.006 & 0.468  \\
Sharpe ratio            & \textbf{1.34}   & 0.596  & -0.779 & \textbf{1.248}  & -0.467 & -0.596 & \textbf{1.548} & -0.013 & 0.943  \\
\bottomrule
\end{tabular}}

\end{table}

Compared to the results for FRNB in Table~\ref{tab: FRNB_all}, the performance metrics for FREB in Table~\ref{tab: FREB_all} are generally weaker across all tick-size groups. While the opportunistic cluster $\phi_2$ still yields the highest SR in both the small-tick and large-tick groups (0.23 and 0.438, respectively), these values are substantially lower than their FRNB counterparts (1.34 and 1.548, respectively). In the medium-tick group, the directional cluster $\phi_1$ again performs best with a SR of 0.531, but this too is below the 1.248 observed for FRNB. These results confirm that OFIs, while still informative, are more effective for short-horizon targets such as FRNB than for longer-horizon targets such as FREB.

\begin{table}[h!]
\caption{Performance metric results for \textbf{FREB} across \textbf{all events} for small-tick, medium-tick, and large-tick stocks in the test dataset. The highest SR within each tick-size group is highlighted.}
\label{tab: FREB_all}
\resizebox{\linewidth}{!}{
\begin{tabular}{l|lll|lll|lll}
\toprule
\multicolumn{1}{l}{\textbf{All events}}  &\multicolumn{3}{c}{\textbf{Small-tick stocks}}  &  \multicolumn{3}{c}{\textbf{Medium-tick stocks}} &  \multicolumn{3}{c}{\textbf{Large-tick stocks}}  \\
\cmidrule(lr){2-4} \cmidrule(lr){5-7}\cmidrule(lr){8-10}
\multicolumn{1}{l}{\textbf{FREB}}  &\multicolumn{1}{c}{\textbf{Opportunistic}}  &  \multicolumn{2}{c}{\textbf{No cluster}} &  \multicolumn{1}{c}{\textbf{Directional}}  &  \multicolumn{2}{c}{\textbf{No cluster}} &\multicolumn{1}{c}{\textbf{Opportunistic}}  &  \multicolumn{2}{c}{\textbf{No cluster}} \\
\cmidrule(lr){2-2} \cmidrule(lr){3-4}\cmidrule(lr){5-5}\cmidrule(lr){6-7}\cmidrule(lr){8-8}\cmidrule(lr){9-10}

   & $OFI^{C}(\phi_{2})$ & $OFI^{S}(\phi_{*})$ & $OFI^{C}(\phi_{*})$ & $OFI^{S}(\phi_{1})$ & $OFI^{S}(\phi_{*})$ & $OFI^{C}(\phi_{*})$ & $OFI^{C}(\phi_{2})$ & $OFI^{S}(\phi_{*})$ & $OFI^{C}(\phi_{*})$ \\
\midrule

E{[}Return{]}           & 0.035  & -0.358 & -0.352 & 0.08   & -0.069 & -0.3   & 0.066  & -0.342 & -0.134 \\
Volatility              & 0.15   & 0.15   & 0.15   & 0.15   & 0.15   & 0.15   & 0.15   & 0.15   & 0.15   \\
Downside deviation      & 0.111  & 0.117  & 0.124  & 0.089  & 0.088  & 0.098  & 0.121  & 0.112  & 0.086  \\
Maximum drawdown        & -0.087 & -0.174 & -0.202 & -0.075 & -0.088 & -0.211 & -0.094 & -0.182 & -0.105 \\
Sortino ratio           & 0.312  & -3.073 & -2.836 & 0.894  & -0.785 & -3.072 & 0.545  & -3.043 & -1.557 \\
Calmar ratio            & 0.397  & -2.06  & -1.742 & 1.061  & -0.787 & -1.423 & 0.699  & -1.879 & -1.278 \\
Hit rate                & 0.56   & 0.431  & 0.457  & 0.517  & 0.448  & 0.448  & 0.5    & 0.431  & 0.474  \\
Avg. profit / avg. loss & 0.814  & 0.879  & 0.794  & 1.021  & 1.137  & 0.89   & 1.088  & 0.881  & 0.969  \\
PnL per trade           & 0.114  & -1.185 & -1.164 & 0.263  & -0.229 & -0.993 & 0.217  & -1.131 & -0.444 \\
Sharpe ratio            & \textbf{0.23}   & -2.389 & -2.346 & \textbf{0.531}  & -0.462 & -2.002 & \textbf{0.438}  & -2.28  & -0.895 \\
\bottomrule
\end{tabular}}

\end{table}

Overall, the findings from Tables~\ref{tab: FRNB_all} and~\ref{tab: FREB_all} affirm the robustness of \texttt{ClusterLOB} across different return horizons and tick-size groups. The \texttt{ClusterLOB} adapts well to varying tick-size environments, generalizes effectively out-of-sample, and provides meaningful improvements over benchmark models. By uncovering latent structure in trader behavior and aligning clusters with distinct predictive regimes, \texttt{ClusterLOB} offers a powerful and interpretable approach to high frequency trading forecasting. We also provide detailed performance results for the add, cancel, and trade event types across different tick-size groups. In line with the work of \cite{sitaru2023order}, integrating event type information through the decomposition of OFI does not lead to improved performance in the forward-looking price impact, the results are presented in Appendix~\ref{appendix}.

\section{Conclusion} \label{sec: Conclusion}

In this study, we propose \texttt{ClusterLOB}, a robust algorithm for analyzing high-frequency LOB data to uncover latent trader behaviors and enhance short-term return predictability. By clustering MBO data based on six time-dependent features using the K-means++ algorithm, we classify market activity into three clusters: directional, opportunistic, and market-making traders. These clusters reflect distinct behavioral patterns and interactions with the market microstructure, providing a foundation for more nuanced signal construction.

We compute OFIs in 30-minute buckets and evaluate their predictive power across a selection of small-tick, medium-tick, and large-tick stocks. Empirical results demonstrate that signals derived from clusters consistently outperform benchmark models without clustering in all event types, especially in terms of SR and other risk-adjusted metrics. We further assess the robustness of \texttt{ClusterLOB} by decomposing performance across different event types (add, cancel, and trade), confirming that the \texttt{ClusterLOB} remains effective across a wide range of market conditions.

Overall, \texttt{ClusterLOB} advances our understanding of market microstructure by isolating meaningful trader behaviors from noisy high-frequency data. The framework not only enhances interpretability but also provides practical improvements for return forecasting and alpha generation, with broad implications for algorithmic trading and quantitative finance.

Several promising directions remain for future exploration. First, we aim to extend the clustering methodology to incorporate temporally adaptive or online clustering techniques, allowing for the identification of dynamic shifts in trader behavior over time. Second, integrating additional features such as order duration, queue position, and market state indicators could further enrich the trader behavioral clusters. Finally, applying \texttt{ClusterLOB} across multiple asset classes or international markets may help uncover universal patterns in order flow behavior, broadening its applicability.

\newpage
\printbibliography

\newpage
\appendix
\section{Appendix} \label{appendix}

\subsection{Add events performance for different tick-size groups}
\label{sec: add}

\begin{table}[h!]
\caption{Performance metric results for \textbf{FRNB} across \textbf{add events} for small-tick, medium-tick, and large-tick stocks in the test dataset. The highest SR within each tick-size group is highlighted.}
\label{tab: FRNB_add}
\resizebox{\linewidth}{!}{
\begin{tabular}{l|lll|lll|lll}
\toprule
\multicolumn{1}{l}{\textbf{Add events}}  &\multicolumn{3}{c}{\textbf{Small-tick stocks}}  &  \multicolumn{3}{c}{\textbf{Medium-tick stocks}} &  \multicolumn{3}{c}{\textbf{Large-tick stocks}}  \\
\cmidrule(lr){2-4} \cmidrule(lr){5-7}\cmidrule(lr){8-10}
\multicolumn{1}{l}{\textbf{FRNB}}  &\multicolumn{1}{c}{\textbf{Market-making}}  &  \multicolumn{2}{c}{\textbf{No cluster}} &  \multicolumn{1}{c}{\textbf{Directional}}  &  \multicolumn{2}{c}{\textbf{No cluster}} &\multicolumn{1}{c}{\textbf{Market-making}}  &  \multicolumn{2}{c}{\textbf{No cluster}} \\
\cmidrule(lr){2-2} \cmidrule(lr){3-4}\cmidrule(lr){5-5}\cmidrule(lr){6-7}\cmidrule(lr){8-8}\cmidrule(lr){9-10}

   & $OFI^{C}(\phi_{3})$ & $OFI^{S}(\phi_{*})$ & $OFI^{C}(\phi_{*})$ & $OFI^{S}(\phi_{1})$ & $OFI^{S}(\phi_{*})$ & $OFI^{C}(\phi_{*})$ & $OFI^{C}(\phi_{3})$ & $OFI^{S}(\phi_{*})$ & $OFI^{C}(\phi_{*})$ \\
\midrule

E{[}Return{]}           & 0.506  & 0.334 & 0.081  & 0.123  & -0.037 & 0.113  & -0.302 & 0.077  & 0.062 \\
Volatility              & 0.15   & 0.15  & 0.15   & 0.15   & 0.15   & 0.15   & 0.15   & 0.15   & 0.15  \\
Downside deviation      & 0.089  & 0.113 & 0.078  & 0.081  & 0.086  & 0.082  & 0.128  & 0.087  & 0.093 \\
Maximum drawdown        & -0.072 & -0.07 & -0.104 & -0.065 & -0.126 & -0.067 & -0.156 & -0.083 & -0.06 \\
Sortino ratio           & 5.71   & 2.953 & 1.042  & 1.518  & -0.433 & 1.377  & -2.357 & 0.887  & 0.67  \\
Calmar ratio            & 7.03   & 4.773 & 0.78   & 1.89   & -0.295 & 1.686  & -1.939 & 0.926  & 1.041 \\
Hit rate                & 0.569  & 0.578 & 0.474  & 0.466  & 0.474  & 0.483  & 0.466  & 0.474  & 0.491 \\
Avg. profit / avg. loss & 1.299  & 1.063 & 1.209  & 1.33   & 1.063  & 1.228  & 0.799  & 1.213  & 1.113 \\
PnL per trade           & 1.674  & 1.105 & 0.268  & 0.406  & -0.123 & 0.373  & -1     & 0.254  & 0.207 \\
Sharpe ratio            & \textbf{3.374}  & 2.228 & 0.541  & \textbf{0.819}  & -0.248 & 0.753  & -2.017 & \textbf{0.513}  & 0.416 \\
\bottomrule
\end{tabular}}

\end{table}

\begin{table}[h!]
\caption{Performance metric results for \textbf{FREB} across \textbf{add events} for small-tick, medium-tick, and large-tick stocks in the test dataset. The highest SR within each tick-size group is highlighted.}
\label{tab: FREB_add}
\resizebox{\linewidth}{!}{
\begin{tabular}{l|lll|lll|lll}
\toprule
\multicolumn{1}{l}{\textbf{Add events}}  &\multicolumn{3}{c}{\textbf{Small-tick stocks}}  &  \multicolumn{3}{c}{\textbf{Medium-tick stocks}} &  \multicolumn{3}{c}{\textbf{Large-tick stocks}}  \\
\cmidrule(lr){2-4} \cmidrule(lr){5-7}\cmidrule(lr){8-10}
\multicolumn{1}{l}{\textbf{FREB}}  &\multicolumn{1}{c}{\textbf{Market-making}}  &  \multicolumn{2}{c}{\textbf{No cluster}} &  \multicolumn{1}{c}{\textbf{Opportunistic}}  &  \multicolumn{2}{c}{\textbf{No cluster}} &\multicolumn{1}{c}{\textbf{Market-making}}  &  \multicolumn{2}{c}{\textbf{No cluster}} \\
\cmidrule(lr){2-2} \cmidrule(lr){3-4}\cmidrule(lr){5-5}\cmidrule(lr){6-7}\cmidrule(lr){8-8}\cmidrule(lr){9-10}

   & $OFI^{C}(\phi_{2})$ & $OFI^{S}(\phi_{*})$ & $OFI^{C}(\phi_{*})$ & $OFI^{S}(\phi_{1})$ & $OFI^{S}(\phi_{*})$ & $OFI^{C}(\phi_{*})$ & $OFI^{C}(\phi_{1})$ & $OFI^{S}(\phi_{*})$ & $OFI^{C}(\phi_{*})$ \\
\midrule
                        
E{[}Return{]}           & 0.147  & -0.12  & -0.106 & 0.395  & -0.032 & -0.234 & 0.045  & -0.244 & -0.109 \\
Volatility              & 0.15   & 0.15   & 0.15   & 0.15   & 0.15   & 0.15   & 0.15   & 0.15   & 0.15   \\
Downside deviation      & 0.093  & 0.102  & 0.114  & 0.072  & 0.095  & 0.104  & 0.148  & 0.092  & 0.074  \\
Maximum drawdown        & -0.072 & -0.112 & -0.101 & -0.046 & -0.105 & -0.135 & -0.064 & -0.138 & -0.08  \\
Sortino ratio           & 1.583  & -1.171 & -0.928 & 5.445  & -0.332 & -2.247 & 0.306  & -2.655 & -1.476 \\
Calmar ratio            & 2.035  & -1.071 & -1.046 & 8.577  & -0.3   & -1.736 & 0.706  & -1.767 & -1.358 \\
Hit rate                & 0.466  & 0.431  & 0.466  & 0.56   & 0.466  & 0.44   & 0.543  & 0.414  & 0.414  \\
Avg. profit / avg. loss & 1.351  & 1.149  & 1.008  & 1.287  & 1.108  & 0.977  & 0.894  & 1.062  & 1.228  \\
PnL per trade           & 0.485  & -0.397 & -0.349 & 1.305  & -0.104 & -0.775 & 0.149  & -0.806 & -0.359 \\
Sharpe ratio            & \textbf{0.977}  & -0.8   & -0.704 & \textbf{2.63}   & -0.21  & -1.563 & \textbf{0.301}  & -1.626 & -0.724 \\
\bottomrule
\end{tabular}}

\end{table}

\clearpage
\newpage
\subsection{Cancel events performance for different tick-size groups}
\label{sec: cancel}

\begin{table}[h!]
\caption{Performance metric results for \textbf{FRNB} across \textbf{cancel events} for small-tick, medium-tick, and large-tick stocks in the test dataset. The highest SR within each tick-size group is highlighted.}
\label{tab: FRNB_cancel}
\resizebox{\linewidth}{!}{
\begin{tabular}{l|lll|lll|lll}
\toprule
\multicolumn{1}{l}{\textbf{Cancel events}}  &\multicolumn{3}{c}{\textbf{Small-tick stocks}}  &  \multicolumn{3}{c}{\textbf{Medium-tick stocks}} &  \multicolumn{3}{c}{\textbf{Large-tick stocks}}  \\
\cmidrule(lr){2-4} \cmidrule(lr){5-7}\cmidrule(lr){8-10}
\multicolumn{1}{l}{\textbf{FRNB}}  &\multicolumn{1}{c}{\textbf{Opportunistic}}  &  \multicolumn{2}{c}{\textbf{No cluster}} &  \multicolumn{1}{c}{\textbf{Directional}}  &  \multicolumn{2}{c}{\textbf{No cluster}} &\multicolumn{1}{c}{\textbf{Directional}}  &  \multicolumn{2}{c}{\textbf{No cluster}} \\
\cmidrule(lr){2-2} \cmidrule(lr){3-4}\cmidrule(lr){5-5}\cmidrule(lr){6-7}\cmidrule(lr){8-8}\cmidrule(lr){9-10}

 & $OFI_{\phi_{2}}^{C}$ & $OFI_{\phi_{*}}^{S}$ & $OFI_{\phi_{*}}^{C}$ & $OFI_{\phi_{1}}^{C}$ & $OFI_{\phi_{*}}^{S}$ & $OFI_{\phi_{*}}^{C}$ & $OFI_{\phi_{1}}^{S}$ & $OFI_{\phi_{*}}^{S}$ & $OFI_{\phi_{*}}^{C}$ \\
\midrule

E{[}Return{]}           & -0.193 & -0.357 & -0.503 & 0.074  & -0.171 & -0.125 & 0.085 & 0.007  & 0.061  \\
Volatility              & 0.15   & 0.15   & 0.15   & 0.15   & 0.15   & 0.15   & 0.15  & 0.15   & 0.15   \\
Downside deviation      & 0.098  & 0.107  & 0.119  & 0.101  & 0.127  & 0.134  & 0.095 & 0.098  & 0.118  \\
Maximum drawdown        & -0.167 & -0.208 & -0.229 & -0.067 & -0.166 & -0.16  & -0.06 & -0.099 & -0.093 \\
Sortino ratio           & -1.971 & -3.331 & -4.228 & 0.734  & -1.342 & -0.929 & 0.891 & 0.074  & 0.516  \\
Calmar ratio            & -1.158 & -1.717 & -2.195 & 1.104  & -1.031 & -0.781 & 1.412 & 0.074  & 0.657  \\
Hit rate                & 0.5    & 0.474  & 0.5    & 0.534  & 0.491  & 0.5    & 0.5   & 0.491  & 0.526  \\
Avg. profit / avg. loss & 0.808  & 0.739  & 0.567  & 0.949  & 0.847  & 0.86   & 1.096 & 1.043  & 0.968  \\
PnL per trade           & -0.64  & -1.181 & -1.662 & 0.245  & -0.566 & -0.413 & 0.28  & 0.024  & 0.202  \\
Sharpe ratio            & \textbf{-1.29}  & -2.381 & -3.351 & \textbf{0.493}  & -1.141 & -0.833 & \textbf{0.565} & 0.049  & 0.407  \\
\bottomrule
\end{tabular}}

\end{table}

\begin{table}[h!]
\caption{Performance metric results for \textbf{FREB} across \textbf{cancel events} for small-tick, medium-tick, and large-tick stocks in the test dataset. The highest SR within each tick-size group is highlighted.}
\label{tab: FREB_cancel}
\resizebox{\linewidth}{!}{
\begin{tabular}{l|lll|lll|lll}
\toprule
\multicolumn{1}{l}{\textbf{Cancel events}}  &\multicolumn{3}{c}{\textbf{Small-tick stocks}}  &  \multicolumn{3}{c}{\textbf{Medium-tick stocks}} &  \multicolumn{3}{c}{\textbf{Large-tick stocks}}  \\
\cmidrule(lr){2-4} \cmidrule(lr){5-7}\cmidrule(lr){8-10}
\multicolumn{1}{l}{\textbf{FREB}}  &\multicolumn{1}{c}{\textbf{Opportunistic}}  &  \multicolumn{2}{c}{\textbf{No cluster}} &  \multicolumn{1}{c}{\textbf{Directional}}  &  \multicolumn{2}{c}{\textbf{No cluster}} &\multicolumn{1}{c}{\textbf{Market-making}}  &  \multicolumn{2}{c}{\textbf{No cluster}} \\
\cmidrule(lr){2-2} \cmidrule(lr){3-4}\cmidrule(lr){5-5}\cmidrule(lr){6-7}\cmidrule(lr){8-8}\cmidrule(lr){9-10}

   & $OFI^{C}(\phi_{2})$ & $OFI^{S}(\phi_{*})$ & $OFI^{C}(\phi_{*})$ & $OFI^{S}(\phi_{1})$ & $OFI^{S}(\phi_{*})$ & $OFI^{C}(\phi_{*})$ & $OFI^{C}(\phi_{3})$ & $OFI^{S}(\phi_{*})$ & $OFI^{C}(\phi_{*})$ \\
\midrule

E{[}Return{]}           & -0.253 & -0.313 & -0.338 & -0.164 & -0.038 & 0.017  & 0.016  & 0.076  & -0.065 \\
Volatility              & 0.15   & 0.15   & 0.15   & 0.15   & 0.15   & 0.15   & 0.15   & 0.15   & 0.15   \\
Downside deviation      & 0.117  & 0.121  & 0.124  & 0.126  & 0.116  & 0.121  & 0.181  & 0.17   & 0.155  \\
Maximum drawdown        & -0.156 & -0.197 & -0.171 & -0.132 & -0.171 & -0.137 & -0.073 & -0.072 & -0.108 \\
Sortino ratio           & -2.156 & -2.575 & -2.726 & -1.298 & -0.324 & 0.141  & 0.087  & 0.444  & -0.42  \\
Calmar ratio            & -1.62  & -1.586 & -1.978 & -1.239 & -0.22  & 0.125  & 0.215  & 1.049  & -0.601 \\
Hit rate                & 0.431  & 0.474  & 0.457  & 0.517  & 0.569  & 0.578  & 0.595  & 0.586  & 0.534  \\
Avg. profit / avg. loss & 0.972  & 0.764  & 0.782  & 0.771  & 0.725  & 0.746  & 0.698  & 0.786  & 0.799  \\
PnL per trade           & -0.836 & -1.033 & -1.119 & -0.541 & -0.124 & 0.057  & 0.052  & 0.25   & -0.214 \\
Sharpe ratio            & \textbf{-1.684} & -2.083 & -2.255 & -1.09  & -0.251 & \textbf{0.114}  & 0.105  & \textbf{0.503}  & -0.432 \\
\bottomrule
\end{tabular}}

\end{table}

\clearpage
\newpage
\subsection{Trade events performance for different tick-size groups}
\label{sec: trade}

\begin{table}[h!]
\caption{Performance metric results for \textbf{FRNB} across \textbf{trade events} for small-tick, medium-tick, and large-tick stocks in the test dataset. The highest SR within each tick-size group is highlighted.}
\label{tab: FRNB_trade}
\resizebox{\linewidth}{!}{
\begin{tabular}{l|lll|lll|lll}
\toprule
\multicolumn{1}{l}{\textbf{Trade events}}  &\multicolumn{3}{c}{\textbf{Small-tick stocks}}  &  \multicolumn{3}{c}{\textbf{Medium-tick stocks}} &  \multicolumn{3}{c}{\textbf{Large-tick stocks}}  \\
\cmidrule(lr){2-4} \cmidrule(lr){5-7}\cmidrule(lr){8-10}
\multicolumn{1}{l}{\textbf{FRNB}}  &\multicolumn{1}{c}{\textbf{Opportunistic}}  &  \multicolumn{2}{c}{\textbf{No cluster}} &  \multicolumn{1}{c}{\textbf{Directional}}  &  \multicolumn{2}{c}{\textbf{No cluster}} &\multicolumn{1}{c}{\textbf{Opportunistic}}  &  \multicolumn{2}{c}{\textbf{No cluster}} \\
\cmidrule(lr){2-2} \cmidrule(lr){3-4}\cmidrule(lr){5-5}\cmidrule(lr){6-7}\cmidrule(lr){8-8}\cmidrule(lr){9-10}

   & $OFI^{C}(\phi_{2})$ & $OFI^{S}(\phi_{*})$ & $OFI^{C}(\phi_{*})$ & $OFI^{S}(\phi_{1})$ & $OFI^{S}(\phi_{*})$ & $OFI^{C}(\phi_{*})$ & $OFI^{C}(\phi_{2})$ & $OFI^{S}(\phi_{*})$ & $OFI^{C}(\phi_{*})$ \\
\midrule

E{[}Return{]}           & 0.098  & 0.035  & 0.102  & -0.09  & -0.134 & -0.053 & 0.07   & 0.121 & 0.211  \\
Volatility              & 0.15   & 0.15   & 0.15   & 0.15   & 0.15   & 0.15   & 0.15   & 0.15  & 0.15   \\
Downside deviation      & 0.106  & 0.09   & 0.088  & 0.11   & 0.114  & 0.122  & 0.092  & 0.08  & 0.096  \\
Maximum drawdown        & -0.061 & -0.119 & -0.062 & -0.154 & -0.154 & -0.132 & -0.092 & -0.07 & -0.084 \\
Sortino ratio           & 0.922  & 0.392  & 1.16   & -0.815 & -1.177 & -0.434 & 0.765  & 1.506 & 2.194  \\
Calmar ratio            & 1.609  & 0.298  & 1.642  & -0.582 & -0.872 & -0.401 & 0.764  & 1.724 & 2.517  \\
Hit rate                & 0.569  & 0.517  & 0.543  & 0.509  & 0.474  & 0.509  & 0.517  & 0.543 & 0.526  \\
Avg. profit / avg. loss & 0.844  & 0.969  & 0.937  & 0.877  & 0.956  & 0.909  & 1.005  & 0.95  & 1.127  \\
PnL per trade           & 0.325  & 0.117  & 0.337  & -0.296 & -0.444 & -0.175 & 0.232  & 0.399 & 0.699  \\
Sharpe ratio            & 0.654  & 0.236  & \textbf{0.679}  & -0.597 & -0.896 & \textbf{-0.353} & 0.469  & 0.805 & \textbf{1.41}   \\
\bottomrule
\end{tabular}}

\end{table}

\begin{table}[h!]
\caption{Performance metric results for \textbf{FREB} across \textbf{trade events} for small-tick, medium-tick, and large-tick stocks in the test dataset. The highest SR within each tick-size group is highlighted.}
\label{tab: FREB_trade}
\resizebox{\linewidth}{!}{
\begin{tabular}{l|lll|lll|lll}
\toprule
\multicolumn{1}{l}{\textbf{Trade events}}  &\multicolumn{3}{c}{\textbf{Small-tick stocks}}  &  \multicolumn{3}{c}{\textbf{Medium-tick stocks}} &  \multicolumn{3}{c}{\textbf{Large-tick stocks}}  \\
\cmidrule(lr){2-4} \cmidrule(lr){5-7}\cmidrule(lr){8-10}
\multicolumn{1}{l}{\textbf{FREB}}  &\multicolumn{1}{c}{\textbf{Opportunistic}}  &  \multicolumn{2}{c}{\textbf{No cluster}} &  \multicolumn{1}{c}{\textbf{Market-making}}  &  \multicolumn{2}{c}{\textbf{No cluster}} &\multicolumn{1}{c}{\textbf{Opportunistic}}  &  \multicolumn{2}{c}{\textbf{No cluster}} \\
\cmidrule(lr){2-2} \cmidrule(lr){3-4}\cmidrule(lr){5-5}\cmidrule(lr){6-7}\cmidrule(lr){8-8}\cmidrule(lr){9-10}

   & $OFI^{C}(\phi_{2})$ & $OFI^{S}(\phi_{*})$ & $OFI^{C}(\phi_{*})$ & $OFI^{S}(\phi_{3})$ & $OFI^{S}(\phi_{*})$ & $OFI^{C}(\phi_{*})$ & $OFI^{C}(\phi_{2})$ & $OFI^{S}(\phi_{*})$ & $OFI^{C}(\phi_{*})$ \\
\midrule

E{[}Return{]}           & 0.492  & 0.352  & 0.409  & 0.204  & 0.176 & 0.195  & 0.068  & -0.064 & 0.21   \\
Volatility              & 0.15   & 0.15   & 0.15   & 0.15   & 0.15  & 0.15   & 0.15   & 0.15   & 0.15   \\
Downside deviation      & 0.079  & 0.08   & 0.09   & 0.116  & 0.077 & 0.092  & 0.11   & 0.156  & 0.139  \\
Maximum drawdown        & -0.036 & -0.063 & -0.072 & -0.083 & -0.08 & -0.113 & -0.088 & -0.086 & -0.062 \\
Sortino ratio           & 6.22   & 4.397  & 4.528  & 1.761  & 2.29  & 2.114  & 0.616  & -0.413 & 1.512  \\
Calmar ratio            & 13.664 & 5.588  & 5.676  & 2.459  & 2.2   & 1.726  & 0.769  & -0.747 & 3.391  \\
Hit rate                & 0.56   & 0.509  & 0.543  & 0.586  & 0.552 & 0.595  & 0.509  & 0.5    & 0.56   \\
Avg. profit / avg. loss & 1.416  & 1.413  & 1.353  & 0.908  & 1.003 & 0.856  & 1.043  & 0.915  & 1.018  \\
PnL per trade           & 1.627  & 1.164  & 1.351  & 0.675  & 0.582 & 0.645  & 0.224  & -0.212 & 0.695  \\
Sharpe ratio            & \textbf{3.279}  & 2.347  & 2.725  & \textbf{1.361}  & 1.174 & 1.3    & 0.451  & -0.428 & \textbf{1.402}  \\
\bottomrule
\end{tabular}}

\end{table}

\clearpage
\newpage
\subsection{Small-tick stocks performance for different event types}
\label{sec: small}

\begin{table}[htbp]
\centering
\caption{The average cumulative PnL of \textbf{FRNB} for \textbf{small-tick} stocks, rescaled to a target volatility of 15\%. From top to bottom, this corresponds to all, add, cancel, and trade event types.}
\label{tab: small_tick_top_FRNB}

  \centering
    \begin{tabular}{p{0.1cm}p{6.8cm}p{6.8cm}}
        \toprule
&\multicolumn{1}{c}{\textbf{Training}} &  \multicolumn{1}{c}{\textbf{Test}} \\
        \midrule
    \vspace{-2.5cm}
    \multirow{1}{*}{\rotatebox[origin=c]{90}{\textbf{All events}}}
    &
      \includegraphics[width=\linewidth,trim=0cm 0cm 0cm 0cm,clip]{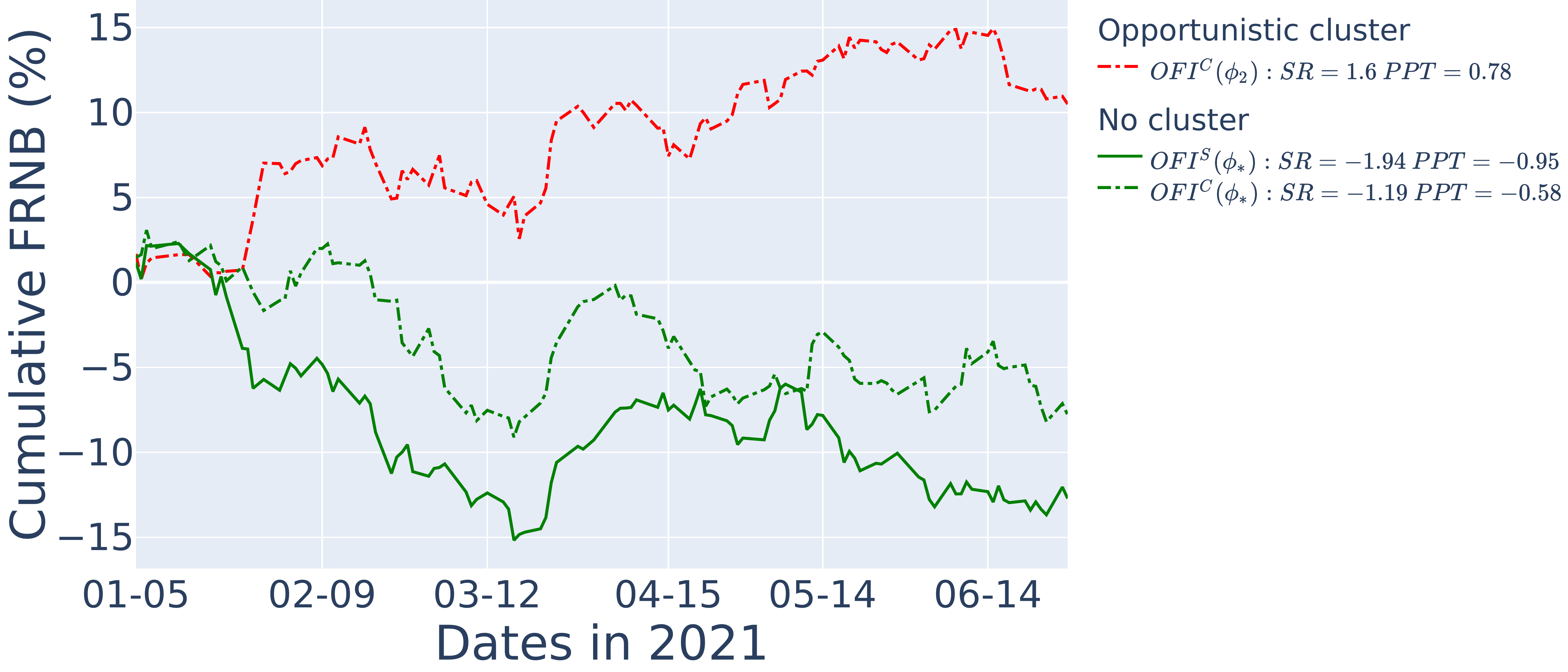}
    &
          \includegraphics[width=\linewidth,trim=0cm 0cm 0cm 0cm,clip]{plots/small/test_top_FRNB_A.pdf}
      \\
    \midrule
        \vspace{-2.5cm}
        \multirow{1}{*}{\rotatebox[origin=c]{90}{\textbf{Add events}}}
    &
      \includegraphics[width=\linewidth,trim=0cm 0cm 0cm 0cm,clip]{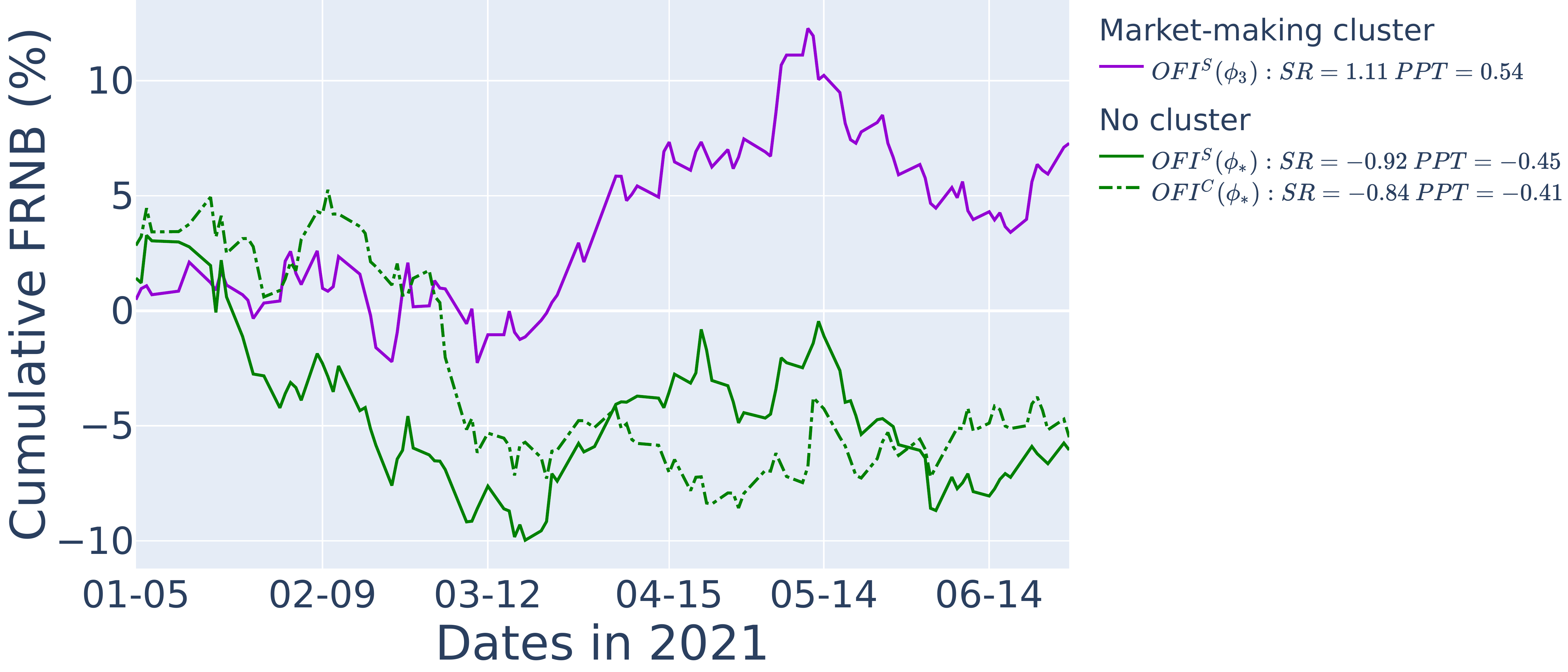}
    &
          \includegraphics[width=\linewidth,trim=0cm 0cm 0cm 0cm,clip]{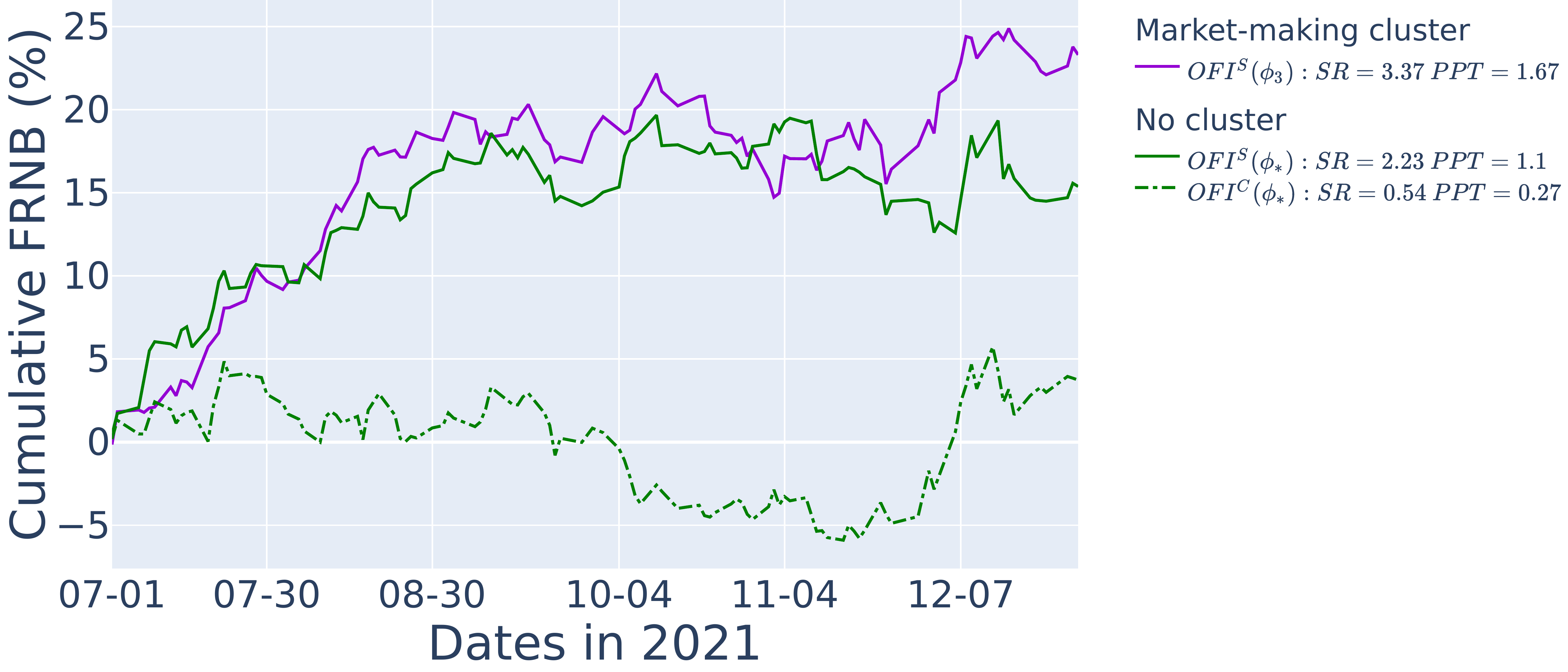}
    \\
    \midrule
    \vspace{-2.5cm}
        \multirow{1}{*}{\rotatebox[origin=c]{90}{\textbf{Cancel events}}}
    &
      \includegraphics[width=\linewidth,trim=0cm 0cm 0cm 0cm,clip]{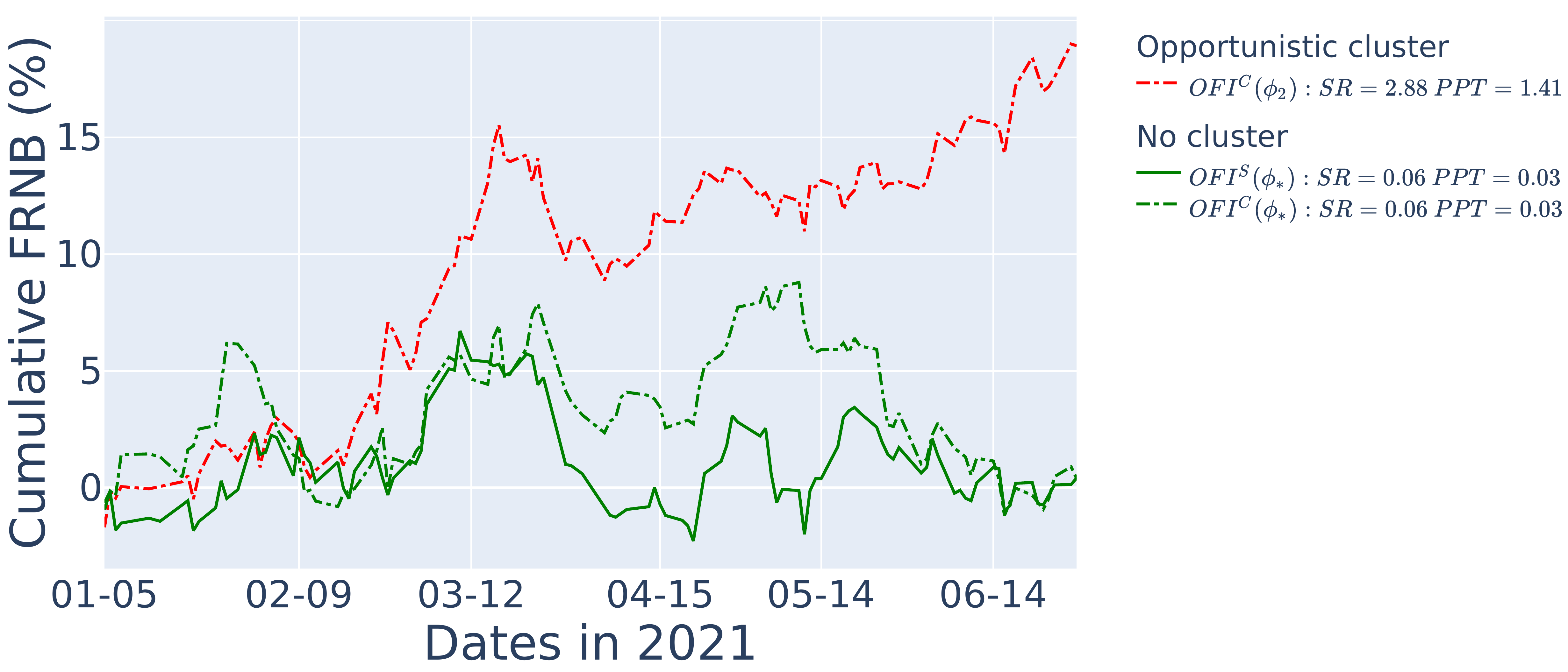}
    &
          \includegraphics[width=\linewidth,trim=0cm 0cm 0cm 0cm,clip]{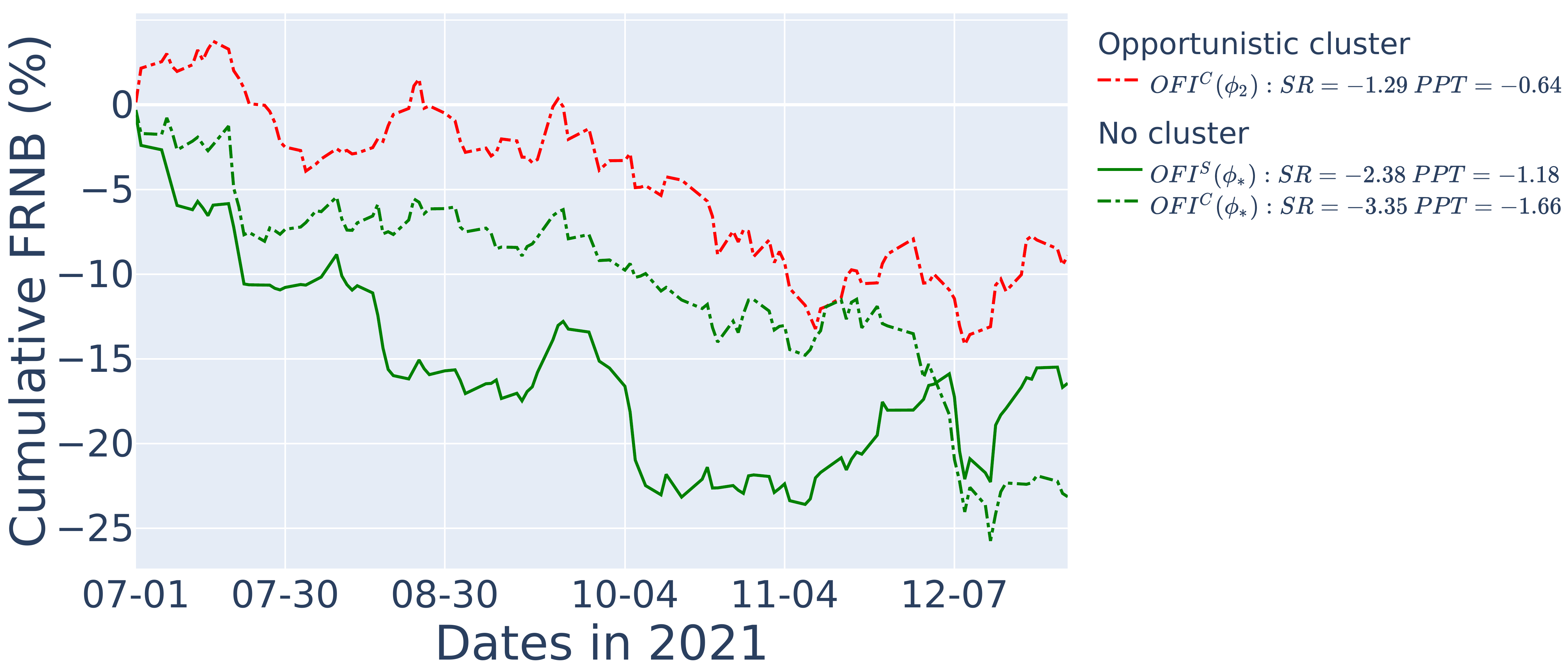}
        \\
    \midrule
    \vspace{-2.5cm}
        \multirow{1}{*}{\rotatebox[origin=c]{90}{\textbf{Trade events}}}
    &
      \includegraphics[width=\linewidth,trim=0cm 0cm 0cm 0cm,clip]{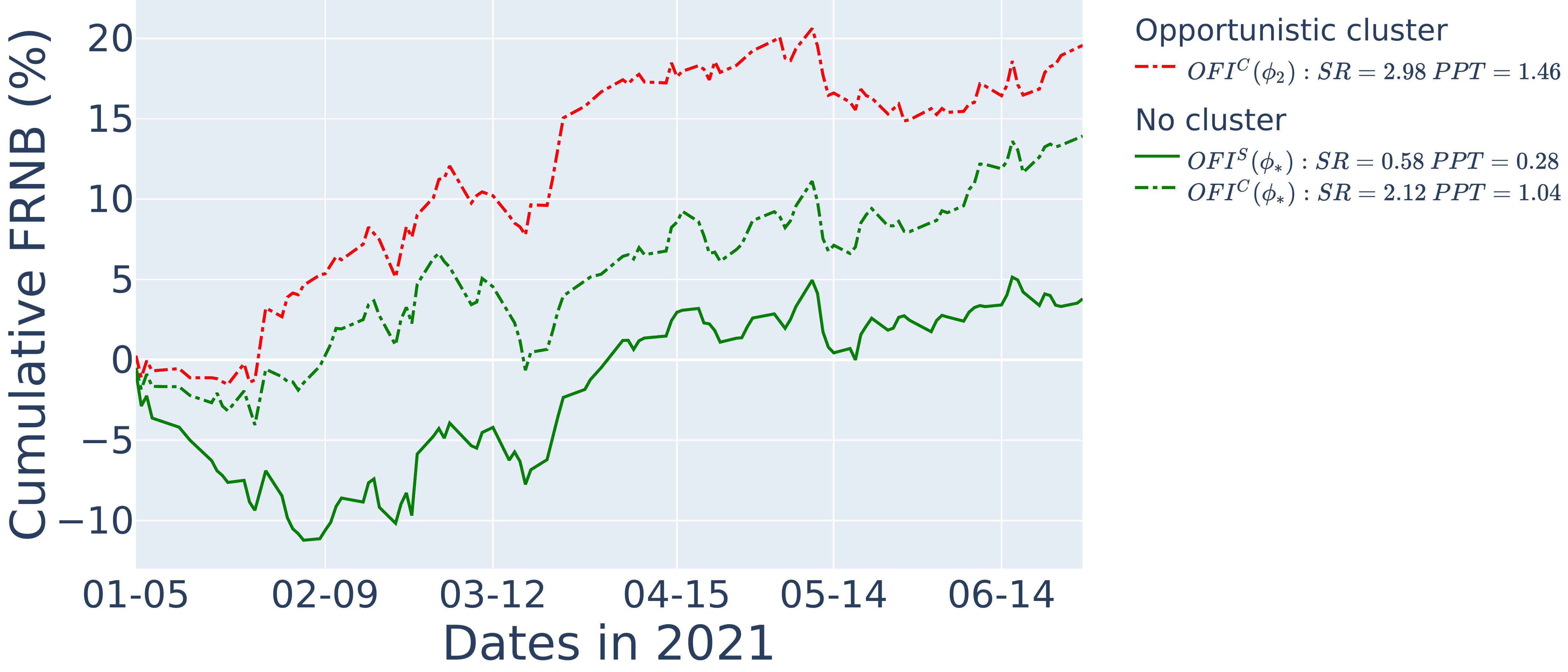}
    &
          \includegraphics[width=\linewidth,trim=0cm 0cm 0cm 0cm,clip]{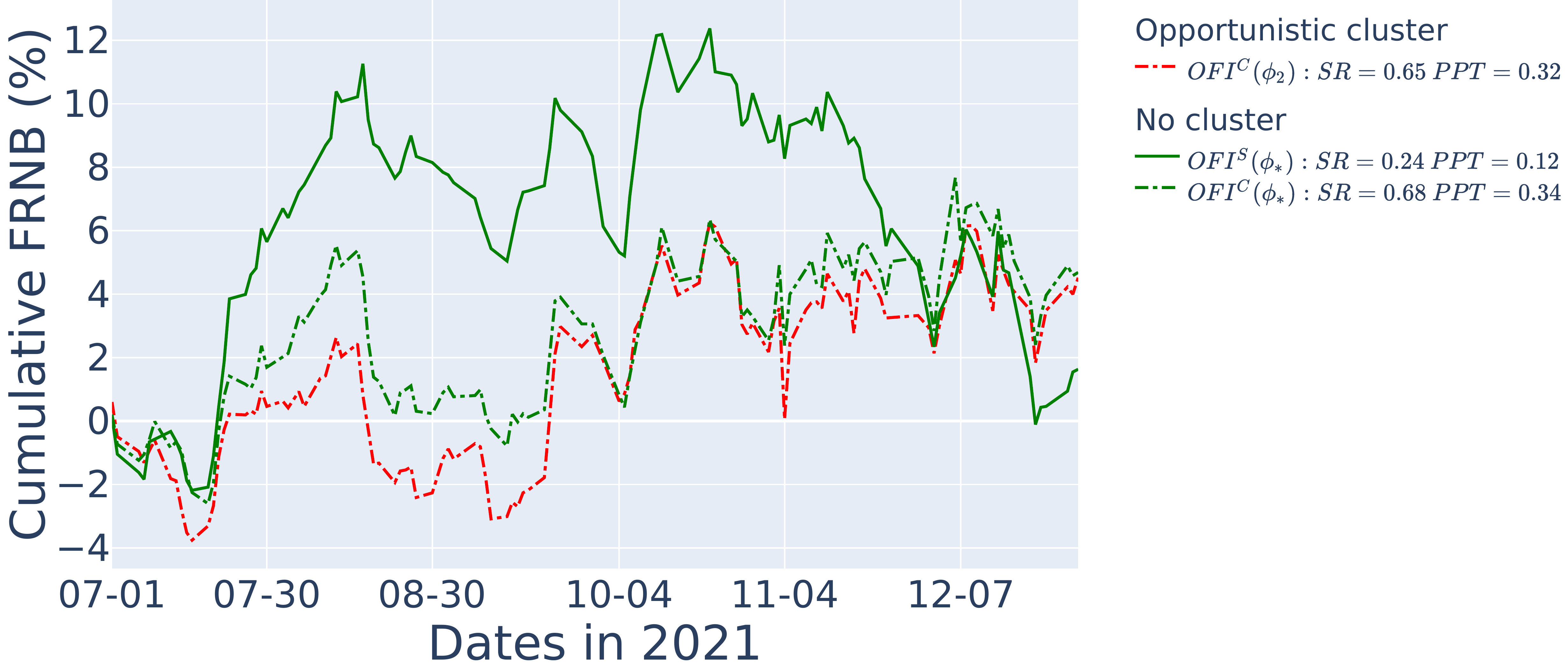} 
          \\
    \bottomrule

    \end{tabular}

\end{table}

\clearpage
\newpage

\begin{table}[htbp]
\centering
\caption{The average cumulative PnL of \textbf{FREB} for \textbf{small-tick} stocks, rescaled to a target volatility of 15\%. From top to bottom, this corresponds to all, add, cancel, and trade event types.}
\label{tab: small_tick_top_FREB}

  \centering
    \begin{tabular}{p{0.1cm}p{6.8cm}p{6.8cm}}
        \toprule
&\multicolumn{1}{c}{\textbf{Training}} &  \multicolumn{1}{c}{\textbf{Test}} \\
        \midrule
   \vspace{-2.5cm}
    \multirow{1}{*}{\rotatebox[origin=c]{90}{\textbf{All events}}}
    &
      \includegraphics[width=\linewidth,trim=0cm 0cm 0cm 0cm,clip]{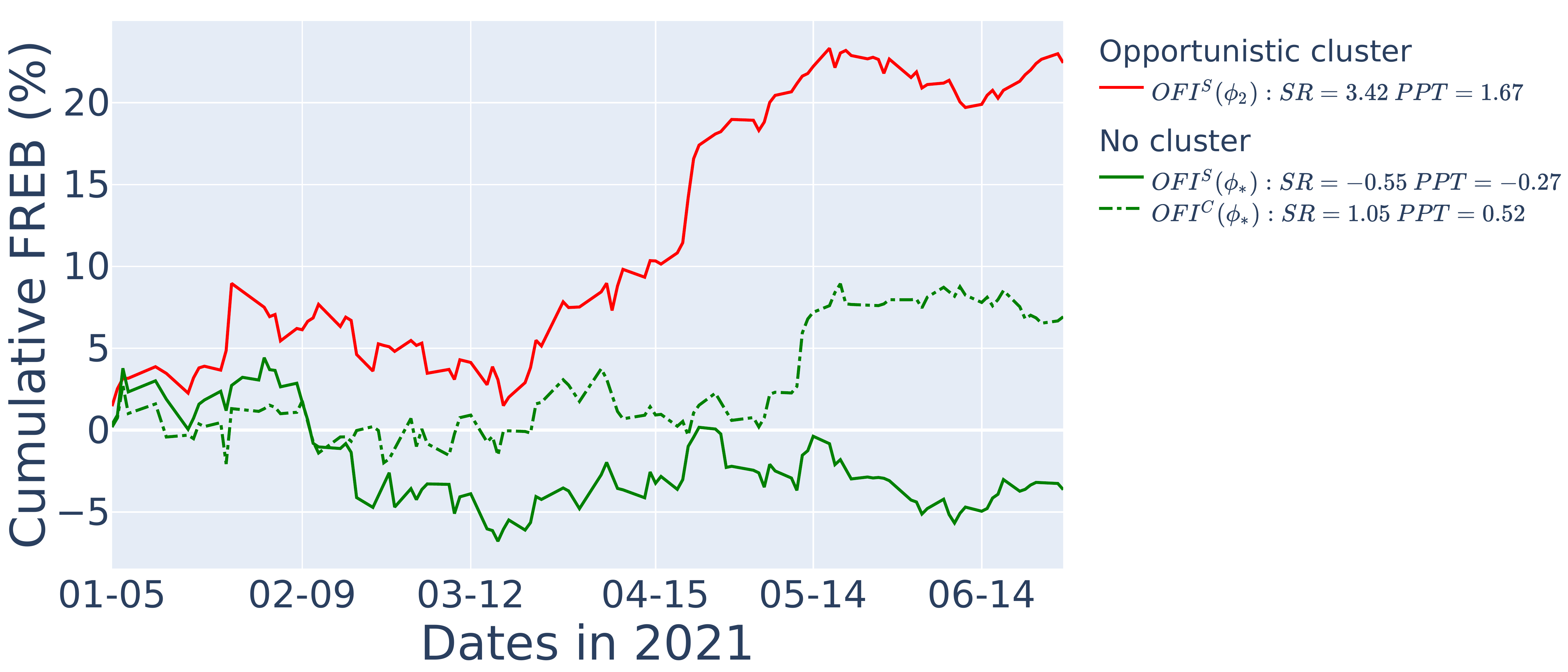}
    &
          \includegraphics[width=\linewidth,trim=0cm 0cm 0cm 0cm,clip]{plots/small/test_top_FREB_A.pdf}
      \\
    \midrule
        \vspace{-2.5cm}
        \multirow{1}{*}{\rotatebox[origin=c]{90}{\textbf{Add events}}}
    &
      \includegraphics[width=\linewidth,trim=0cm 0cm 0cm 0cm,clip]{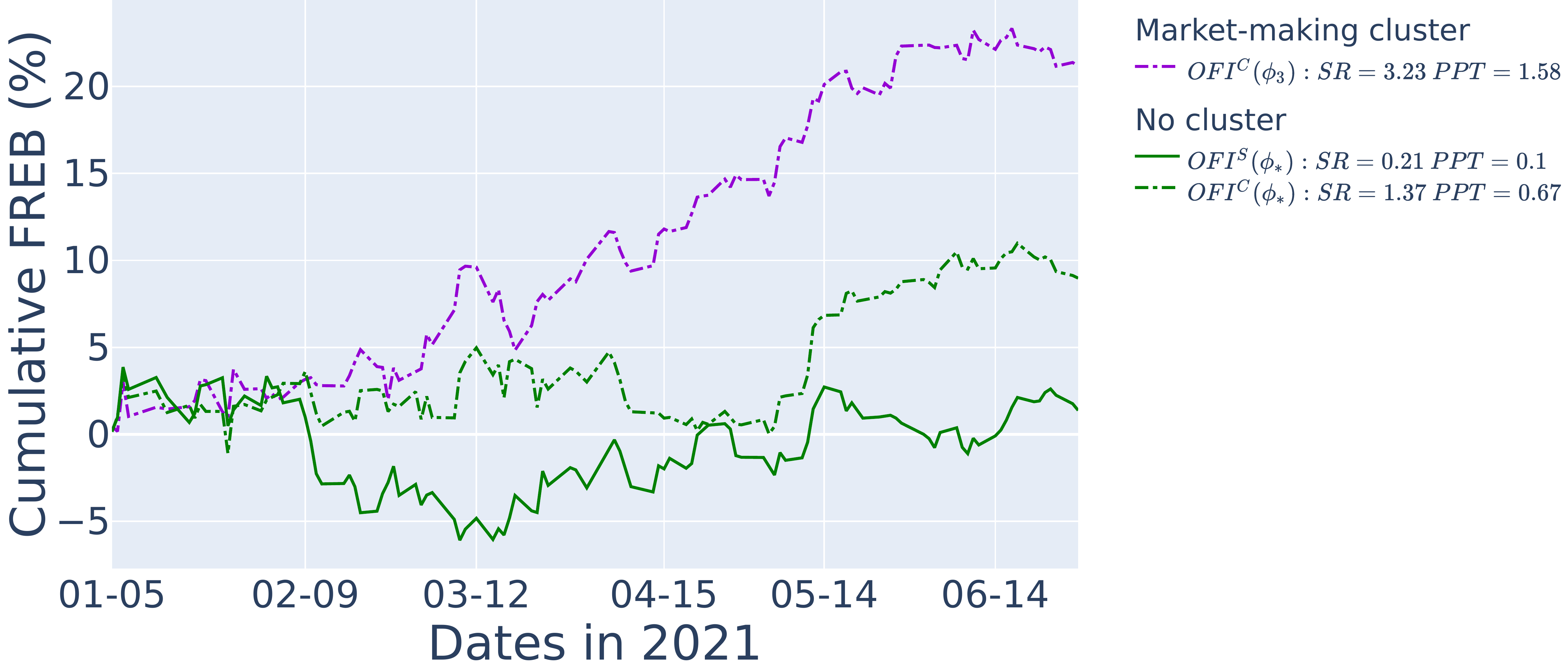}
    &
          \includegraphics[width=\linewidth,trim=0cm 0cm 0cm 0cm,clip]{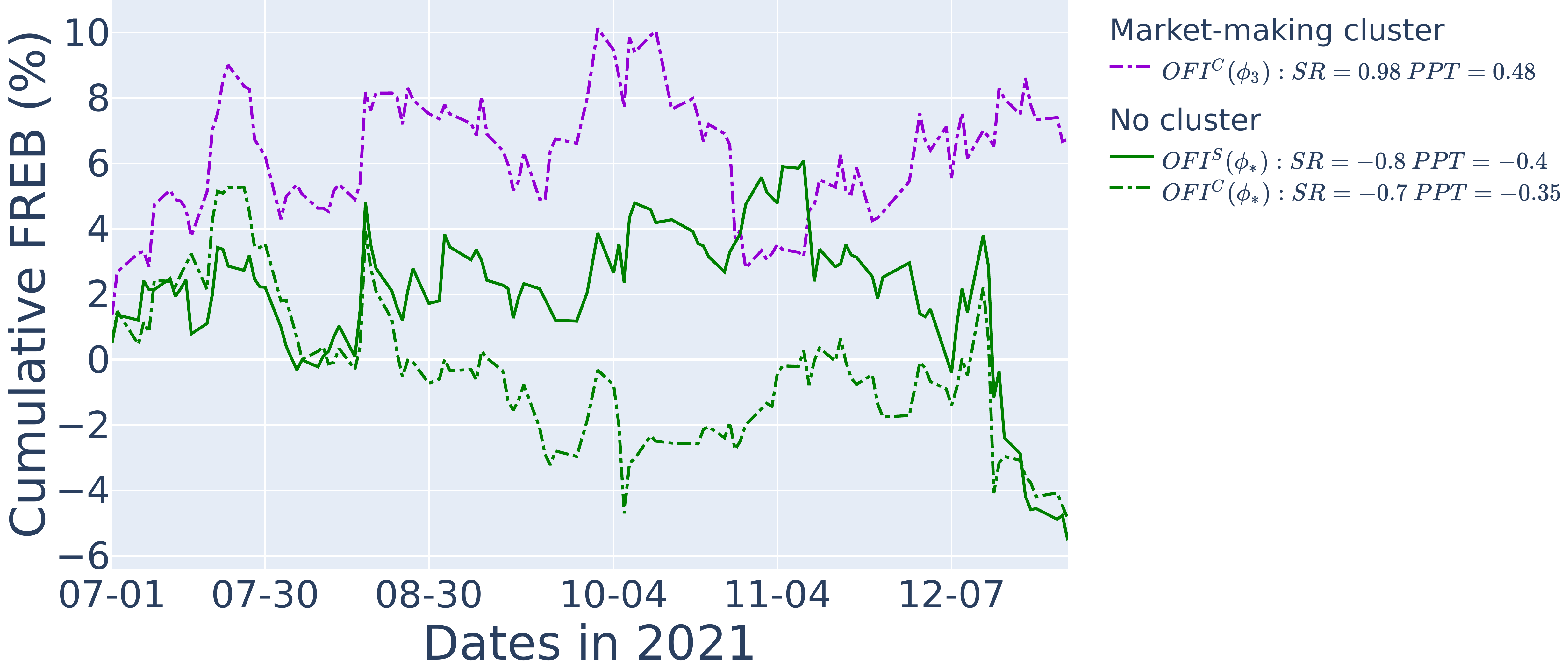}
    \\
    \midrule
    \vspace{-2.5cm}
        \multirow{1}{*}{\rotatebox[origin=c]{90}{\textbf{Cancel events}}}
    &
      \includegraphics[width=\linewidth,trim=0cm 0cm 0cm 0cm,clip]{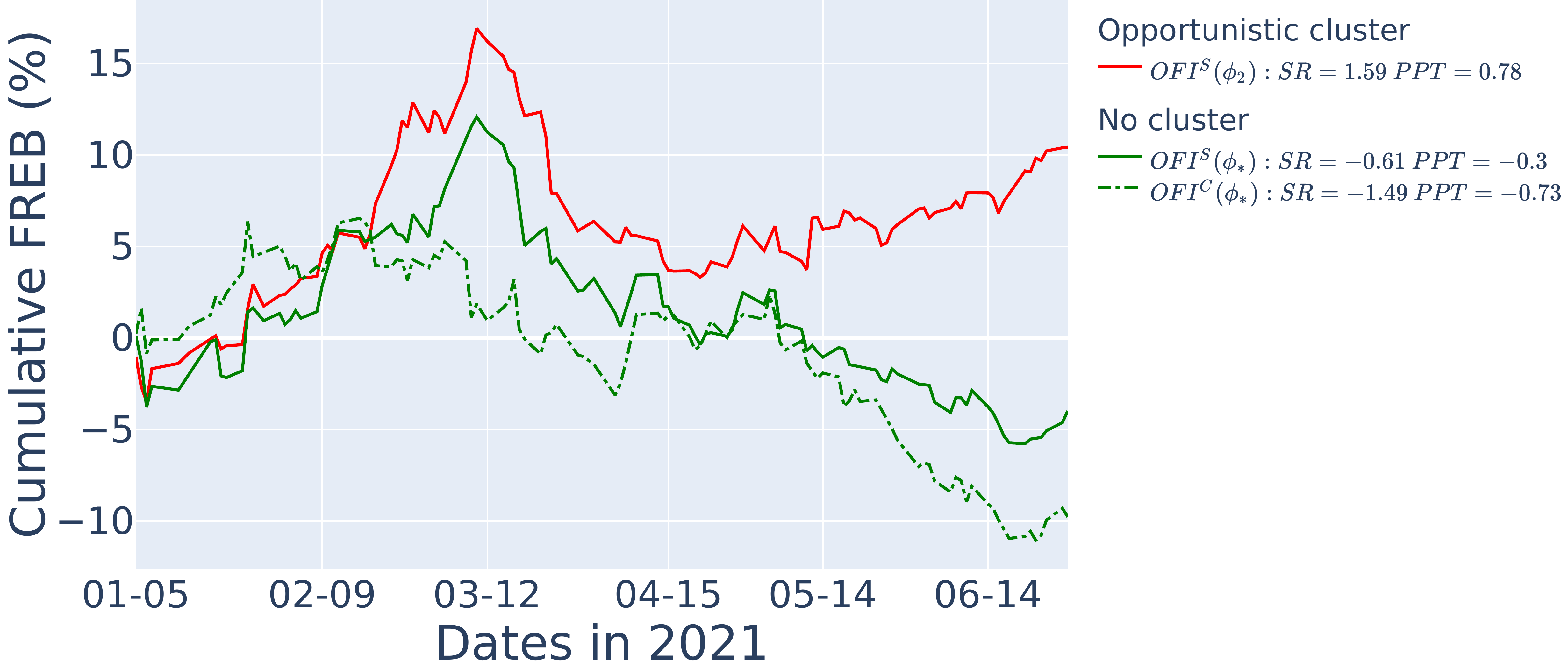}
    &
          \includegraphics[width=\linewidth,trim=0cm 0cm 0cm 0cm,clip]{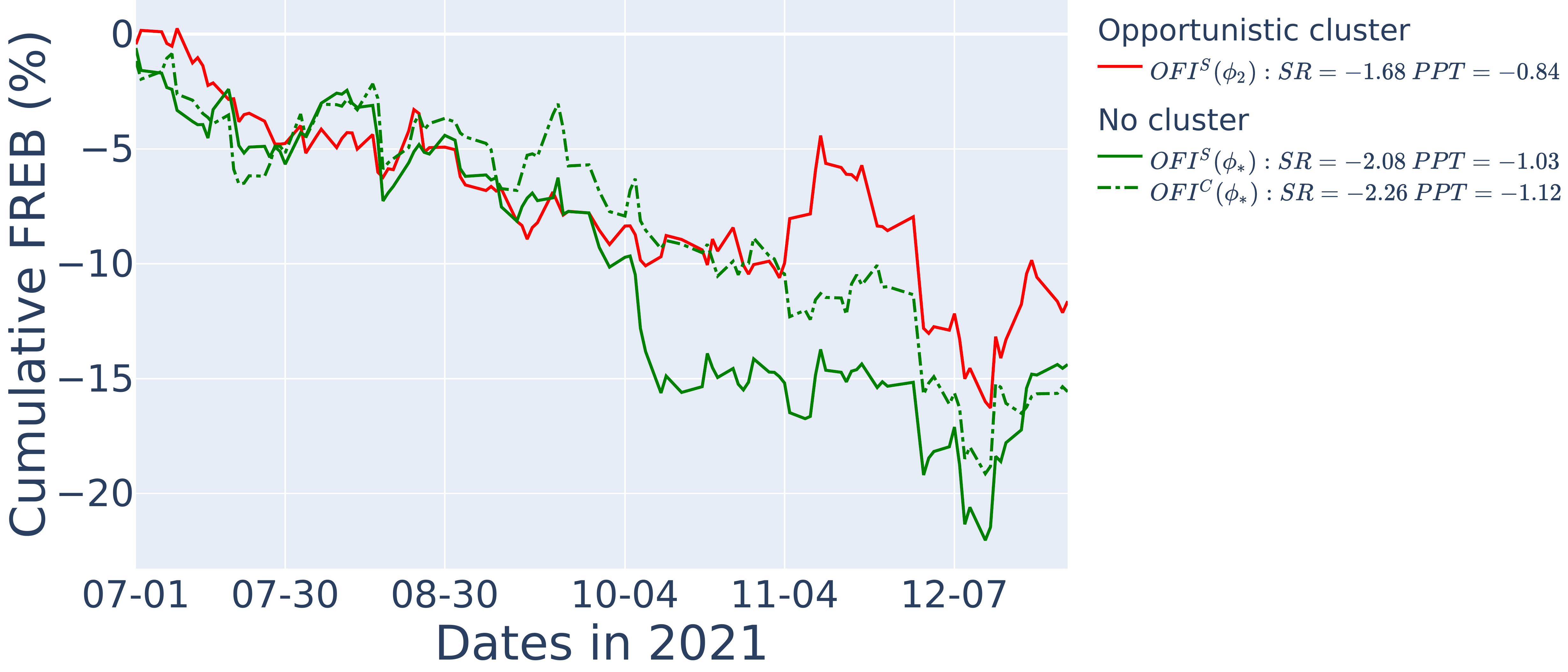}
        \\
    \midrule
  \vspace{-2.5cm}
        \multirow{1}{*}{\rotatebox[origin=c]{90}{\textbf{Trade events}}}
    &
      \includegraphics[width=\linewidth,trim=0cm 0cm 0cm 0cm,clip]{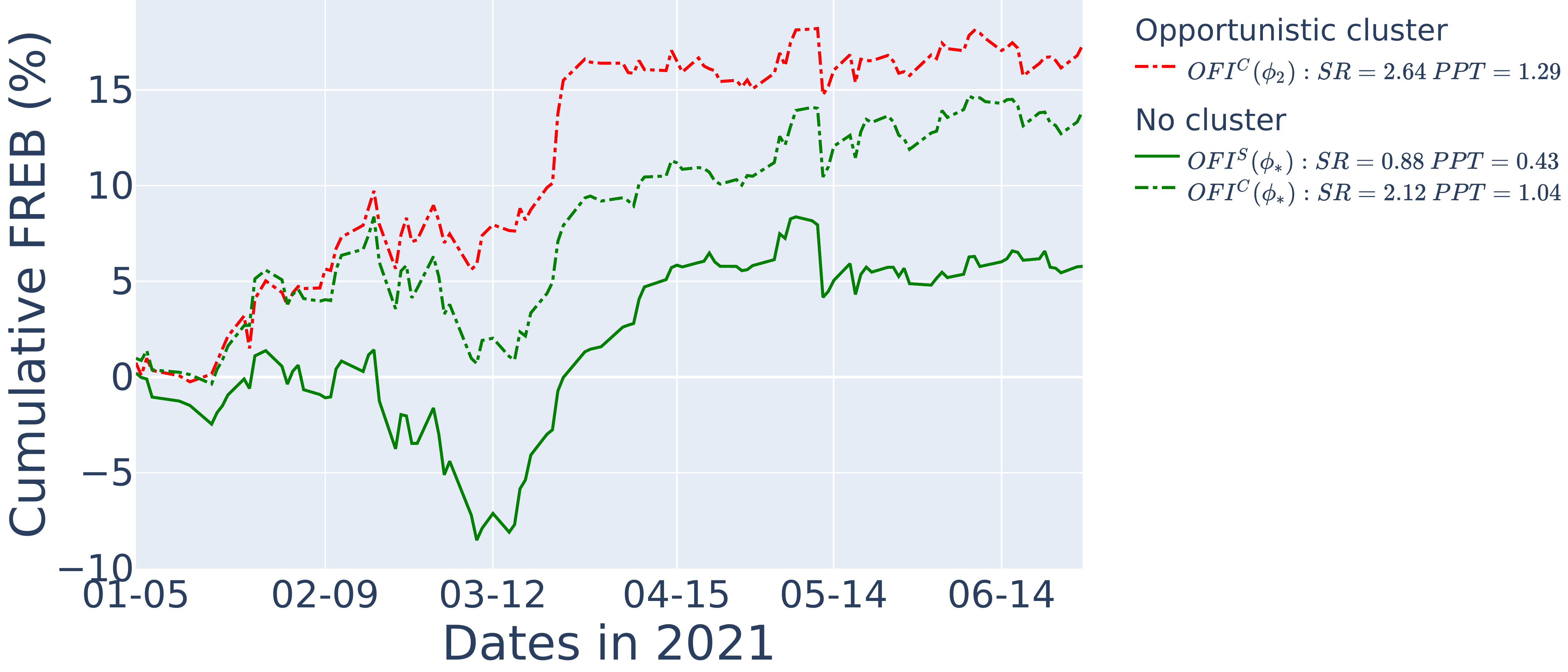}
    &
          \includegraphics[width=\linewidth,trim=0cm 0cm 0cm 0cm,clip]{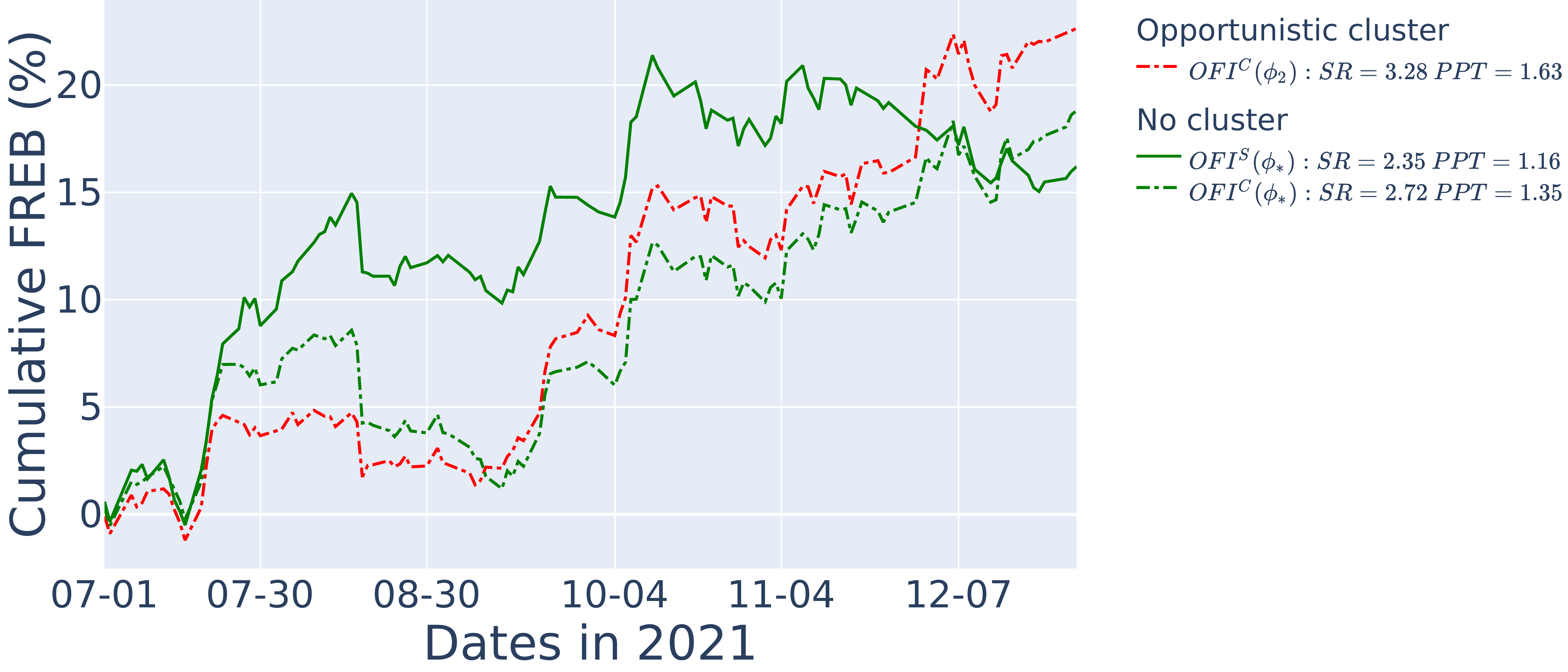} 
          \\
\bottomrule

    \end{tabular}

\end{table}

\clearpage
\newpage
\subsection{Medium-tick stocks performance for different event types}
\label{sec: medium}

\begin{table}[htbp]
\centering
\caption{The average cumulative PnL of \textbf{FRNB} for \textbf{medium-tick} stocks, rescaled to a target volatility of 15\%. From top to bottom, this corresponds to all, add, cancel, and trade event types.}
\label{tab: medium_tick_FRNB}
  \centering
    \begin{tabular}{p{0.1cm}p{6.8cm}p{6.8cm}}
        \toprule
&\multicolumn{1}{c}{\textbf{Training}} &  \multicolumn{1}{c}{\textbf{Test}} \\
        \midrule
   \vspace{-2.5cm}
    \multirow{1}{*}{\rotatebox[origin=c]{90}{\textbf{All events}}}
    &
      \includegraphics[width=\linewidth,trim=0cm 0cm 0cm 0cm,clip]{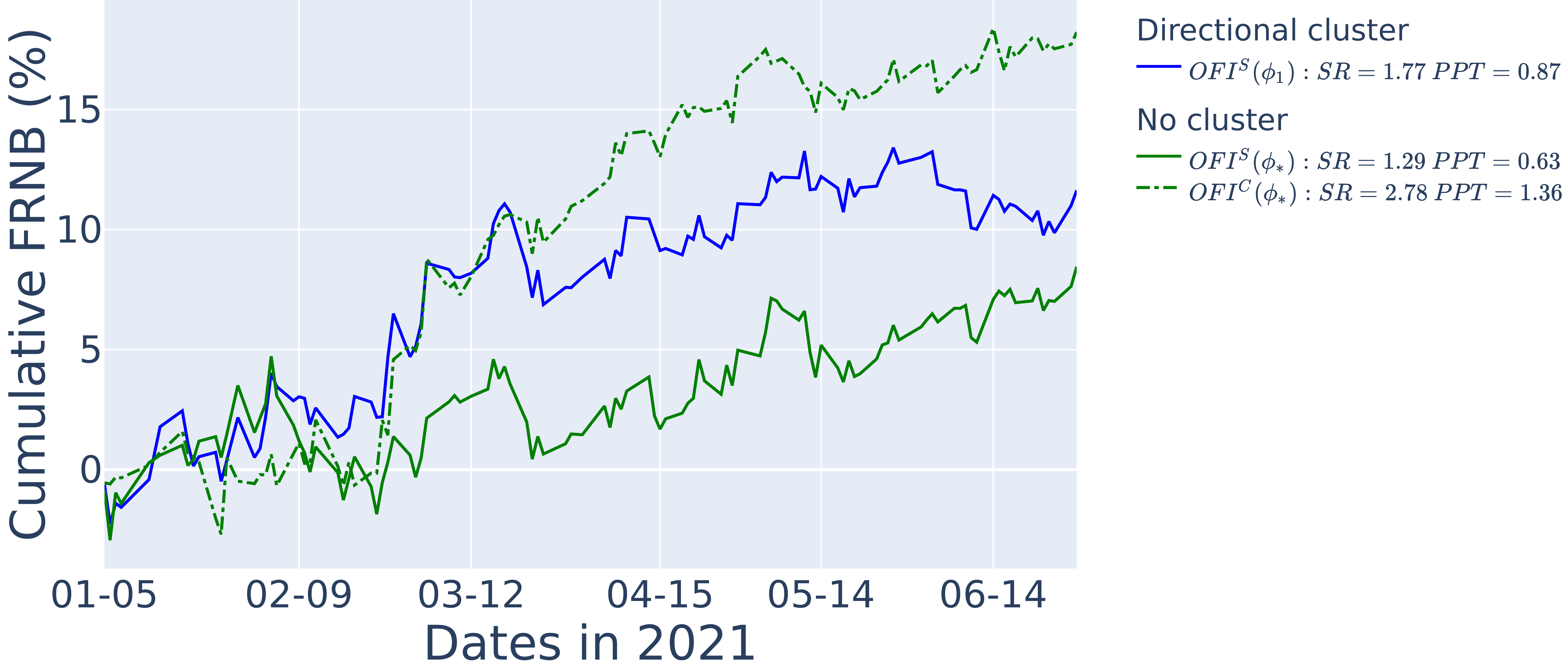}
    &
          \includegraphics[width=\linewidth,trim=0cm 0cm 0cm 0cm,clip]{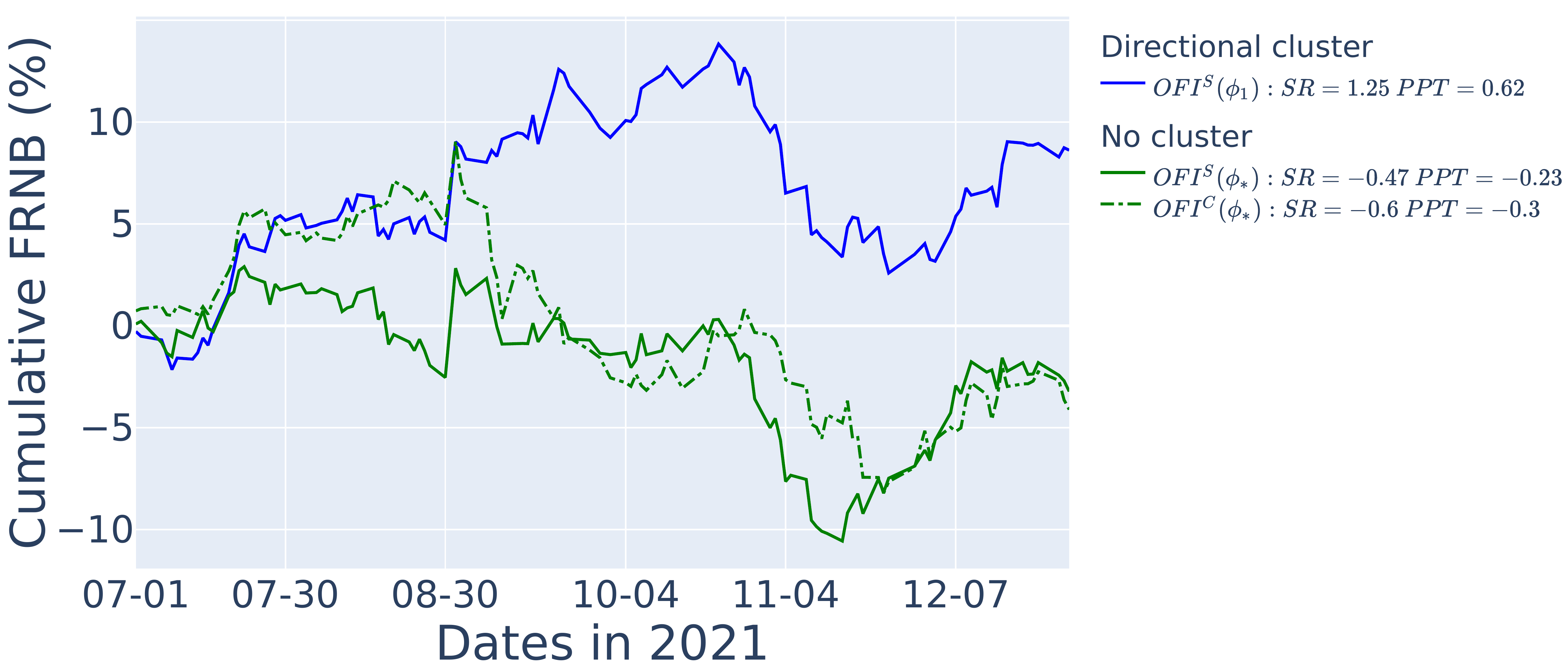}
      \\
    \midrule
        \vspace{-2.5cm}
        \multirow{1}{*}{\rotatebox[origin=c]{90}{\textbf{Add events}}}
    &
      \includegraphics[width=\linewidth,trim=0cm 0cm 0cm 0cm,clip]{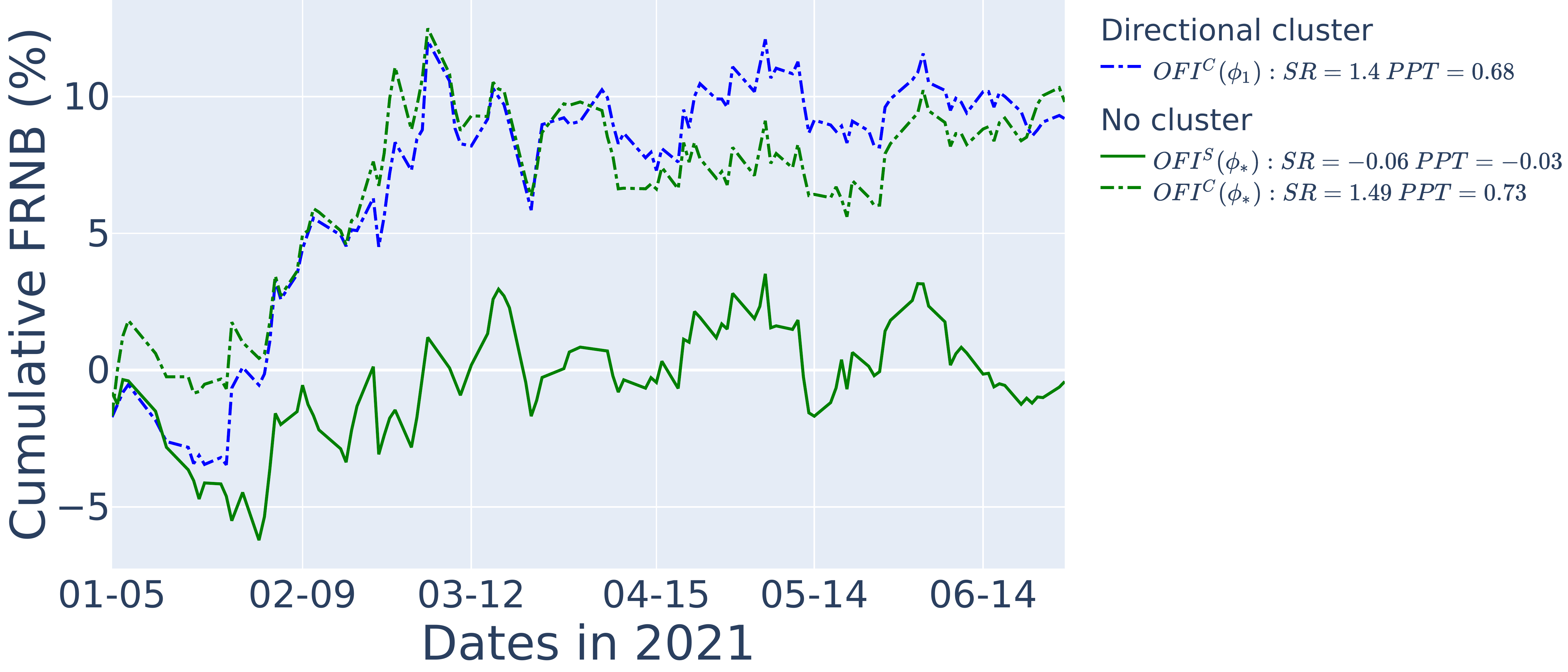}
    &
          \includegraphics[width=\linewidth,trim=0cm 0cm 0cm 0cm,clip]{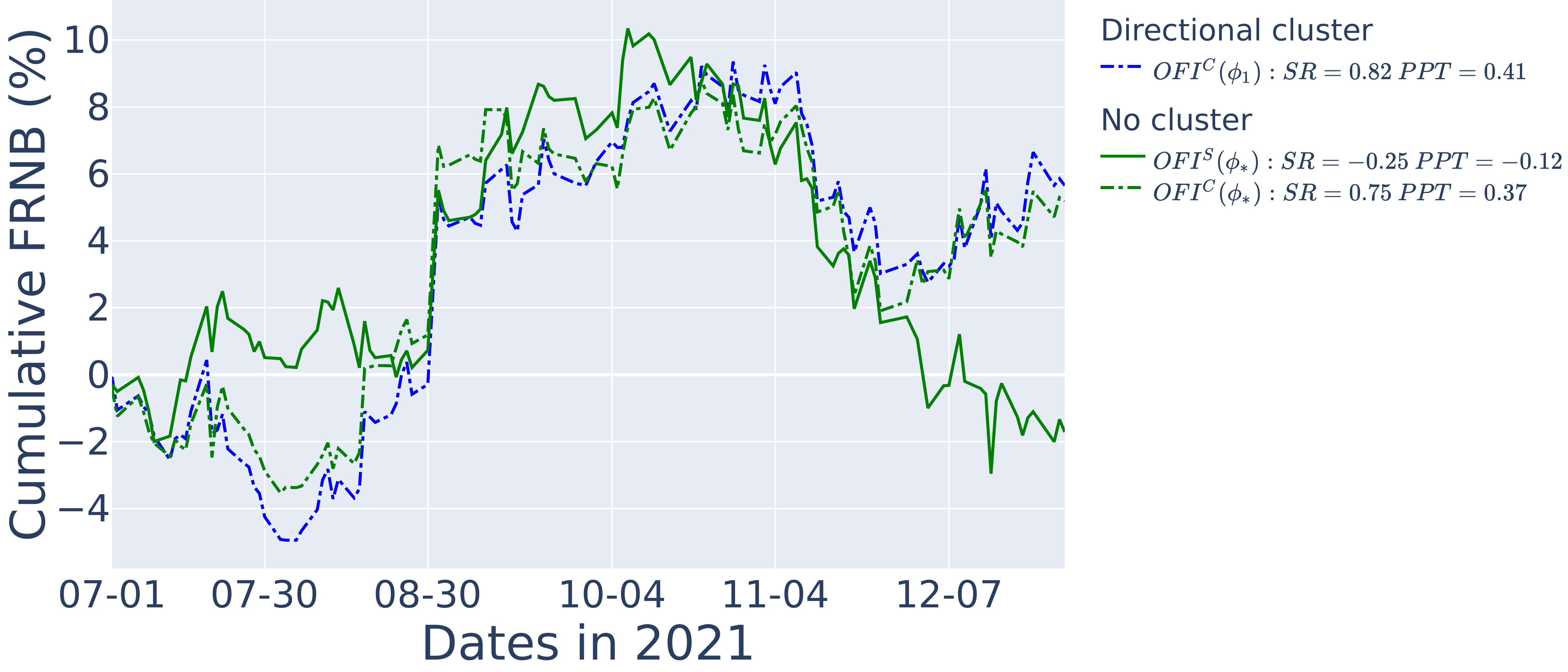}
    \\
    \midrule
   \vspace{-2.5cm}
        \multirow{1}{*}{\rotatebox[origin=c]{90}{\textbf{Cancel events}}}
    &
      \includegraphics[width=\linewidth,trim=0cm 0cm 0cm 0cm,clip]{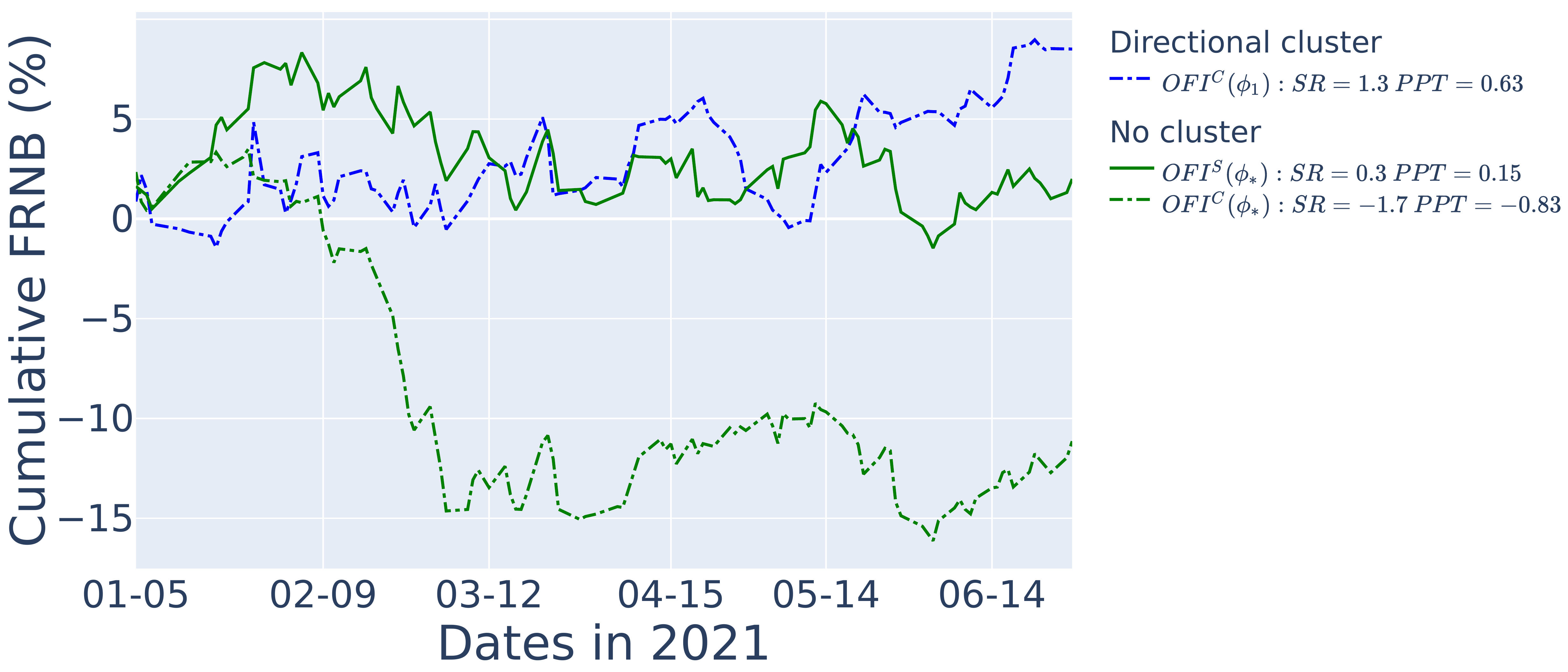}
    &
          \includegraphics[width=\linewidth,trim=0cm 0cm 0cm 0cm,clip]{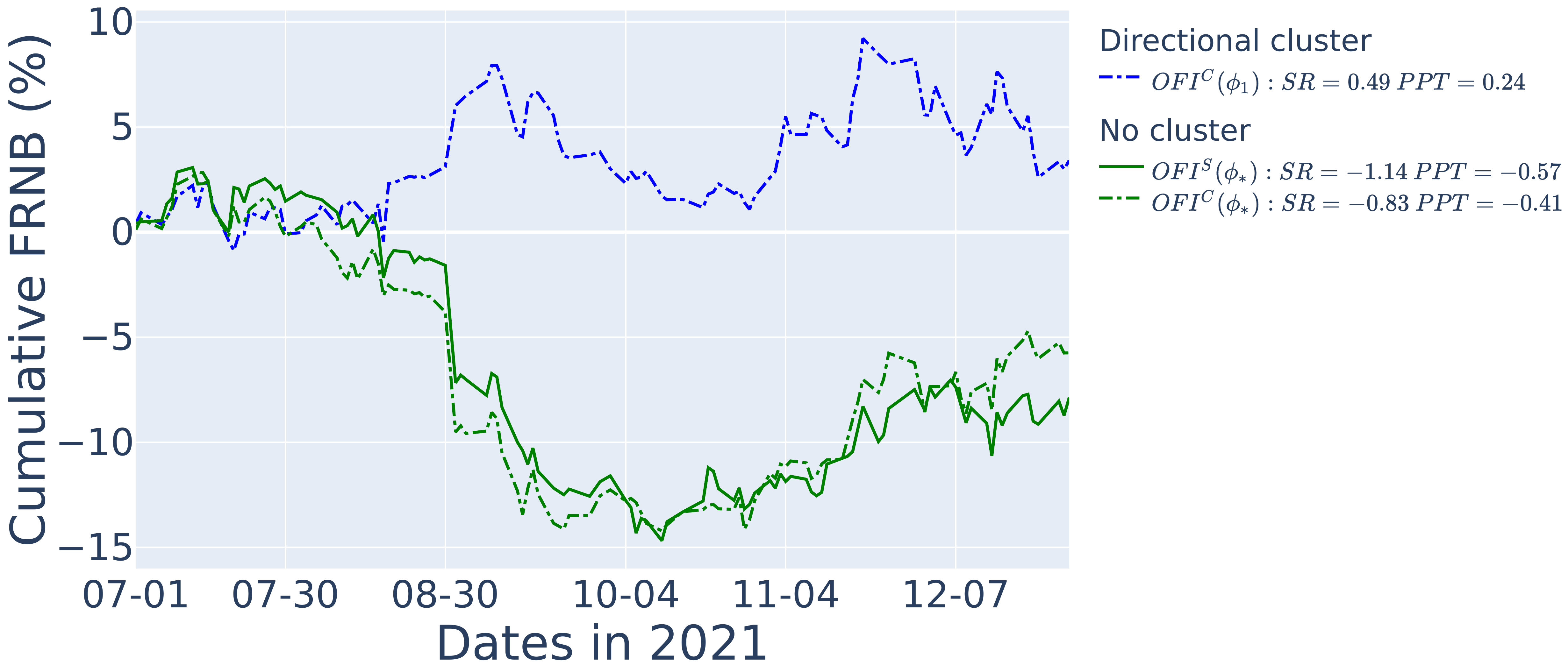}
        \\
    \midrule
  \vspace{-2.5cm}
        \multirow{1}{*}{\rotatebox[origin=c]{90}{\textbf{Trade events}}}
    &
      \includegraphics[width=\linewidth,trim=0cm 0cm 0cm 0cm,clip]{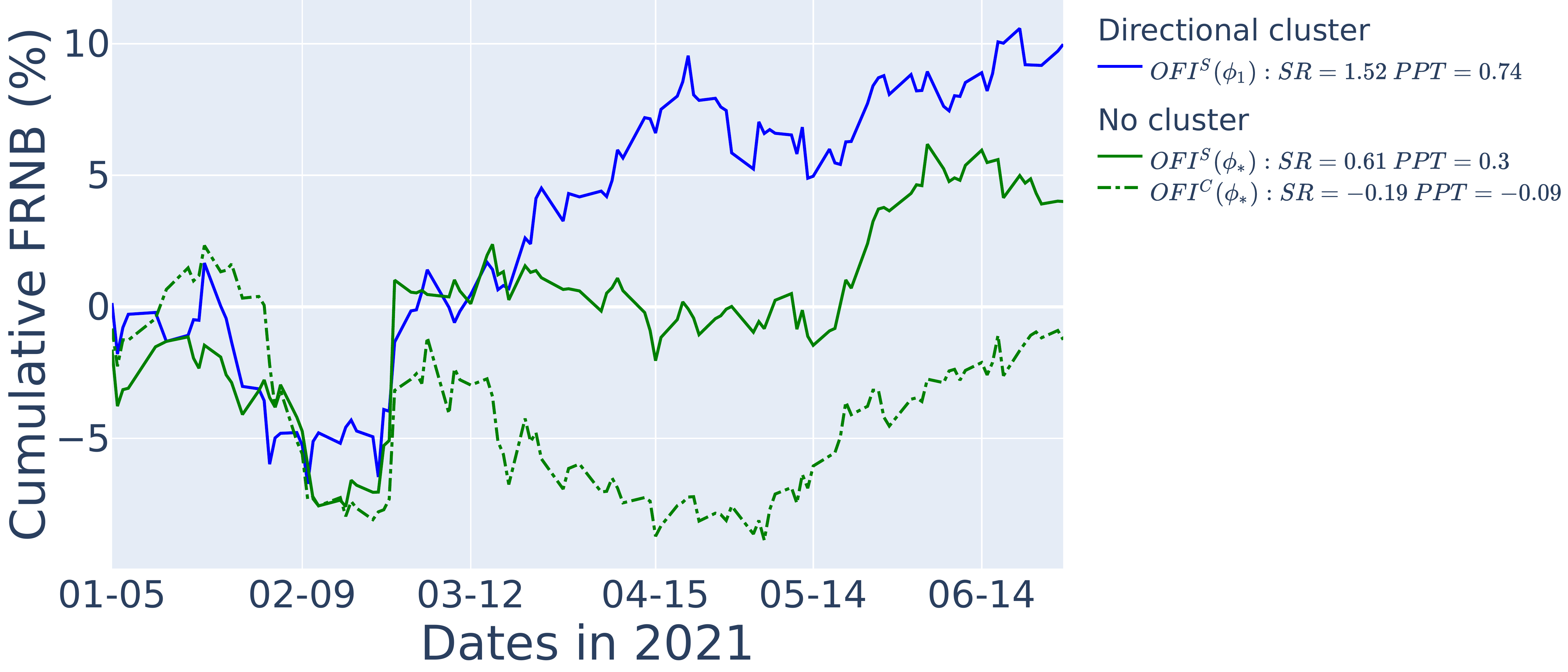}
    &
          \includegraphics[width=\linewidth,trim=0cm 0cm 0cm 0cm,clip]{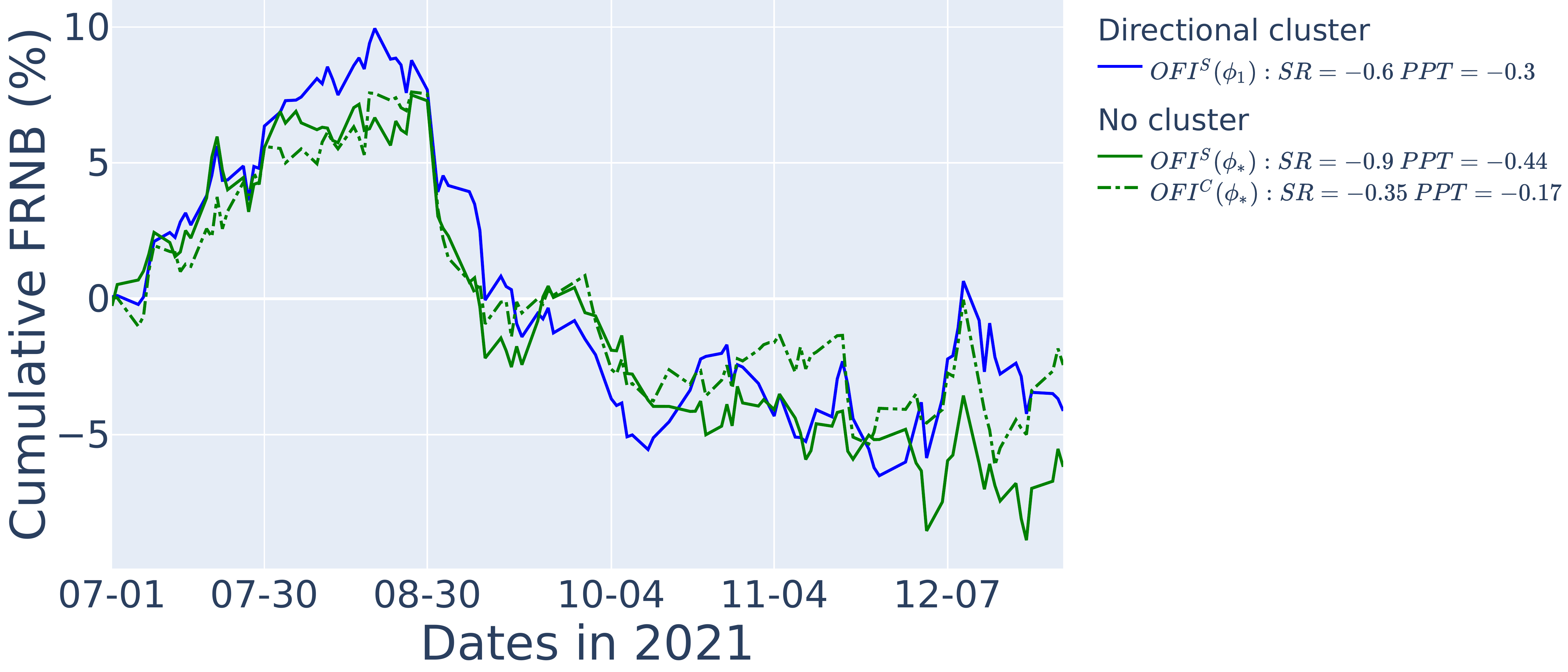} 
          \\
\bottomrule

    \end{tabular}

\end{table}

\clearpage
\newpage

\begin{table}[htbp]
\centering
\caption{The average cumulative PnL of \textbf{FREB} for \textbf{medium-tick} stocks, rescaled to a target volatility of 15\%. From top to bottom, this corresponds to all, add, cancel, and trade event types.}
\label{tab: medium_tick_FREB}
  \centering
    \begin{tabular}{p{0.1cm}p{6.8cm}p{6.8cm}}
        \toprule
&\multicolumn{1}{c}{\textbf{Training}} &  \multicolumn{1}{c}{\textbf{Test}} \\
        \midrule
    \vspace{-2.5cm}
    \multirow{1}{*}{\rotatebox[origin=c]{90}{\textbf{All events}}}
    &
      \includegraphics[width=\linewidth,trim=0cm 0cm 0cm 0cm,clip]{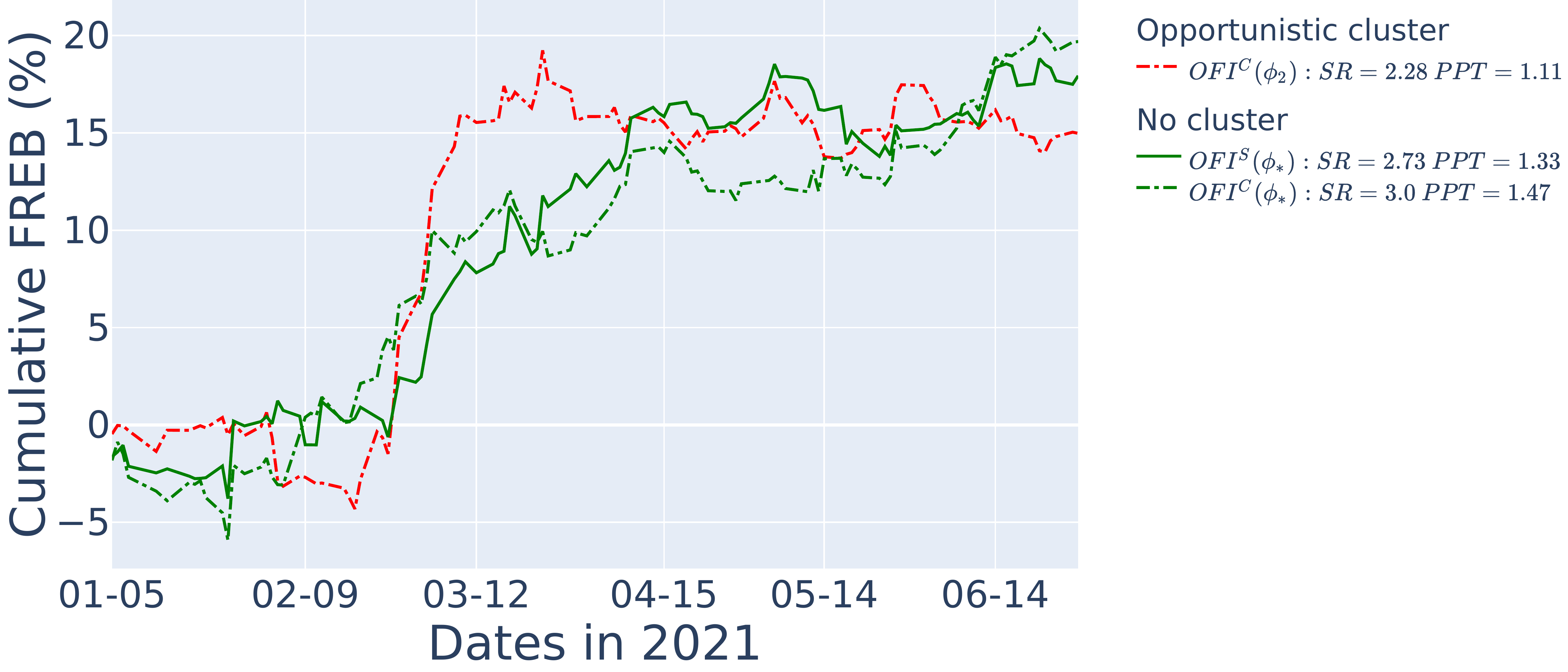}
    &
          \includegraphics[width=\linewidth,trim=0cm 0cm 0cm 0cm,clip]{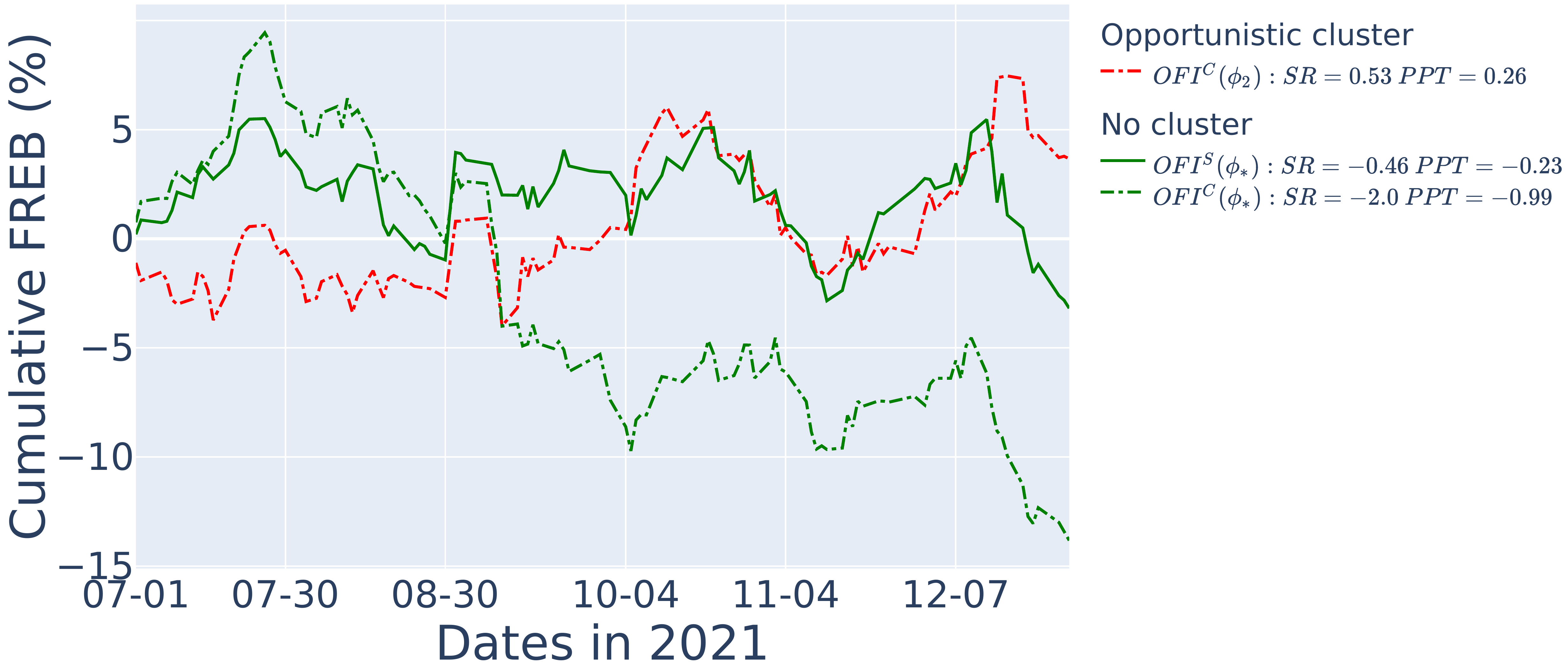}
      \\
    \midrule
        \vspace{-2.5cm}
        \multirow{1}{*}{\rotatebox[origin=c]{90}{\textbf{Add events}}}
    &
      \includegraphics[width=\linewidth,trim=0cm 0cm 0cm 0cm,clip]{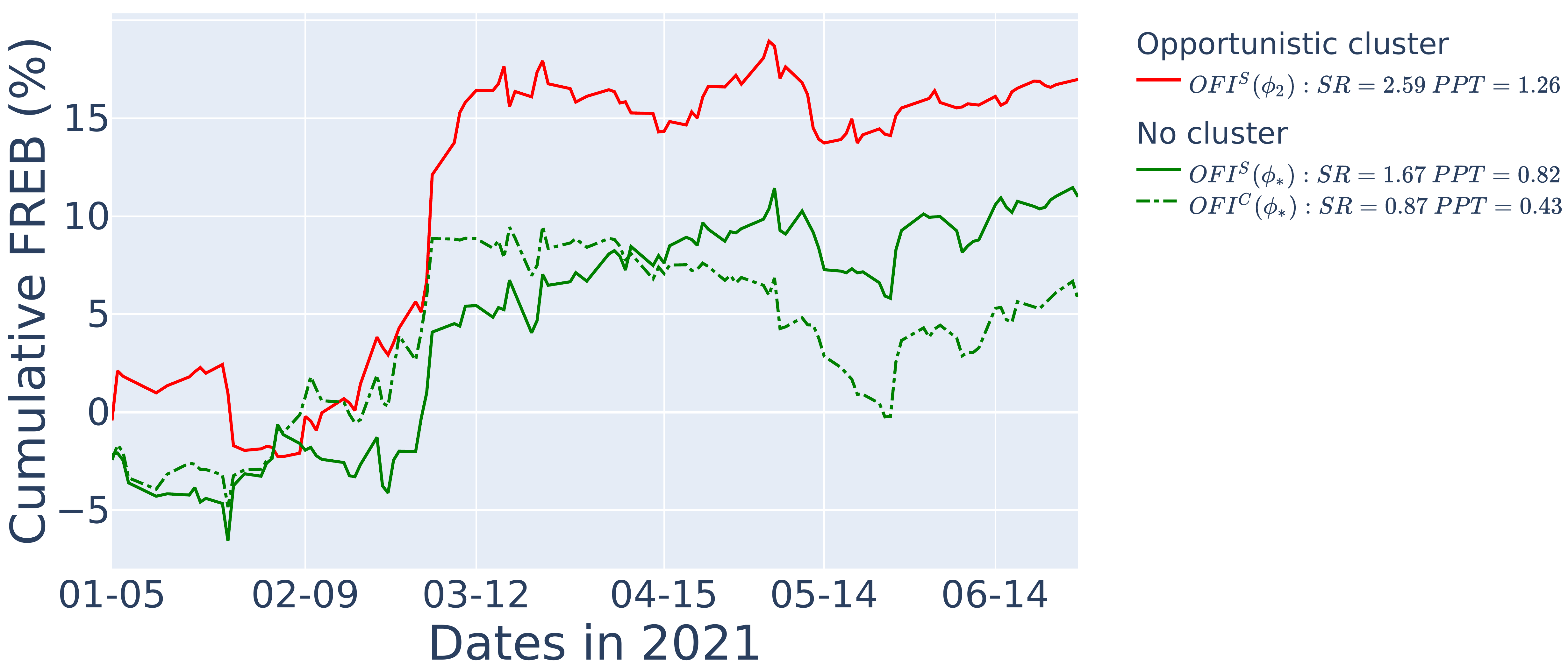}
    &
          \includegraphics[width=\linewidth,trim=0cm 0cm 0cm 0cm,clip]{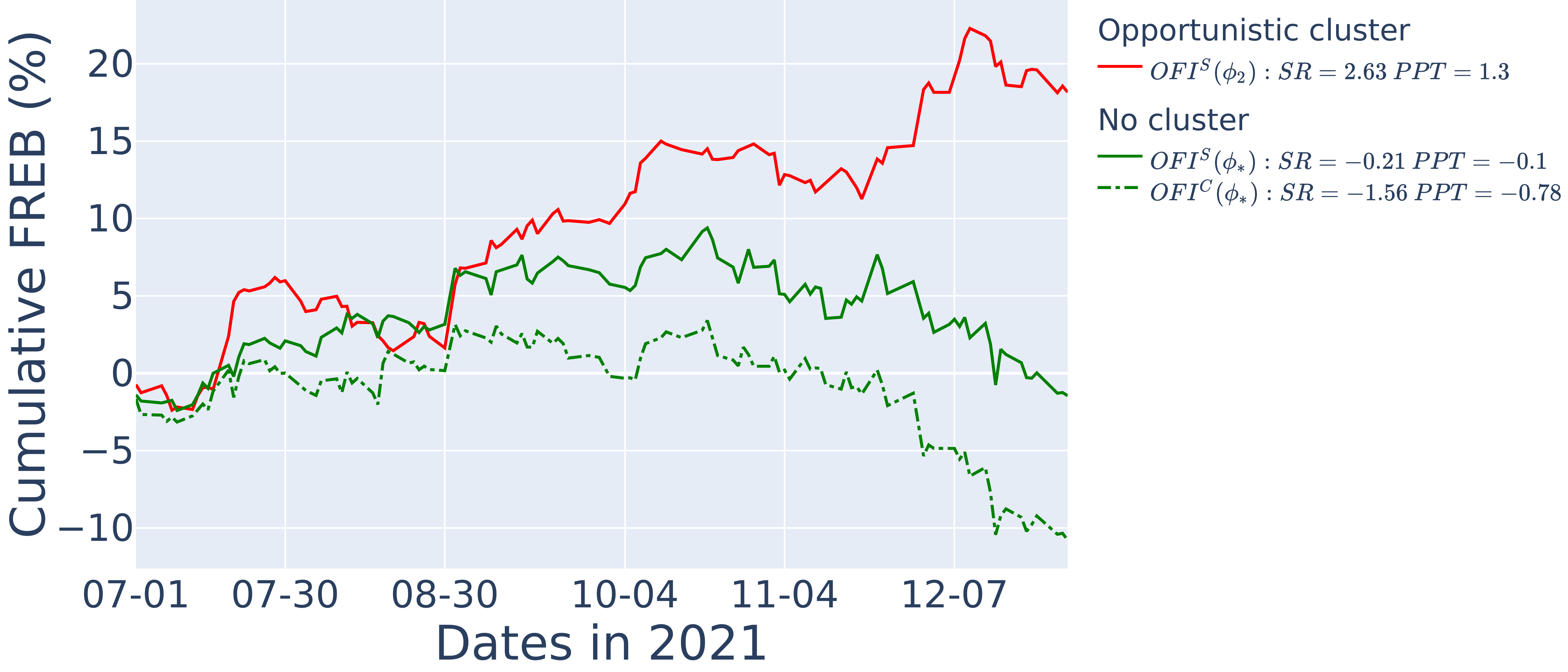}
    \\
    \midrule
    \vspace{-2.5cm}
        \multirow{1}{*}{\rotatebox[origin=c]{90}{\textbf{Cancel events}}}
    &
      \includegraphics[width=\linewidth,trim=0cm 0cm 0cm 0cm,clip]{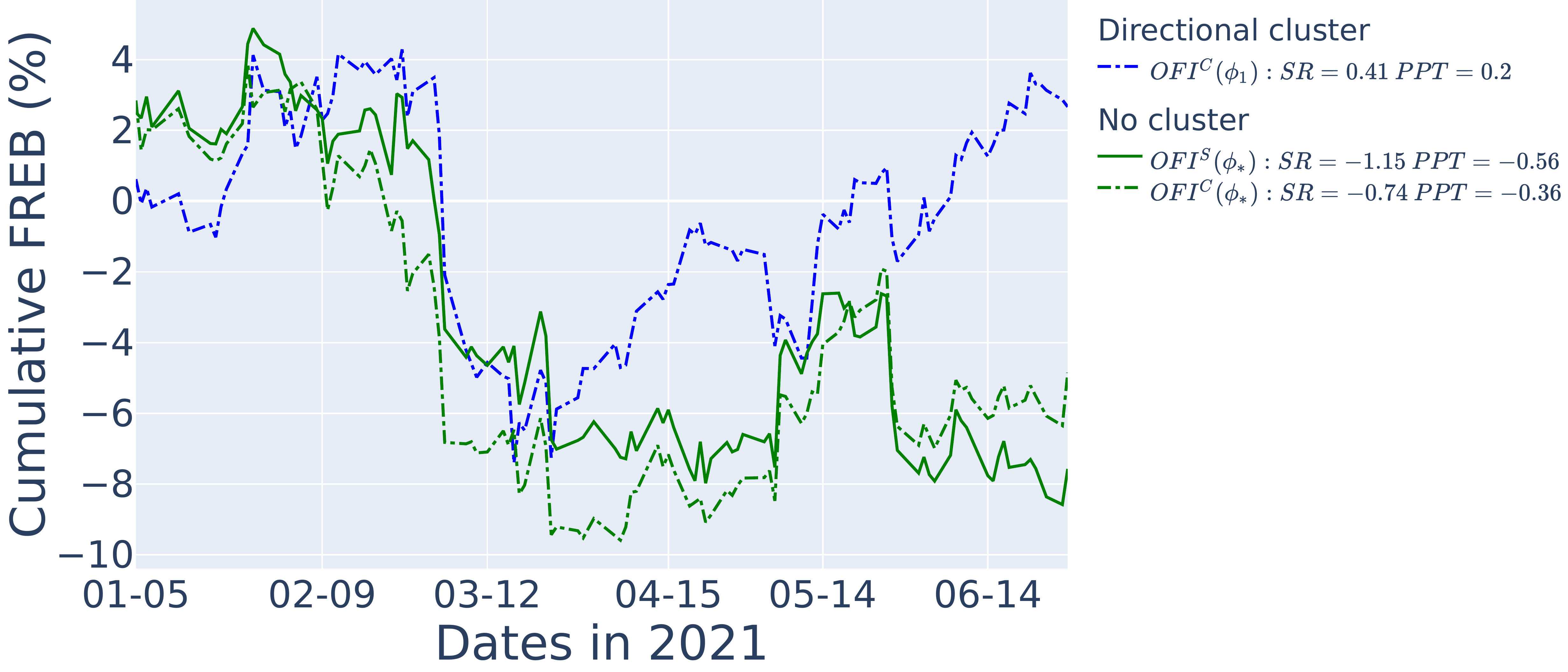}
    &
          \includegraphics[width=\linewidth,trim=0cm 0cm 0cm 0cm,clip]{plots/medium/test_top_FRNB_D.pdf}
        \\
    \midrule
    \vspace{-2.5cm}
        \multirow{1}{*}{\rotatebox[origin=c]{90}{\textbf{Trade events}}}
    &
      \includegraphics[width=\linewidth,trim=0cm 0cm 0cm 0cm,clip]{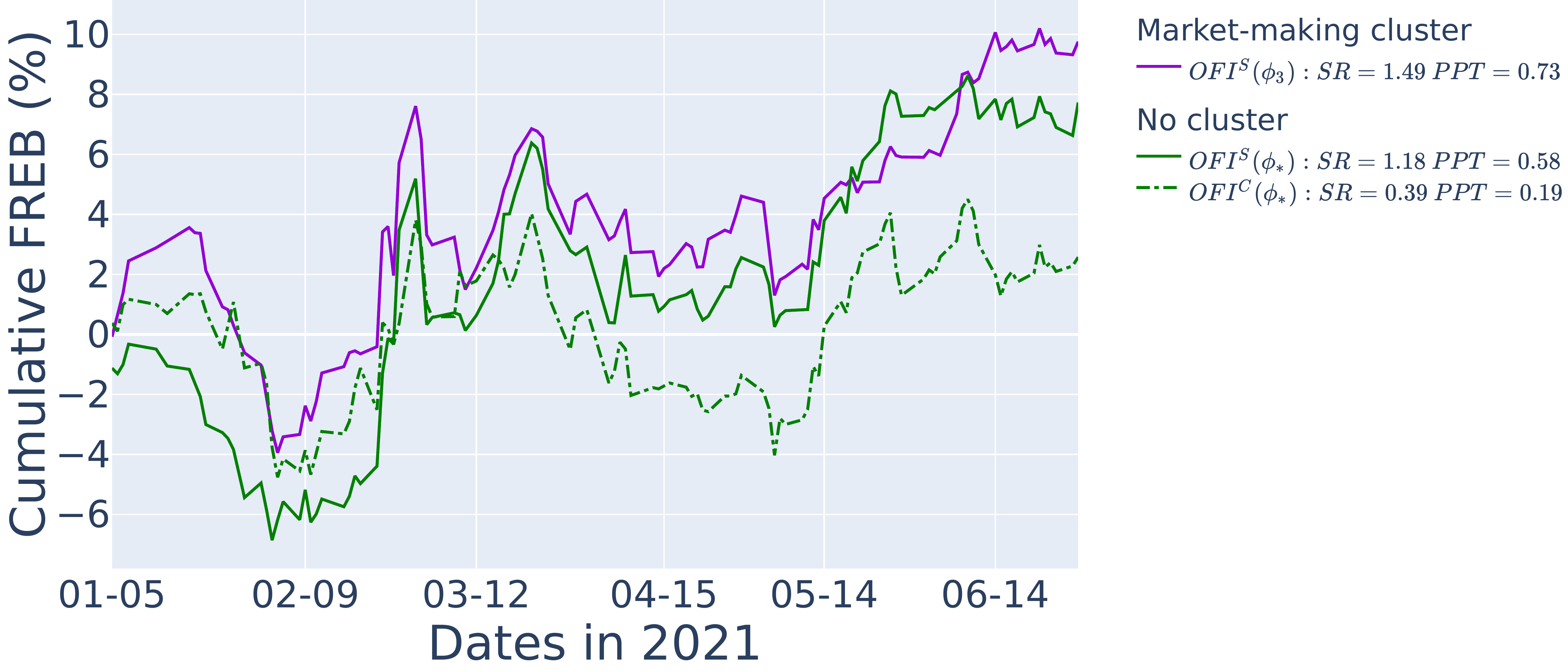}
    &
          \includegraphics[width=\linewidth,trim=0cm 0cm 0cm 0cm,clip]{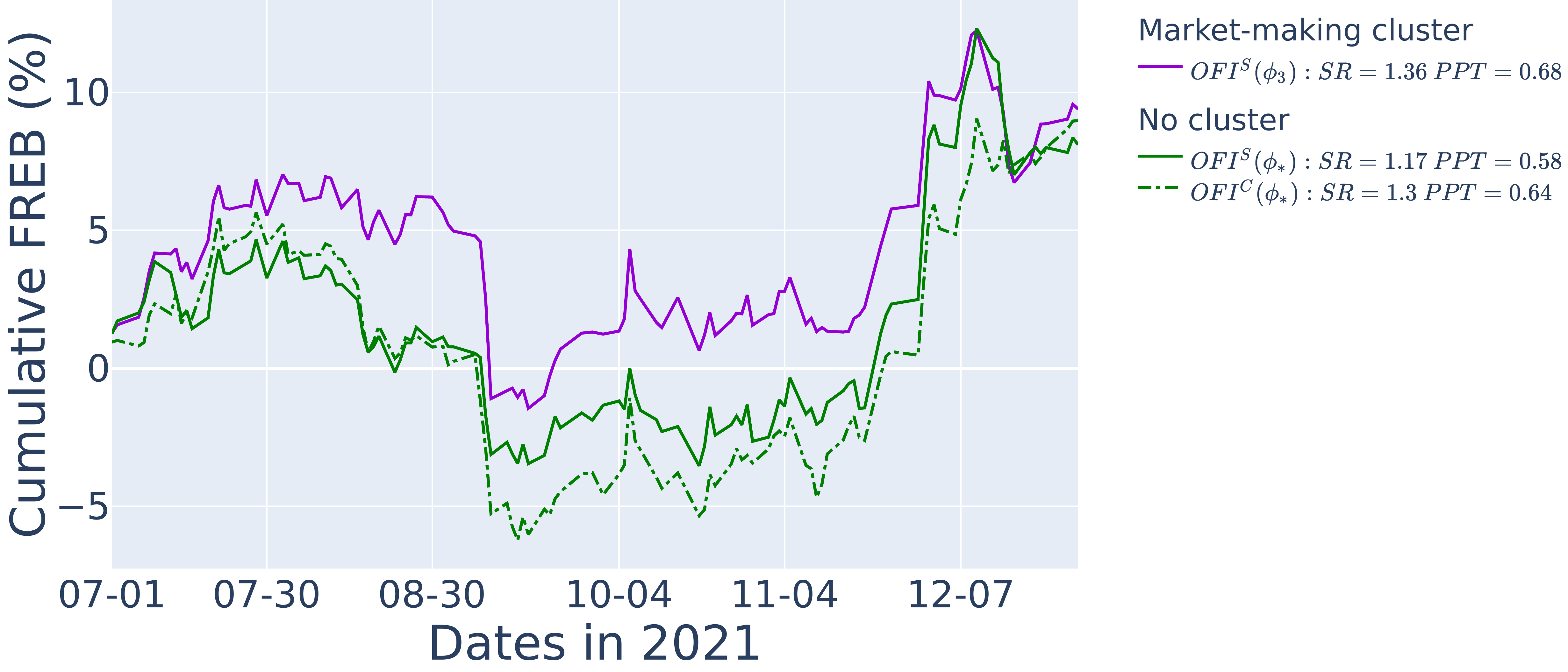} 
          \\
\bottomrule

    \end{tabular}

\end{table}

\clearpage
\newpage
\subsection{Large-tick stocks performance for different event types}
\label{sec: large}

\begin{table}[htbp]
\centering
\caption{The average cumulative PnL of \textbf{FRNB} for \textbf{large-tick} stocks, rescaled to a target volatility of 15\%. From top to bottom, this corresponds to all, add, cancel, and trade event types.}
\label{tab: large_tick_FRNB}
  \centering
    \begin{tabular}{p{0.1cm}p{6.8cm}p{6.8cm}}
        \toprule
&\multicolumn{1}{c}{\textbf{Training}} &  \multicolumn{1}{c}{\textbf{Test}} \\
        \midrule
    \vspace{-2.5cm}
    \multirow{1}{*}{\rotatebox[origin=c]{90}{\textbf{All events}}}
    &
      \includegraphics[width=\linewidth,trim=0cm 0cm 0cm 0cm,clip]{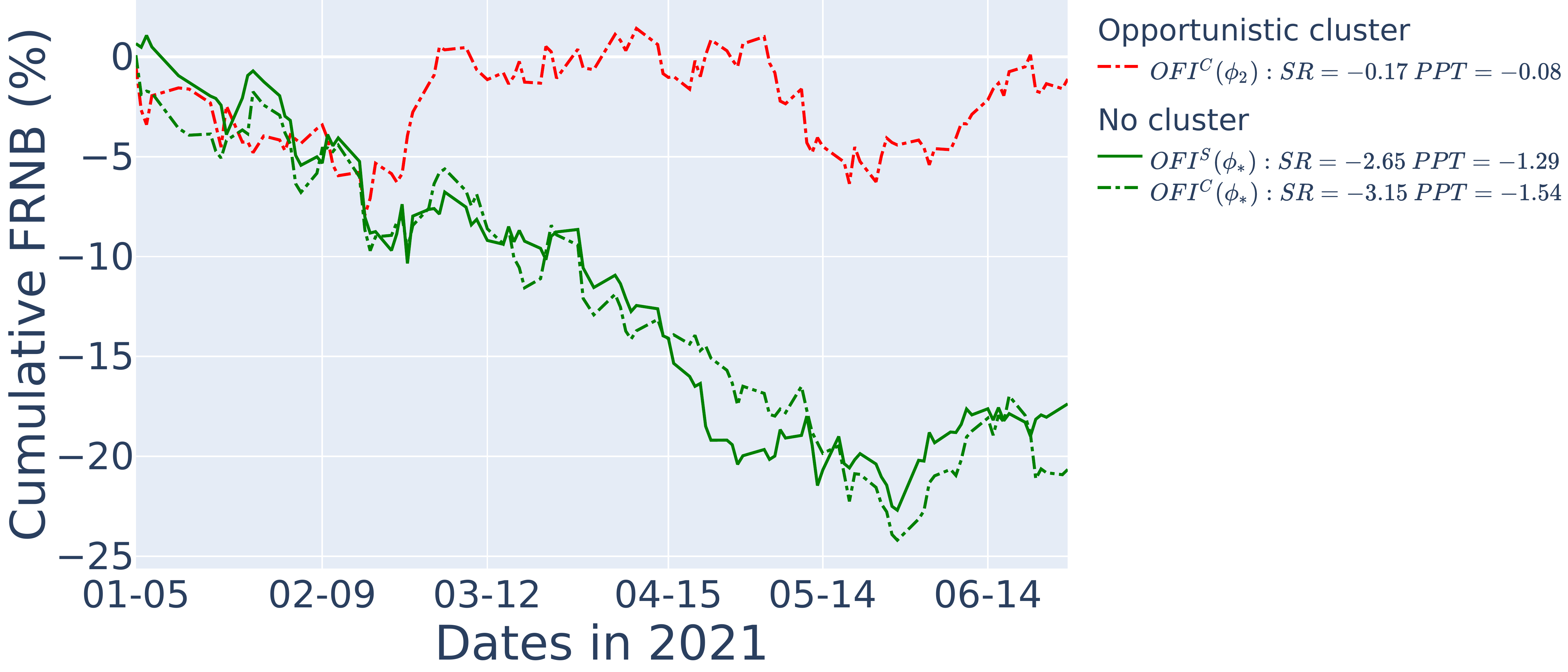}
    &
          \includegraphics[width=\linewidth,trim=0cm 0cm 0cm 0cm,clip]{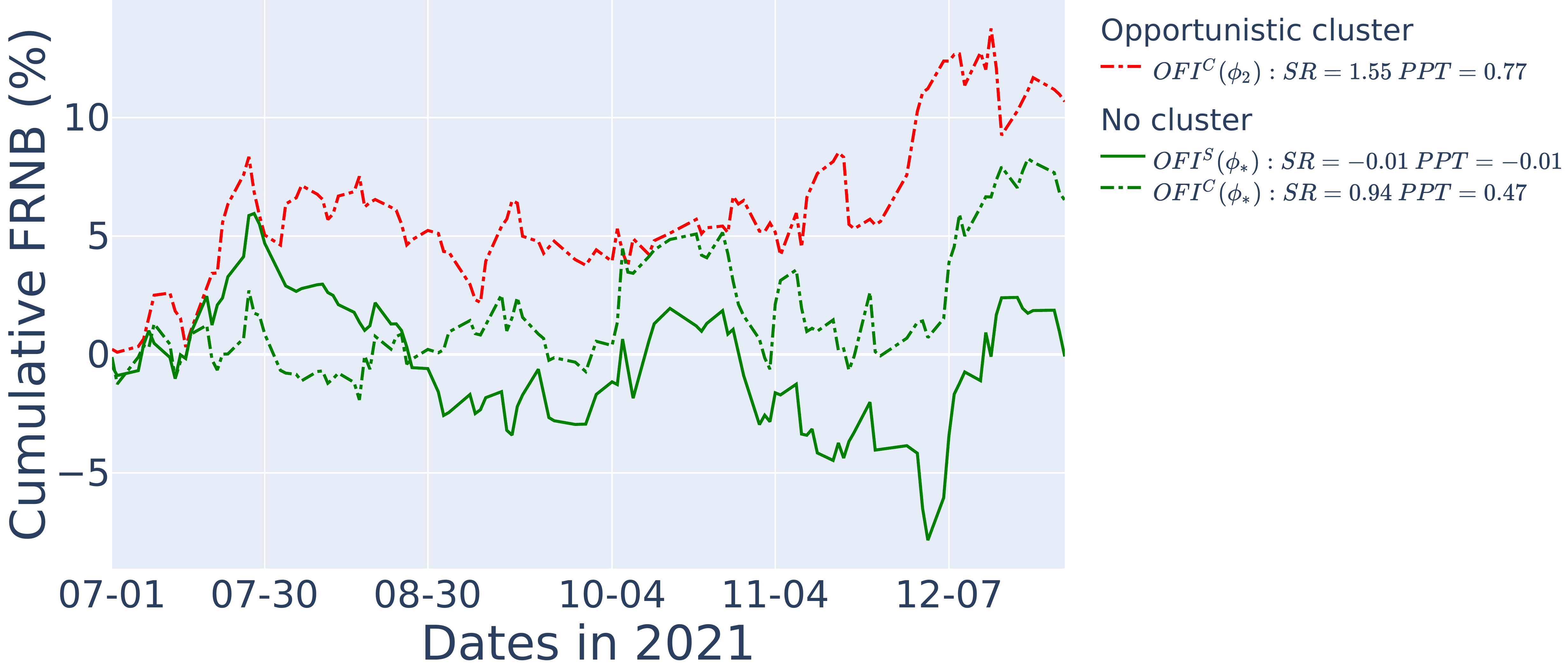}
      \\
    \midrule
        \vspace{-2.5cm}
        \multirow{1}{*}{\rotatebox[origin=c]{90}{\textbf{Add events}}}
    &
      \includegraphics[width=\linewidth,trim=0cm 0cm 0cm 0cm,clip]{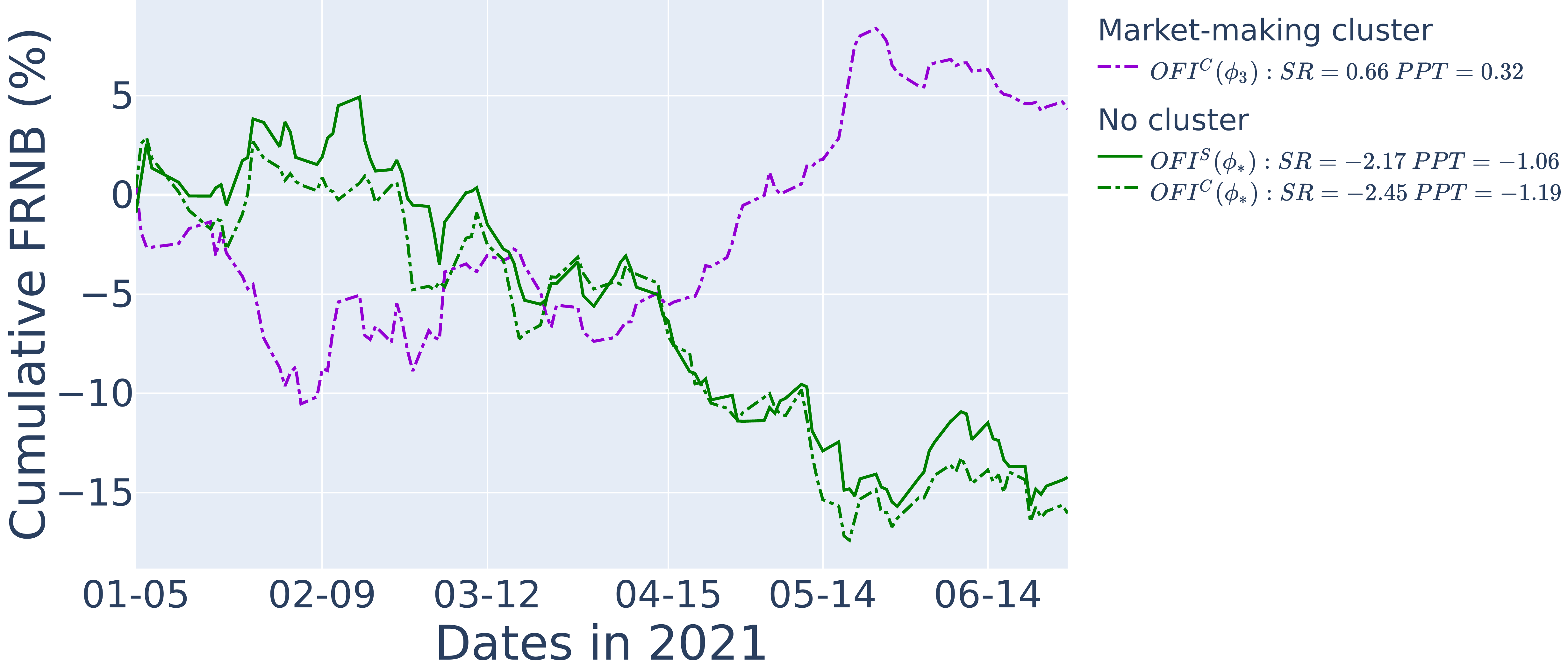}
    &
          \includegraphics[width=\linewidth,trim=0cm 0cm 0cm 0cm,clip]{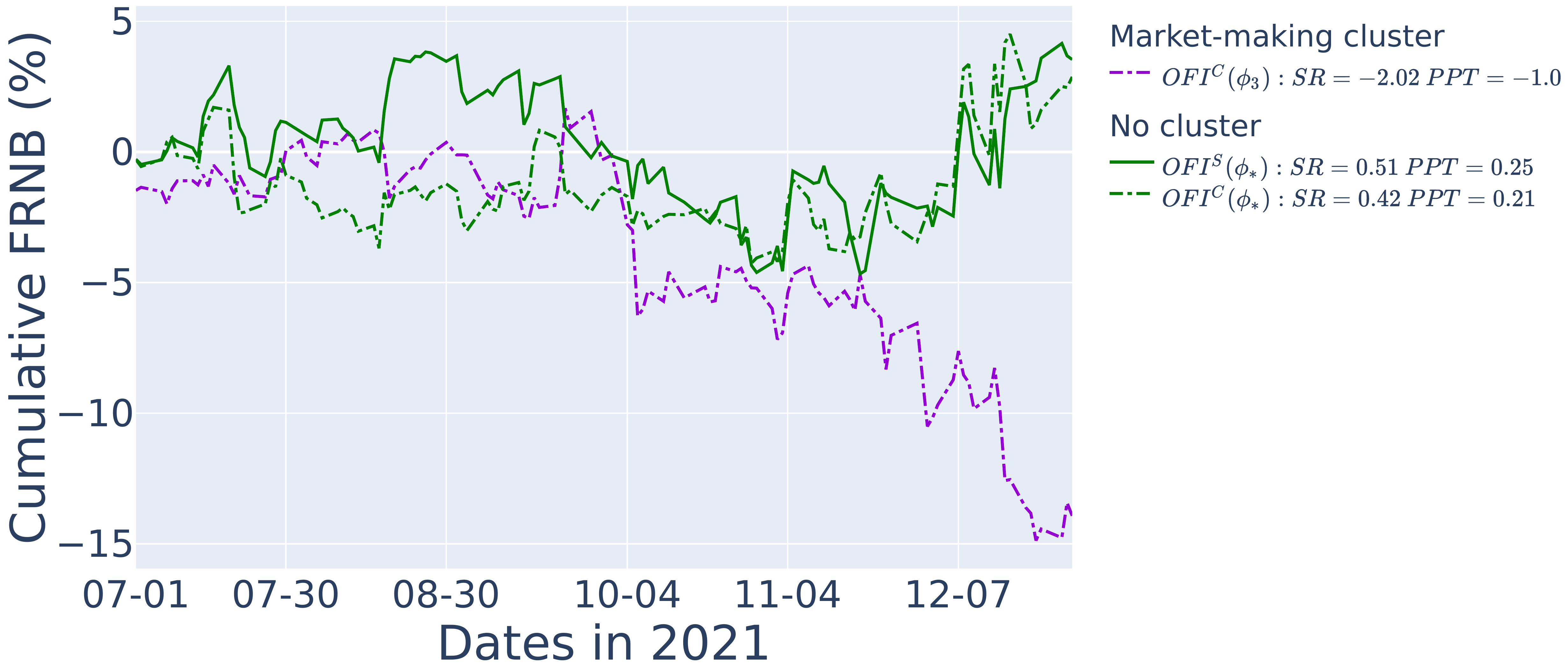}
    \\
    \midrule
   \vspace{-2.5cm}
        \multirow{1}{*}{\rotatebox[origin=c]{90}{\textbf{Cancel events}}}
    &
      \includegraphics[width=\linewidth,trim=0cm 0cm 0cm 0cm,clip]{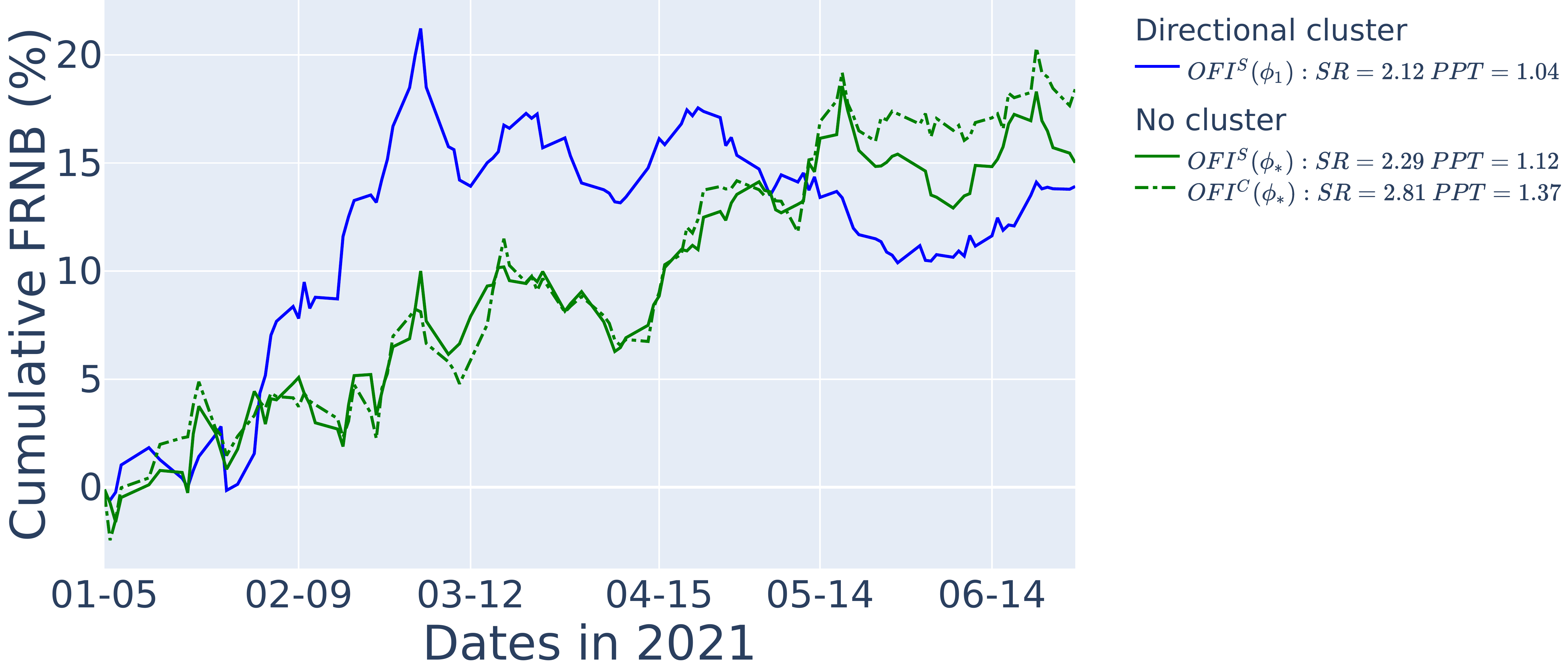}
    &
          \includegraphics[width=\linewidth,trim=0cm 0cm 0cm 0cm,clip]{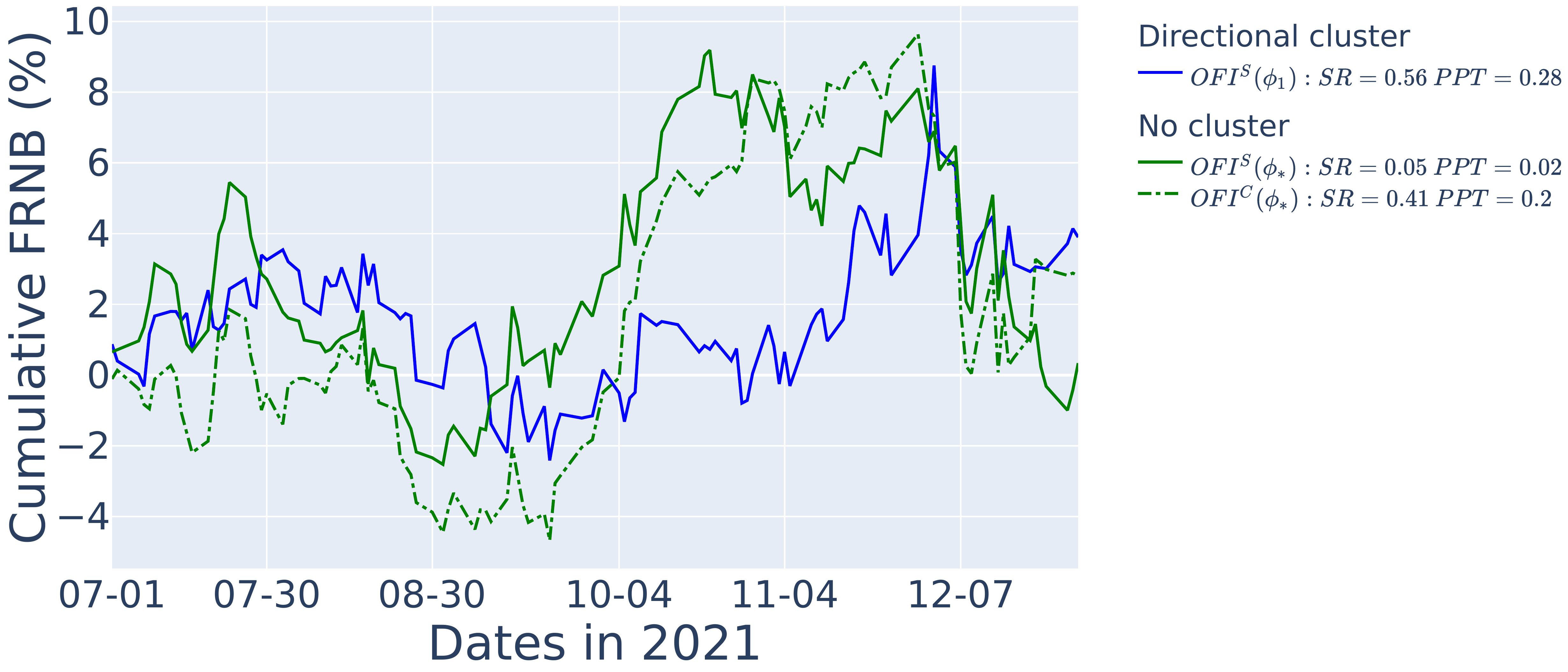}
        \\
    \midrule
    \vspace{-2.5cm}
        \multirow{1}{*}{\rotatebox[origin=c]{90}{\textbf{Trade events}}}
    &
      \includegraphics[width=\linewidth,trim=0cm 0cm 0cm 0cm,clip]{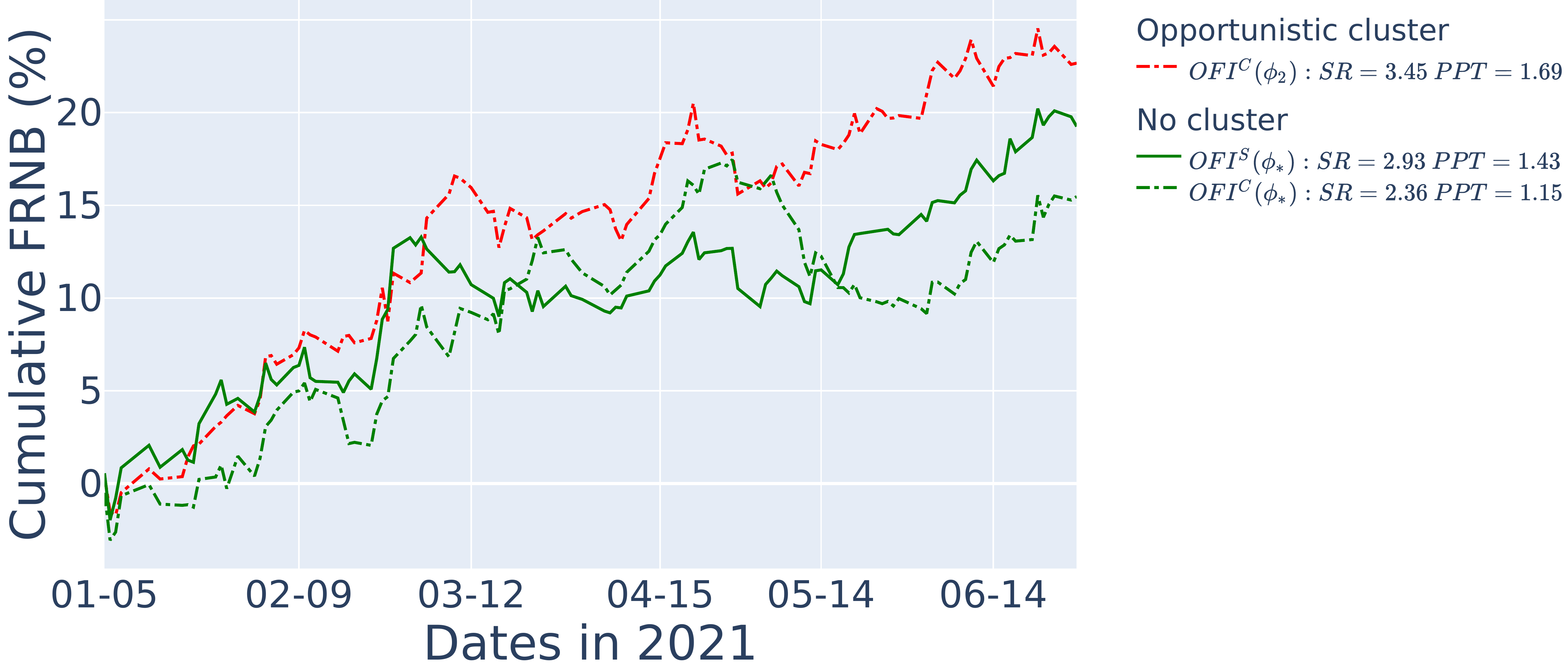}
    &
          \includegraphics[width=\linewidth,trim=0cm 0cm 0cm 0cm,clip]{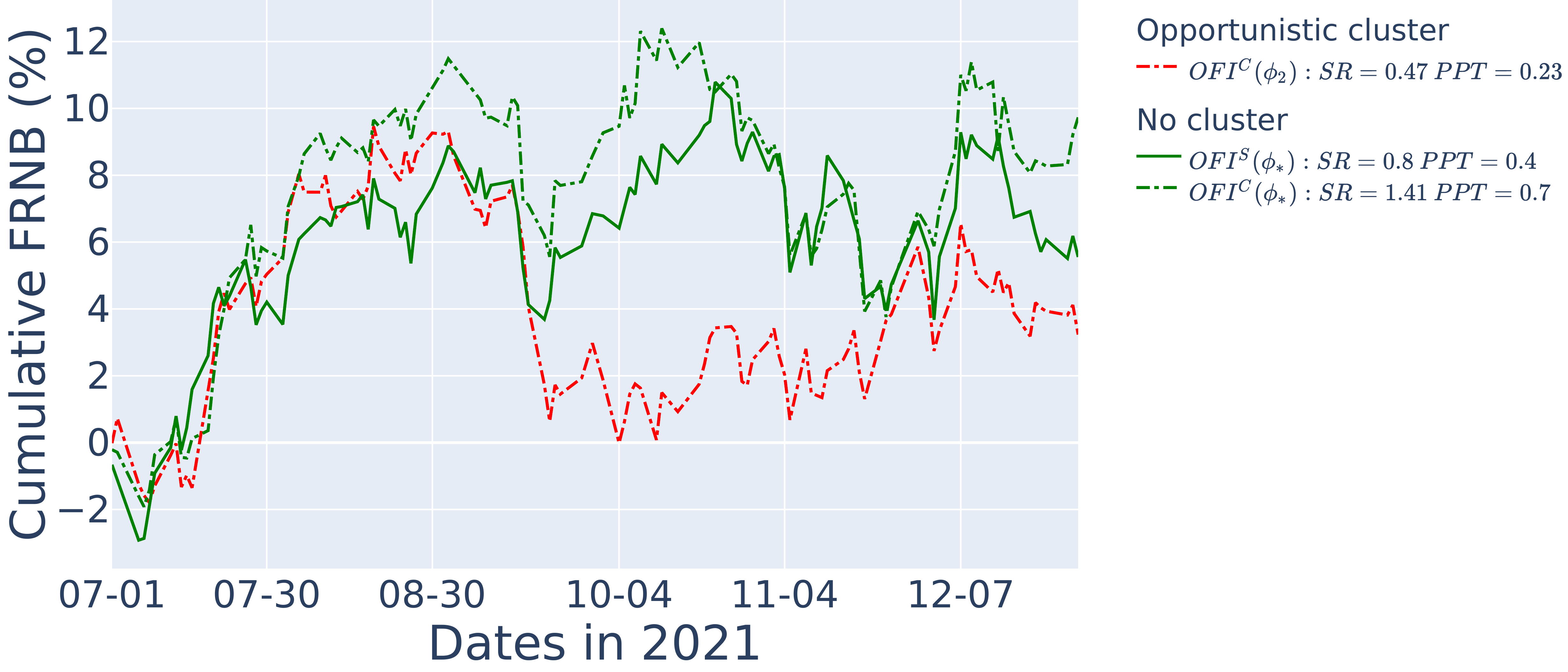} 
          \\
\bottomrule

    \end{tabular}

\end{table}

\clearpage
\newpage
\begin{table}[htbp]
\centering
\caption{The average cumulative PnL of \textbf{FREB} for \textbf{large-tick} stocks, rescaled to a target volatility of 15\%. From top to bottom, this corresponds to all, add, cancel, and trade event types.}
\label{tab: large_tick_FREB}
  \centering
    \begin{tabular}{p{0.1cm}p{6.8cm}p{6.8cm}}
        \toprule
&\multicolumn{1}{c}{\textbf{Training}} &  \multicolumn{1}{c}{\textbf{Test}} \\
        \midrule
    \vspace{-2.5cm}
    \multirow{1}{*}{\rotatebox[origin=c]{90}{\textbf{All events}}}
    &
      \includegraphics[width=\linewidth,trim=0cm 0cm 0cm 0cm,clip]{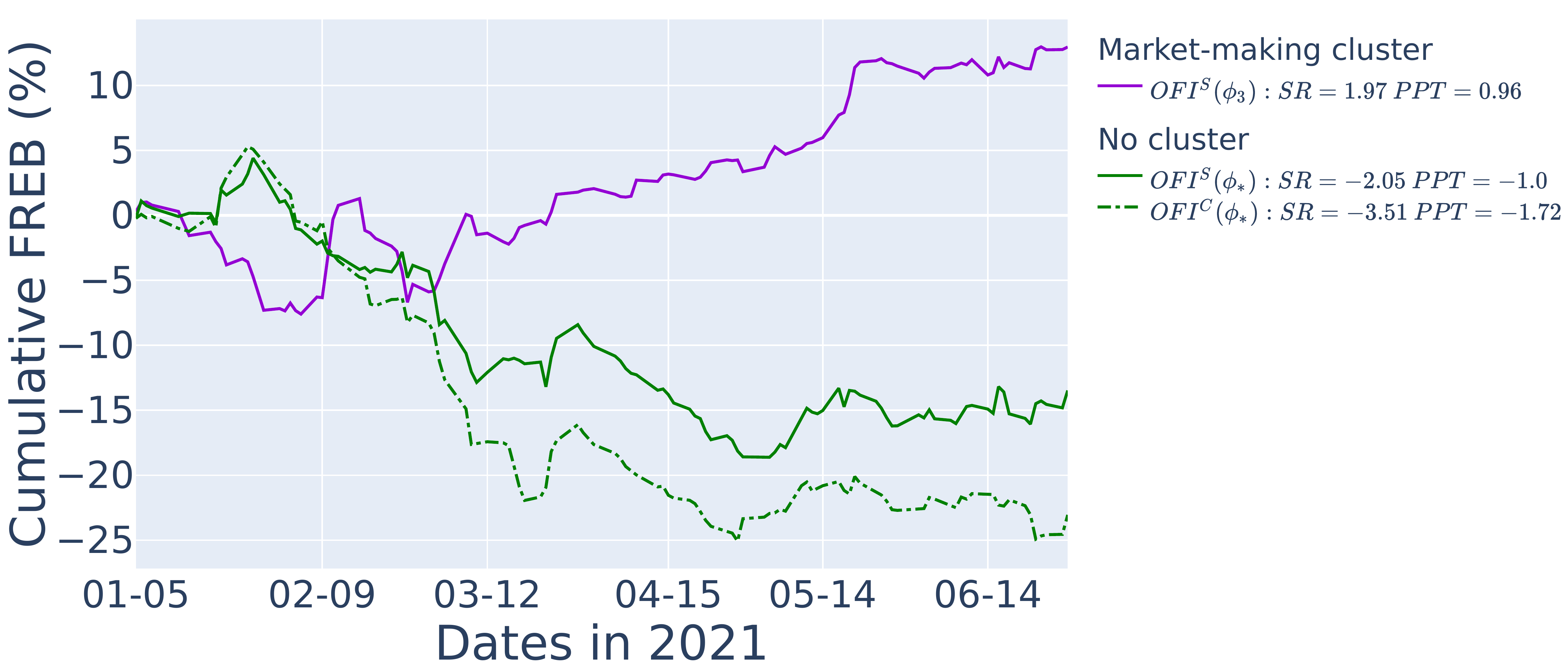}
    &
          \includegraphics[width=\linewidth,trim=0cm 0cm 0cm 0cm,clip]{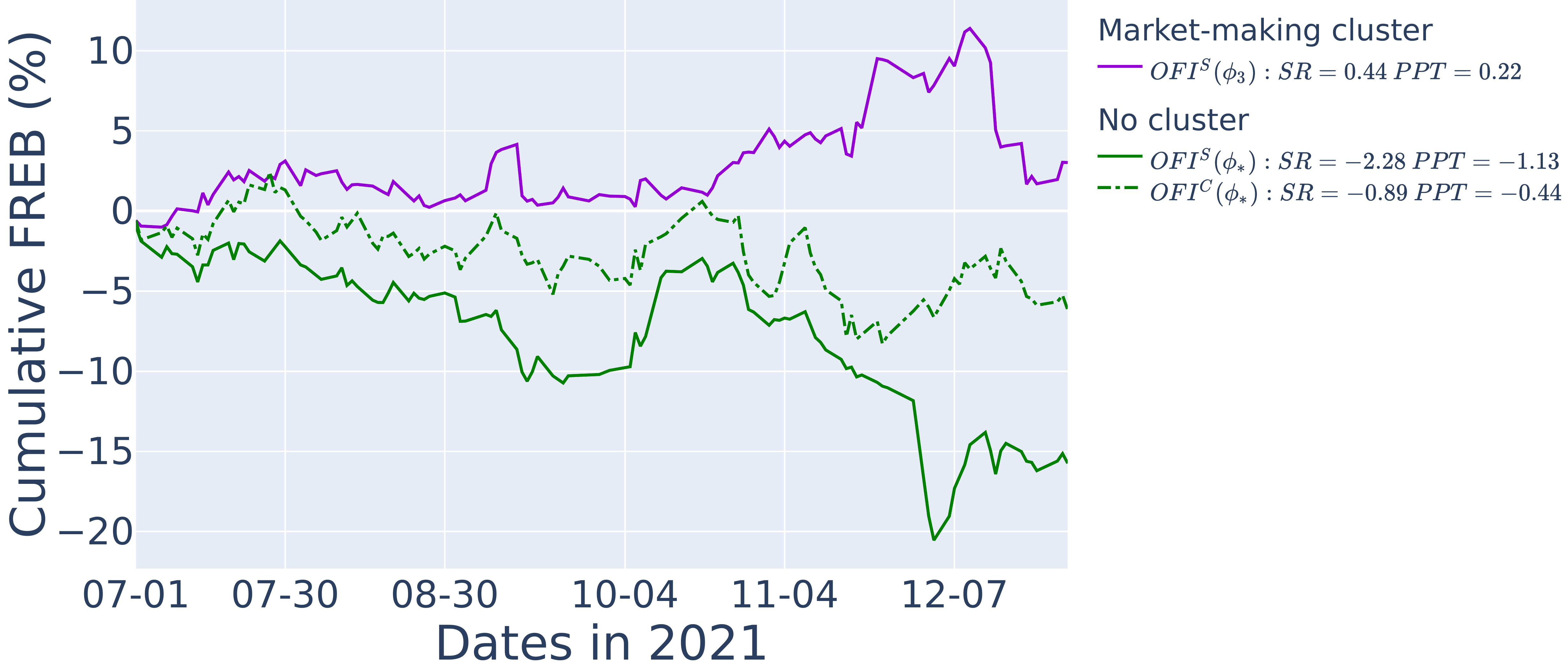}
      \\
    \midrule
        \vspace{-2.5cm}
        \multirow{1}{*}{\rotatebox[origin=c]{90}{\textbf{Add events}}}
    &
      \includegraphics[width=\linewidth,trim=0cm 0cm 0cm 0cm,clip]{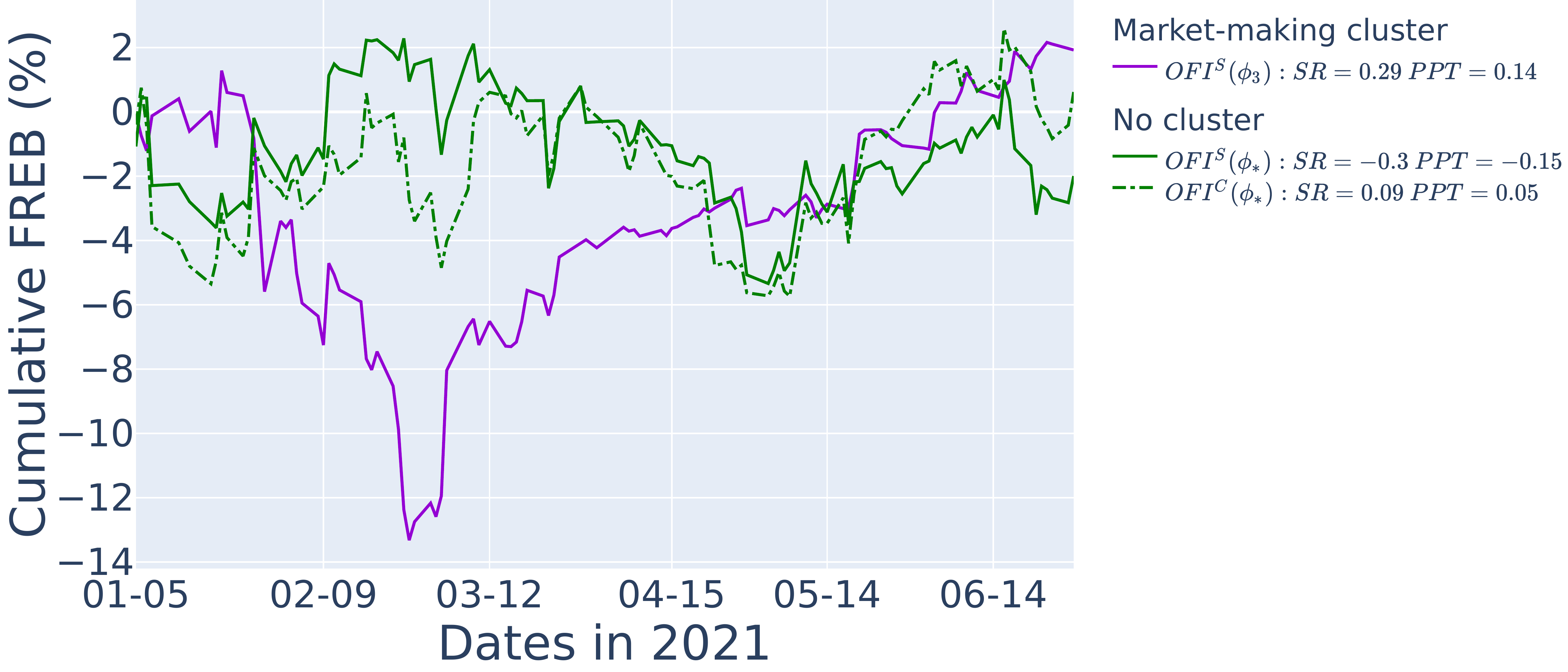}
    &
          \includegraphics[width=\linewidth,trim=0cm 0cm 0cm 0cm,clip]{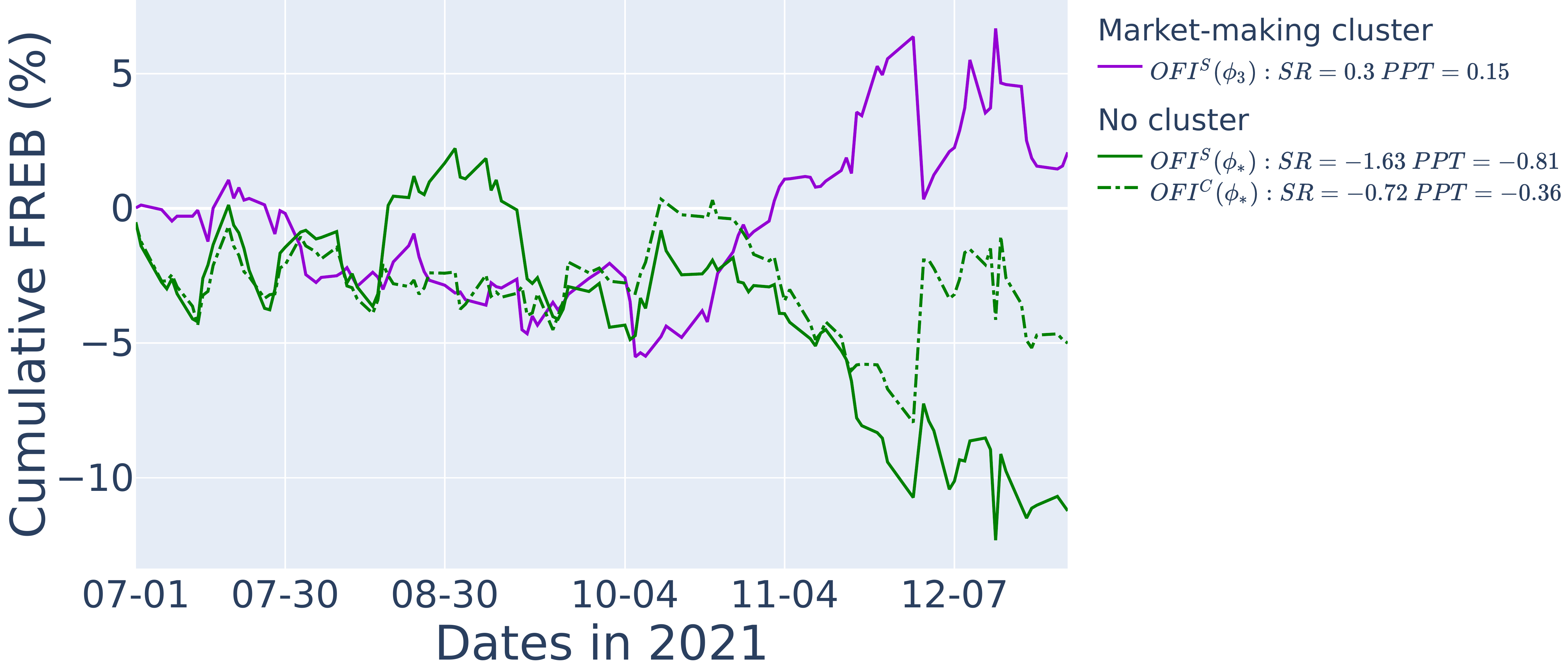}
    \\
    \midrule
    \vspace{-2.5cm}
        \multirow{1}{*}{\rotatebox[origin=c]{90}{\textbf{Cancel events}}}
    &
      \includegraphics[width=\linewidth,trim=0cm 0cm 0cm 0cm,clip]{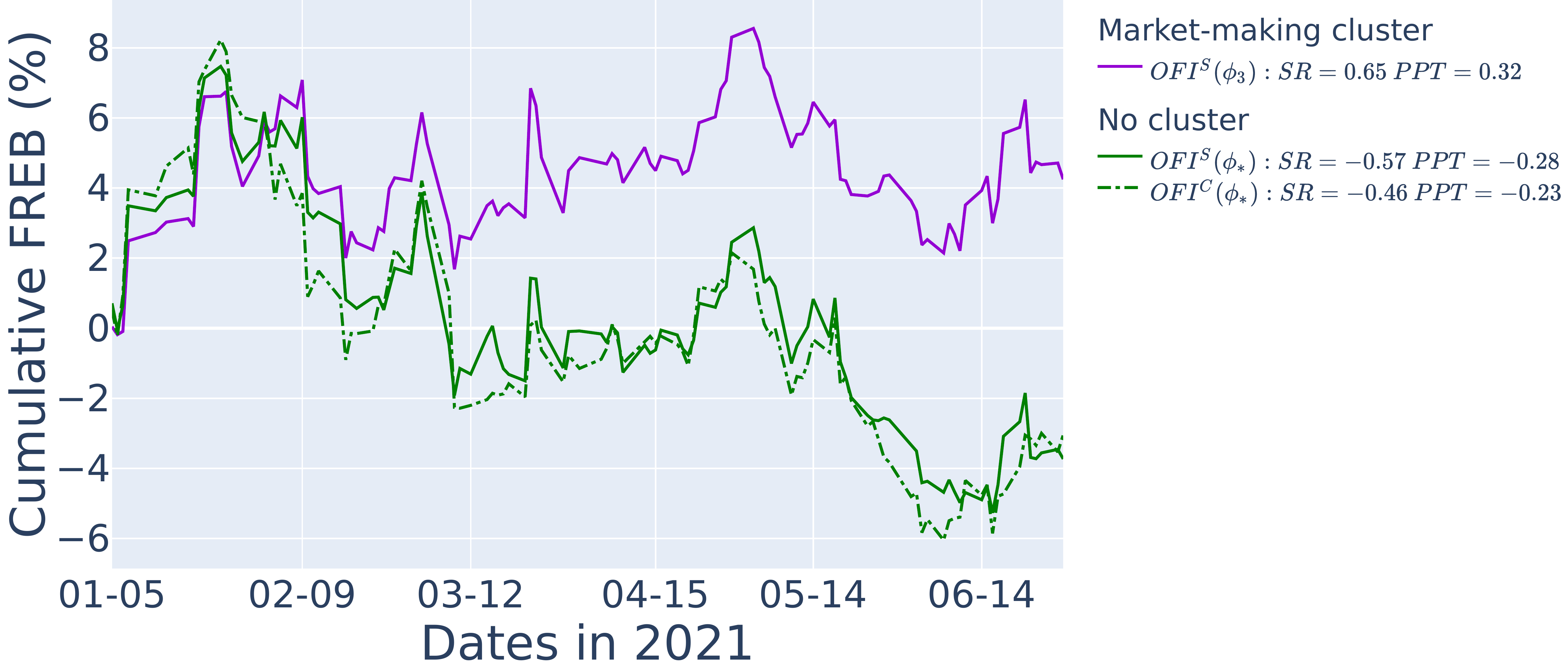}
    &
          \includegraphics[width=\linewidth,trim=0cm 0cm 0cm 0cm,clip]{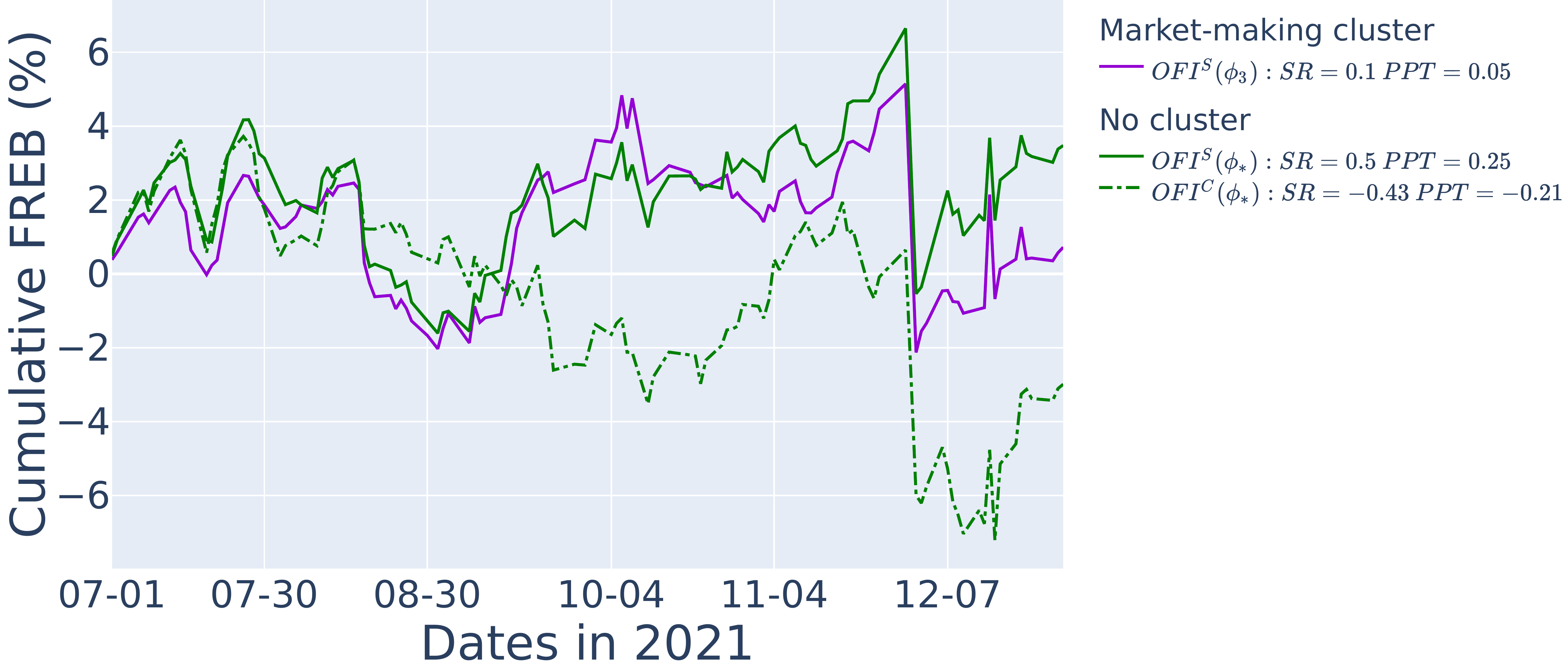}
        \\
    \midrule
    \vspace{-2.5cm}
        \multirow{1}{*}{\rotatebox[origin=c]{90}{\textbf{Trade events}}}
    &
      \includegraphics[width=\linewidth,trim=0cm 0cm 0cm 0cm,clip]{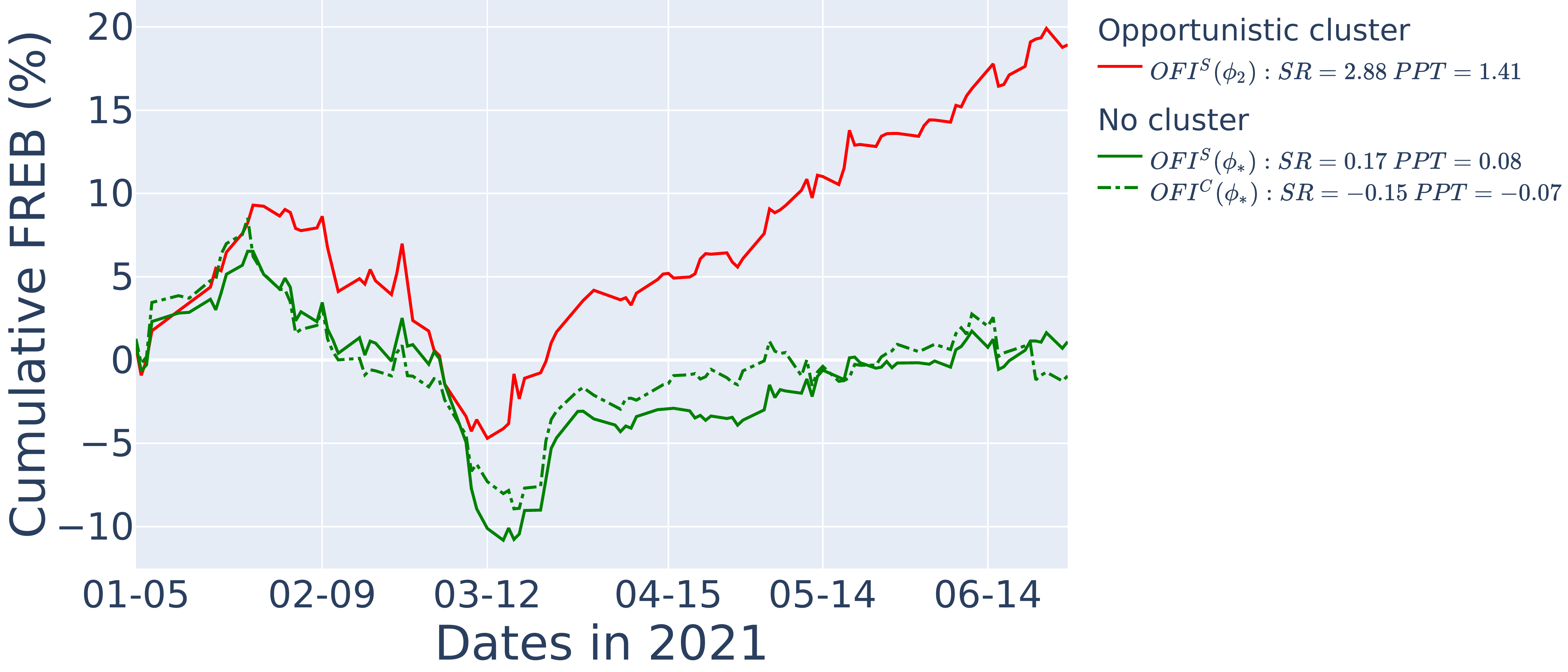}
    &
          \includegraphics[width=\linewidth,trim=0cm 0cm 0cm 0cm,clip]{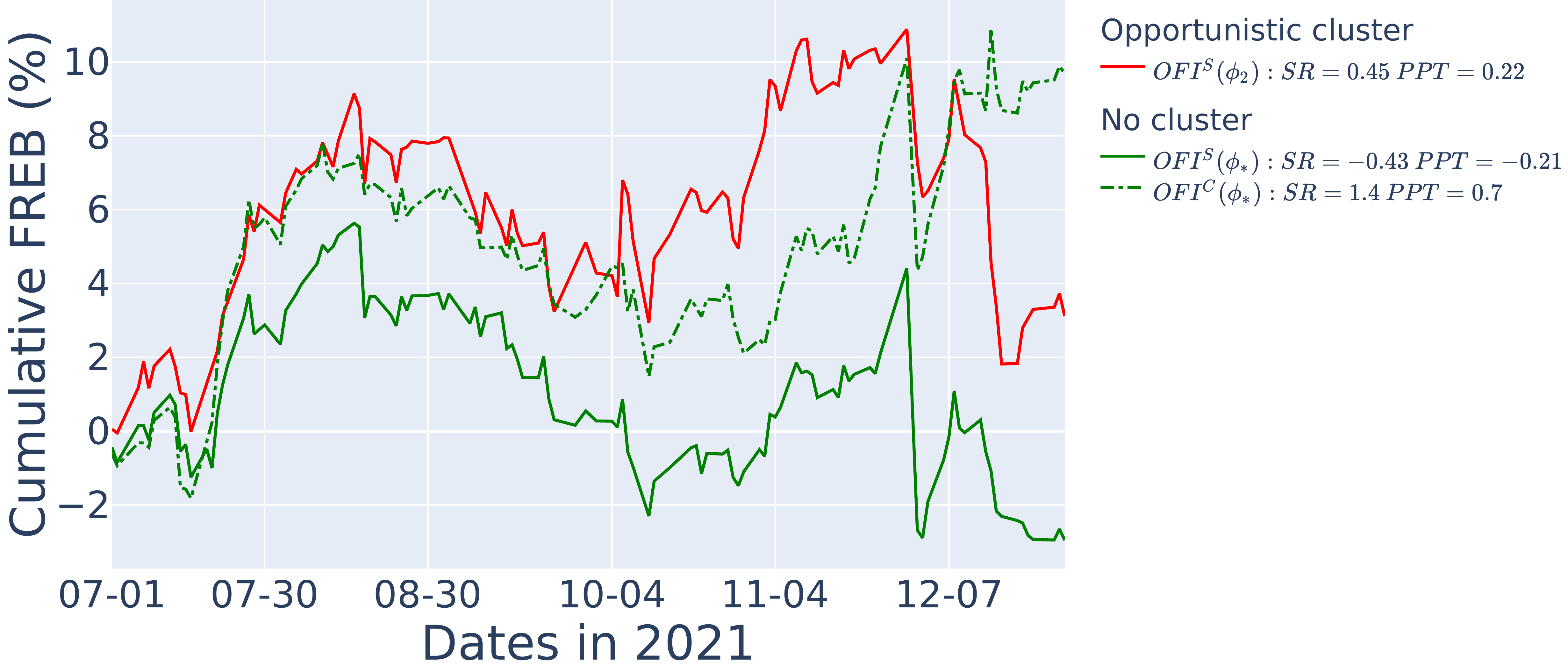} 
                    \\
\bottomrule
    \end{tabular}

\end{table}

\end{document}